\documentclass[fleqn,usenatbib]{mnras}

\usepackage{newtxtext,newtxmath}

\usepackage[T1]{fontenc}
\usepackage{ae,aecompl}
\usepackage{graphicx}	% Including figure files
\usepackage{amsmath}	% Advanced maths commands

\title[Physical properties of SMGs from AS2UDS]{An ALMA survey of the SCUBA-2 CLS UDS field: Physical properties of 707 Sub-millimetre Galaxies}

\author[Dudzevi\v{c}i\={u}t\.{e} et al.]{U.\,Dudzevi\v{c}i\={u}t\.{e}$^{1}$\thanks{E-mail: ugne.dudzeviciute2@durham.ac.uk},
Ian Smail$^{1}$,
A.\,M.\ Swinbank$^{1}$,
S.\,M.\ Stach$^{1}$,
O.\ Almaini$^{2}$,
\newauthor
E.\ da Cunha$^{3,4,5}$,
Fang~Xia An$^{6}$,
V.\ Arumugam$^{7}$,
J.\ Birkin$^{1}$,
A.\,W.\ Blain$^{8}$,
\newauthor
S.\,C.\ Chapman$^{9}$,
C.-C.\ Chen$^{10}$,
C.\,J.\ Conselice$^{2}$,
K.\,E.\,K.\ Coppin$^{11}$,
J.\,S.\ Dunlop$^{12}$,
\newauthor
D.\ Farrah$^{13,14}$,
J.\,E.\ Geach$^{11}$,
B.\ Gullberg$^{1}$,
W.\,G.\,Hartley$^{15}$,
J.\,A.\,Hodge$^{16}$,
\newauthor
R.\,J.\ Ivison$^{10,12}$,
D.\,T.\ Maltby$^{2}$,
D.\ Scott$^{17}$,
C.\,J.\ Simpson$^{18}$,
J.\,M.\ Simpson$^{1}$,
\newauthor
A.\,P.\ Thomson$^{19}$,
F.\ Walter$^{20}$,
J.\,L.\ Wardlow$^{21}$,
A.\ Weiss$^{22}$,
P.\ van der Werf$^{16}$
\\
$^{1}$ Centre for Extragalactic Astronomy, Department of Physics, Durham University, South Road, Durham DH1 3LE, UK\\
$^{2}$ School of Physics and Astronomy, University of Nottingham, University Park, Nottingham, NG7 2RD, UK \\
$^{3}$ International Centre for Radio Astronomy Research, University of Western Australia, 35 Stirling Hwy, Crawley, WA 6009, Australia\\
$^{4}$ Research School of Astronomy and Astrophysics, The Australian National University, Canberra, ACT 2611, Australia\\
$^{5}$ ARC Centre of Excellence for All Sky Astrophysics in 3 Dimensions (ASTRO 3D) \\
$^{6}$ Department of Physics and Astronomy, University of the Western Cape, Robert Sobukwe Road, 7535, South Africa\\
$^{7}$Institut de Radioastronomie Millim\'{e}trique, 300 rue de la Piscine, Domaine Universitaire, 38406 Saint Martin d'H\`{e}res, France\\
$^{8}$ Department of Physics and Astronomy, University of Leicester,University Road, Leicester LE1 7RH, UK\\
$^{9}$ Department of Physics and Atmospheric Science, Dalhousie University Halifax, NS B3H 3J5, Canada\\
$^{10}$ European Southern Observatory, Karl Schwarzschild Strasse 2, Garching, Germany\\
$^{11}$ Centre for Astrophysics Research, School of Physics, Astronomy and Mathematics, University of Hertfordshire, Hatfield AL10 9AB, UK\\
$^{12}$ Institute for Astronomy, University of Edinburgh, Royal Observatory, Blackford Hill, Edinburgh EH9 3HJ, UK\\
$^{13}$ Department of Physics and Astronomy, University of Hawaii, 2505 Correa Road, Honolulu, HI 96822, USA\\
$^{14}$ Institute for Astronomy, 2680 Woodlawn Drive, University of Hawaii, Honolulu, HI 96822, USA\\
$^{15}$ Department of Physics and Astronomy, University College London, Gower Street, London WC1E 6BT\\
$^{16}$ Leiden Observatory, Leiden University, P.O.\ box 9513, NL-2300 RA Leiden, The Netherlands\\
$^{17}$ Department of Physics and Astronomy, University of British Columbia, 6224 Agricultural Road, Vancouver, BC V6T 1Z1, Canada\\
$^{18}$ Gemini Observatory, Hilo, HI 96720, USA\\
$^{19}$ The University of Manchester, Oxford Road, Manchester, M13 9PL, UK\\
$^{20}$ Max-Planck-Institut f\"{u}r Astronomy, K\"{o}nigstuhl 17, 69117 Heidelberg, Germany\\
$^{21}$ Department of Physics, Lancaster University, Lancaster, LA1 4YB, UK\\
$^{22}$ Max-Planck-Institut f\"{u}r Radioastronomie, Auf dem H\"{u}gel 69 D-53121 Bonn, Germany}
\date{Accepted 2020 March 11. Received 2020 March 11; in original form 2019 September 30}

\pubyear{2020}

\begin{document}
\label{firstpage}
\pagerange{\pageref{firstpage}--\pageref{lastpage}}
\maketitle

\begin{abstract}
We analyse the physical properties of a large, homogeneously selected sample of ALMA-located sub-millimetre galaxies (SMGs). This survey, AS2UDS, identified 707 SMGs across the $\sim $\,1\,deg$^2$ field, including $\sim$\,17 per cent which are undetected at $K \gtrsim$\,25.7 mag.  We interpret their UV-to-radio data using {\sc magphys} and determine a median  redshift of $z$\,$=$\,2.61\,$\pm$\,0.08 (1-$\sigma$ range of $z =$\,1.8--3.4) with just $\sim$\,6 per cent at $z>$\,4. Our survey provides a sample of massive dusty galaxies at $z\gtrsim$\,1, with median dust and stellar masses of $M_{\rm d}$\,$=$\,(6.8\,$\pm$\,0.3)\,$\times$\,10$^{8}$\,M$_\odot$ (thus, gas masses of $\sim$\,10$^{11}$\,M$_\odot$) and $M_\ast$\,$=$\,(1.26\,$\pm$\,0.05)\,$\times$\,10$^{11}$\,M$_\odot$. We find no evolution in dust temperature at a constant far-infrared luminosity across $z\sim$\,1.5--4. The gas mass function of our sample increases to $z\sim$\,2--3 and then declines at $z>$\,3. The space density and masses of SMGs suggest that almost all galaxies with $M_\ast\gtrsim $\,3\,$\times$\,10$^{11}$\,M$_\odot$ have passed through an SMG-like phase. The redshift distribution is well fit by a model combining evolution of the gas fraction in halos with the growth of halo mass past a critical threshold of $M_{\rm h}$\,$\sim$\,6\,$\times$10$^{12}$\,M$_\odot$, thus SMGs may represent the highly efficient collapse of gas-rich massive halos. We show that SMGs are broadly consistent with simple homologous systems in the far-infrared, consistent with a centrally illuminated starburst. Our study provides strong support for an evolutionary link between the active, gas-rich SMG population at $z>$\,1 and the formation of massive, bulge-dominated galaxies across the history of the Universe.
\end{abstract}

\begin{keywords}
sub-millimetre: galaxies -- galaxies: high-redshift -- galaxies: starburst 
\end{keywords}

%%%%%%%%%%%%%%%%% BODY OF PAPER %%%%%%%%%%%%%%%%%%

%
%
\section{Introduction} \label{sec:intro}

Analysis of the relative brightness of the extragalactic background in the UV/optical and far-infrared/sub-millimetre suggest that around half of all of the star formation that has occurred over the history of the Universe was obscured by dust \citep[e.g.][]{1996puget}. This far-infrared/sub-millimetre emission is expected to primarily comprise the reprocessing of  UV emission from young, massive stars by dust grains in the interstellar medium (ISM) of distant galaxies, which is re-emitted in the form of far-infrared/sub-millimetre photons as the grains cool.   
Understanding the nature, origin, and evolution of this dust-obscured activity in galaxies is therefore crucial for obtaining a complete understanding of their formation and growth (see \citealt{2014casey} for a review). 

In the local Universe, the most dust-obscured galaxies are also some of the most actively star-forming systems:   ultra-luminous infrared galaxies \citep[ULIRGs;][]{1996sanders&mirabel} with star-formation rates of $\gtrsim$\,100\,M$_\odot$\,yr$^{-1}$. These radiate $\gtrsim$\,95\,per cent of their bolometric luminosity in the mid-/far-infrared as a result of strong dust obscuration of their star-forming regions. These galaxies have relatively faint luminosities in UV/optical wavebands, but
far-infrared luminosities of $L_{\rm IR} \geq$\,10$^{12} $\,L$_\odot$ and hence they are most easily identified locally through surveys in the far-infrared waveband (e.g.\, {\it IRAS} 60$\mu$m). It has been suggested that the high star-formation rates of ULIRGs arise from the concentration of massive molecular gas reservoirs (and thus, high ISM densities and strong dust absorption) in galaxies that are undergoing tidal interactions as a result of mergers \citep{1988sanders}. 

The far-infrared ($\gtrsim$100$\mu$m)  spectral energy distribution (SED) of the dust-reprocessed emission from ULIRGs can be roughly approximated by a modified blackbody.
The rapid decline in the brightness of the source at wavelengths 
beyond the SED peak on the Rayleigh-Jeans tail creates a  strong negative $k$-correction for observations of this population at high redshifts \citep{1991franceschini,1993blain&longair}.  Hence, a dusty galaxy with a fixed far-infrared luminosity and temperature will have an almost constant apparent flux density in the sub-millimetre waveband (which traces rest-frame emission beyond the redshifted peak of the SED) from $z \sim$\,1 to $z \sim$\,7 (see \citealt{2014casey}). As a result, surveys in the sub-millimetre waveband in principle allow us to construct luminosity-limited samples of obscured, star-forming galaxies over a very wide range of cosmic time, spanning the expected
peak activity in galaxy formation at $z\gtrsim$\,1--3 \citep[e.g.][]{2005chapman,2012casey,2013weiss,2014simpson,2017brisbin,2016strandet}.   

Sub-millimetre galaxies (SMGs) with 850-$\mu$m flux densities of $S_{850}\gtrsim$\,1--10\,mJy were first uncovered over 20 years ago using the atmospheric window around 850-$\mu$m with the SCUBA instrument on the James Clerk Maxwell Telescope (JCMT) \citep{1997smail,1998barger,1998hughes,1999eales}. Subsequent studies have suggested they represent a population of particularly dusty, high-infrared luminosity systems ($>$\,10$^{12}$\,L$_\odot$) that are typically found at high redshift ($z\sim$\,1--4). They have large gas reservoirs \citep{1998frayer,2005greve,2013bothwell}, stellar masses of the order of 10$^{11}$\,M$_\odot$ and can reach very high star-formation rates up to (and in some cases in excess of) $\sim$\,1,000\,M$_\odot$\,yr$^{-1}$. SMGs have some observational properties that appear similar to those of local ULIRGs, such as high far-infrared luminosities and star-formation rates; however, their space densities are a factor of $\sim$\,1,000\,$\times$ higher than the comparably luminous local population \citep[e.g.][]{1997smail,2005chapman,2014simpson}.   
Thus, in contrast to the local Universe, these luminous systems are a non-negligible component of the star-forming population at high redshift. Very wide-field surveys with the SPIRE instrument on {\it Herschel} have traced this dusty luminous population, using very large samples, to lower redshifts and lower far-infrared luminosities \citep[e.g.][]{2016bourne}. However, the modest angular resolution of {\it Herschel}/SPIRE and resulting bright confusion limit, at longer far-infrared wavelengths limits its ability to select all but the very brightest (unlensed) sources at the era of peak activity in the obscured population at $z\gtrsim$\,1--2 \citep{2011symeonidis}. Such low-resolution far-infrared-selected samples are also more challenging to analyse owing to the ambiguities in source identification that results from ground-based follow-up to locate counterparts, which is necessarily undertaken at longer wavelengths than the original
far-infrared selection.    

With such high star-formation rates, SMGs can rapidly increase their (apparently already significant) stellar masses on a timescale of just $\sim$\,100\,Myr \citep[e.g.][]{2013bothwell}.    
High star-formation rates and high stellar masses of this population, along with the high metallicities suggested by the significant dust content, have therefore been used to argue that they may be an important phase in the formation of the stellar content of the most massive galaxies in the Universe, being the progenitors of local luminous spheroids and elliptical galaxies \citep{1999lilly,2005chapman,2014simpson}. There have also been suggestions of an evolutionary link with quasi-stellar objects (QSOs) \citep[e.g.][]{2006swinbank,2008wall,2012simpson,2012hickox} due to the similarities in their redshift distributions. More recently these systems have been potentially linked to the formation of compact quiescent galaxies seen at $z\sim$\,1--2 \citep[e.g.][]{2012whitaker, 2014simpson,2014toft} as a result of their short gas depletion timescales. This connection has been strengthened by recent observations in the rest-frame far-infrared that suggest very compact extents of the star-forming regions \citep[][]{2014toft,2015ikarashi,2015simpsona,2019gullberg}. Thus several lines of evidence suggest that SMGs are an important element for constraining the models of massive galaxy formation and evolution.

The pace of progress of our understanding of the nature and properties of the SMG population has accelerated in the last five years, owing to the commissioning of the
Atacama Large Millimetre/Submillimetre Telescope (ALMA). ALMA has enabled high-sensitivity ($\ll$\,1\,mJy rms) and high-angular-resolution ($\lesssim$\,1$''$ FWHM) observations in the sub-/millimetre wavebands of samples of dust-obscured galaxies at high redshifts, including SMGs. In the first few years of operations, ALMA has been used to undertake a number of typically deep continuum surveys of small contiguous fields \citep{2016walter,2016hatsukade,2017dunlop, 2018umehata, 2018hatsukade,2018franco,2018munoz}, with areas of 10s of arcmin$^2$ (including lensing clusters and proto-cluster regions).  These small field studies typically contain sources at flux limits of $S_{870}\simeq$\,0.1--1\,mJy (corresponding to star-formation rates of $\sim$\,10--100\,M$_\odot$\,yr$^{-1}$ or far-infrared luminosities of $\sim$\,(0.5--5)$\times$10$^{11}$\,L$_\odot$) and so provide a valuable link between the bright SMGs seen in the panoramic single-dish surveys and the populations of typically less-actively star-forming galaxies studied in UV/optical-selected surveys.   However, owing to their small areas they do not contain more than a few examples of the brighter SMGs. To efficiently study the brighter sources requires targeted follow-up of sources from panoramic single-dish surveys. Hence, ALMA 
has also been employed to study the dust continuum emission from samples of $\lesssim$\,100 SMGs selected from single-dish surveys at 870 or 1100\,$\mu$m \citep[e.g.][]{2013hodge,2017brisbin,2018cowie}.  The primary goal of these studies has been to first precisely locate the galaxy or galaxies responsible for the sub-millimetre emission in the (low-resolution) single-dish source and to then understand their properties \citep[e.g.][]{2014simpson, 2017brisbin}.  

The first ALMA follow-up of a single-dish sub/millimetre survey was the ALESS survey \citep{2013karim,2013hodge} of a sample of 122 sources with $S_{870}\geq$\,3.5\,mJy selected from the 0.25\,deg$^2$ LABOCA 870-$\mu$m map of the Extended {\it Chandra} Deep Field South (ECDFS) by \cite{2009weiss}. The multi-wavelength properties of 99 SMGs from the robust main sample were analysed using the {\sc magphys} SED modelling code by \cite{2015dacunha} (see also the {\sc magphys} analysis of a similar-sized sample of 1.1-mm selected SMGs in the COSMOS field by \citealt{2017miettinen}). This approach of using a single consistent approach to model the UV/optical and far-infrared emission provides several significant benefits for these dusty and typically very faint galaxies, over previous approaches of independently modelling the UV/optical and far-infrared emission \citep[e.g.][]{2008clements,2018cowie}. In particular, the use in {\sc magphys} of an approximate energy balance formulation between the energy absorbed by dust from the UV/optical and that re-emitted in the far-infrared provides more reliable constraints on the photometric redshifts for the SMGs \citep[e.g.][]{2015dacunha,2017miettinen}. This is particularly critical in order to derive complete and unbiased redshift distributions for flux-limited samples of SMGs, as $\sim$\,20 per cent of SMGs are typically too faint to be detected at wavelengths shortward of the near-infrared \citep[e.g.][]{2014simpson,2018franco} and hence are frequently missing from such analyses.  The energy balance coupling is also expected to improve the derivation of physical properties of these optically faint systems, such as stellar masses and dust attenuation, which are otherwise typically poorly constrained \citep{2011hainline,2011dunlop}.

While the studies by \cite{2015dacunha} and \cite{2017miettinen} have provided improved constraints on the physical parameters of samples of $\sim$\,100 SMGs, the modest size of these samples does not allow for robust analysis of the evolutionary trends in these parameters within the population \citep{2015dacunha}, or to study sub-sets of SMGs, such as the highest-redshift examples \citep{2009coppin,2012swinbank} or those that show signatures of both star-formation and AGN activity \citep{2013wang}.
To fully characterize the population of SMGs and interpret their role in the overall galaxy evolution requires a large, homogeneously selected sample with precisely located sub-millimetre emission from sub/millimetre interferometers.  We have therefore just completed an ALMA study of a complete sample of 716 single-dish sources selected from the SCUBA-2 Cosmology Legacy Survey (S2CLS) 850-$\mu$m map of the UKIDSS UDS field \citep[S2CLS is presented in][]{2017geach}. This targetted ALMA study  -- called AS2UDS  \citep{2019stach} --  used sensitive 870-$\mu$m continuum observations  obtained in Cycles 1, 3, 4 and 5 to precisely locate (to within $\ll$\,0.1$''$) 707 SMGs across the $\sim$\,0.9\,deg$^{2}$ S2CLS--UDS field.  AS2UDS provides the largest homogeneously-selected sample of ALMA-identified SMGs currently available, $\sim$\,6\,$\times$ larger than the largest existing ALMA surveys \citep{2013hodge,2017miettinen}.  

In this paper, we construct the UV-to-radio SEDs of our sample of 707 ALMA-identified SMGs from the AS2UDS survey using a physically motivated model, {\sc magphys} \citep{2015dacunha,2019battisti}. We use the model to interpret the SEDs and so investigate the rest-frame optical (stellar) and infrared (dust) properties of the SMGs.  This sample allows us to both improve the statistics to search for trends within the population \citep[e.g.][]{2018stach,2019stach} and to understand the influence of selection biases on our results and the conclusions of previous studies.  
With a statistically well-constrained and complete understanding of their redshift distribution and physical properties, we are able to address what place the SMG phase takes in the evolution of massive galaxies. Through our paper, we compare our results to samples of both local ULIRGs and near-infrared selected high-redshift field galaxies, which we analyse in a consistent manner to our SMG sample to avoid any systematic uncertainties affecting our conclusions. 

Our paper is structured as follows. In \S\ref{obs} we describe the multi-wavelength observations of the AS2UDS SMGs. In \S\ref{magphys} we describe the SED fitting procedure using {\sc magphys} and test its robustness. We present the results including the redshift distribution, multi-wavelength properties and evolutionary trends of the whole AS2UDS SMG population in \S\ref{analysis}. We discuss the implications of our results in \S\ref{discussion} and present our conclusions in \S\ref{conc}. Unless stated otherwise, we use $\Lambda$CDM cosmology with with $H_0 $\,$=$\,70\,km\,s$^{-1}$\,Mpc$^{-1}$, $\Omega_\Lambda$\,$=$\,0.7, $\Omega_m $\,$=$\,0.3. The AB photometric magnitude system is used throughout.

\section{Observations and sample selection} \label{obs}

In this section, we describe the multi-wavelength photometric data we
use to derive the SED from the UV-to-radio wavelengths for each galaxy in our sample.  From these SEDs, we
aim to derive the physical properties of each SMG (such as their
photometric redshift, star-formation rate, stellar, dust and gas
masses).  To aid the interpretation of our results, we also exploit
the $\sim$\,300,000 $K$-selected field galaxies in the UKIDSS UDS (Almaini et al.\,in
prep.).  We measure the photometry and SEDs for the field galaxies and
SMGs in a consistent manner and describe the sources of these data
and any new photometric measurements below.

\subsection{ALMA}

A detailed description of the ALMA observations, data reduction and
construction of the catalogue for the SMGs in our sample can be found in
\cite{2019stach}.  Briefly, the AS2UDS (defined in \S \ref{sec:intro}) comprises an ALMA follow-up survey of a complete sample of
716 SCUBA-2 sources that are detected at $>$\,4-$\sigma$
($S_{850}\geq$\,3.6\,mJy) in the S2CLS map of the UKIDSS UDS field
\citep{2017geach}. The S2CLS map of the UDS covers an area of
0.96\,deg$^{2}$ with noise level below
1.3\,mJy and a median depth of
$\sigma_{850}$\,=\,0.88\,mJy\,beam$^{-1}$. All 716 SCUBA-2 sources detected in the map were observed
in ALMA Band 7 (344\,GHz or 870\,$\mu$m) between Cycles 1, 3, 4 and 5 (a pilot study of 27 of
the brightest sources observed in Cycle 1 is discussed in
\citealt{2015simpsonb,2017simpson}).  Due to configuration changes
between cycles, the spatial resolution of the data varies in range
0.15--0.5$''$ FWHM, although all of the maps are tapered to 0.5$''$
FWHM for detection purposes \citep[see][for details]{2019stach}.  The
final catalogue contains 708 individual ALMA-identified SMGs spanning
$S_{870}$\,=\,0.6--13.6\,mJy ($>$4.3$\sigma$) corresponding to a 2 per cent false-positive rate.  We remove one bright, strongly lensed source \citep{2011ikarashi} from our analysis and the remaining 707 ALMA-identified SMGs are the focus of this study of the physical properties.

\subsection{Optical and near-/mid-infrared imaging}

\subsubsection{Optical $U$-band to $K$-band photometry}

At the typical redshift of SMGs, $z\sim$\,2.5 \citep[e.g.][]{2005chapman,2014simpson,2017danielson,2017brisbin}, the observed
optical to mid-infrared corresponds to the rest-frame UV/optical/near-infrared,
which is dominated by the (dust-attenuated) stellar continuum
emission, emission lines, and any possible AGN emission.  The
rest-frame UV/optical/near-infrared also includes spectral features that are
important for deriving photometric redshift, in particular, the
photometric redshifts have sensitivity to the Lyman break, Balmer
and/or 4000\AA\ break and, the (rest-frame) 1.6-$\mu$m stellar ``bump''.

To measure the optical/near-infrared photometry for the galaxies in
the UDS, we exploit the panchromatic photometric coverage of this field.
In particular, we utilise the UKIRT Infrared Deep Sky Survey
\citep[UKIDSS:][]{2007lawrence} UDS data release 11 (UKIDSS DR11),
which is a $K$-band selected photometric catalogue (Almaini et al., in
prep.) covering an area of 0.8\,deg$^2$ with a 3-$\sigma$
point-source depth of $K$\,$=$\,25.7\,mag (all photometry in this
section is measured in 2$''$ diameter apertures and has been aperture
corrected, unless otherwise stated).  This $K$-band selected catalogue
has 296,007 sources, of which more than 90 per cent are flagged as galaxies with reliable $K$-band photometry. For any subsequent analysis, we restrict our analysis to 205,910 sources that have no contamination flags.
The UKIDSS survey imaged the UDS field with the UKIRT WFCAM camera in $K$, $H$ and
$J$ bands and the DR11 catalogue also includes the matched photometry in $J$-
and $H$-band to 3-$\sigma$ depths of $J$\,$=$\,26.0 and $H$\,$=$\,25.5.

In addition, $Y$-band photometry was also obtained from the VISTA/VIDEO survey, which has a 3-$\sigma$ depth of 25.1\,mag and $BVRi'z'$-band photometry was obtained from Subaru/Suprimecam imaging,
which has 3-$\sigma$ depths of 28.2, 27.6, 27.5, 27.5, and 26.4\,mag,
respectively. Finally, $U$-band photometry of the UDS field from the
CFHT/Megacam survey is also included in the DR11 catalogue. This
$U$-band imaging reaches a 3-$\sigma$ point-source depth of 27.1\,mag.

To derive the photometry of the ALMA SMGs in the
optical/near-infrared, first, we align the astrometry between the
UKIDSS DR11 catalogue with the ALMA astrometry by matching the positions
of the ALMA SMGs to the $K$-band catalogue, identifying and removing an
offset of $\Delta$RA\,=\,0.1$''$ and $\Delta$Dec\,=\,0.1$''$ in the $K$-band.
We find that 634/707 SMGs lie within the deep regions of the $K$-band image, after excluding regions masked due to noisy edges, artefacts, and bright stars. The two catalogues are then matched using a radius of 0.6$''$ (which has a
false-match rate of 3.5 per cent; see \citealt{2018an} for details). This results in 526/634 SMGs with $K$-band detections (83 per cent). We note that 43 of these sources are within a $K$-band region flagged with possibly contaminated photometry; however, the inclusion of these sources in our analysis does not change any of our conclusions of this study, thus we retain then and flag then in our catalogue. 

Our detection fraction is comparable to, but slightly higher than, the fraction identified in smaller samples of SMGs in other fields, which is likely due to the very deep near-infrared coverage available in the UDS.  
For example, in the ALMA survey of the ECDFS, ALESS -- \cite{2014simpson} show
that 61\,/\,99 (60 per cent) of the ALMA SMGs have $K$-band counterparts to a limit of $K$\,$=$\,24.4.  This is significantly lower than the detection
rate in our UDS survey, although cutting our UDS catalogue at the same
$K$-band limit as the ECDFS results in a detected fraction of 68 per cent.
Similarly, 65 per cent of the ALMA SMGs in the CDFS from \cite{2018cowie}
(which have a median 870-$\mu$m flux of $S_{870}$\,$=$\,1.8\,mJy)
are brighter than $K =$\,24.4.  Finally, \cite{2017brisbin}
identify optical counterparts to 97\,/\,152 (64 per cent) of ALMA-identified
SMGs from a Band 6 (1.2\,mm) survey of AzTEC sources using the public
COSMOS2015 catalogue \citep{2016laigle}, which is equivalent to $K \lesssim$\,24.7,
for the deepest parts.
Thus, our detection rate of 83 per cent of ALMA SMGs with $K$-band
counterparts is consistent with previous surveys but also demonstrates
that even with extremely deep near-infrared imaging, a significant number (17 per cent
or 108 galaxies) are faint or undetected in the near-infrared at $K \geq$\,25.7.

Since SMGs are dominated by high redshift, dusty highly-starforming galaxies, their
observed optical/near-infrared colours are typically red \citep[e.g.][]{1999smail,2004smail}, and so the detection rate as a function of wavelength
drops at shorter wavelengths, reaching just 26 per cent in the $U$-band
(Table \ref{table:photom}).  We will return to a discussion of the detected fraction of
SMGs as a function of wavelength, their colours, and implications on
derived quantities in $\S$\,\ref{der_props}.

\begin{figure*}
  \includegraphics[width=\textwidth]{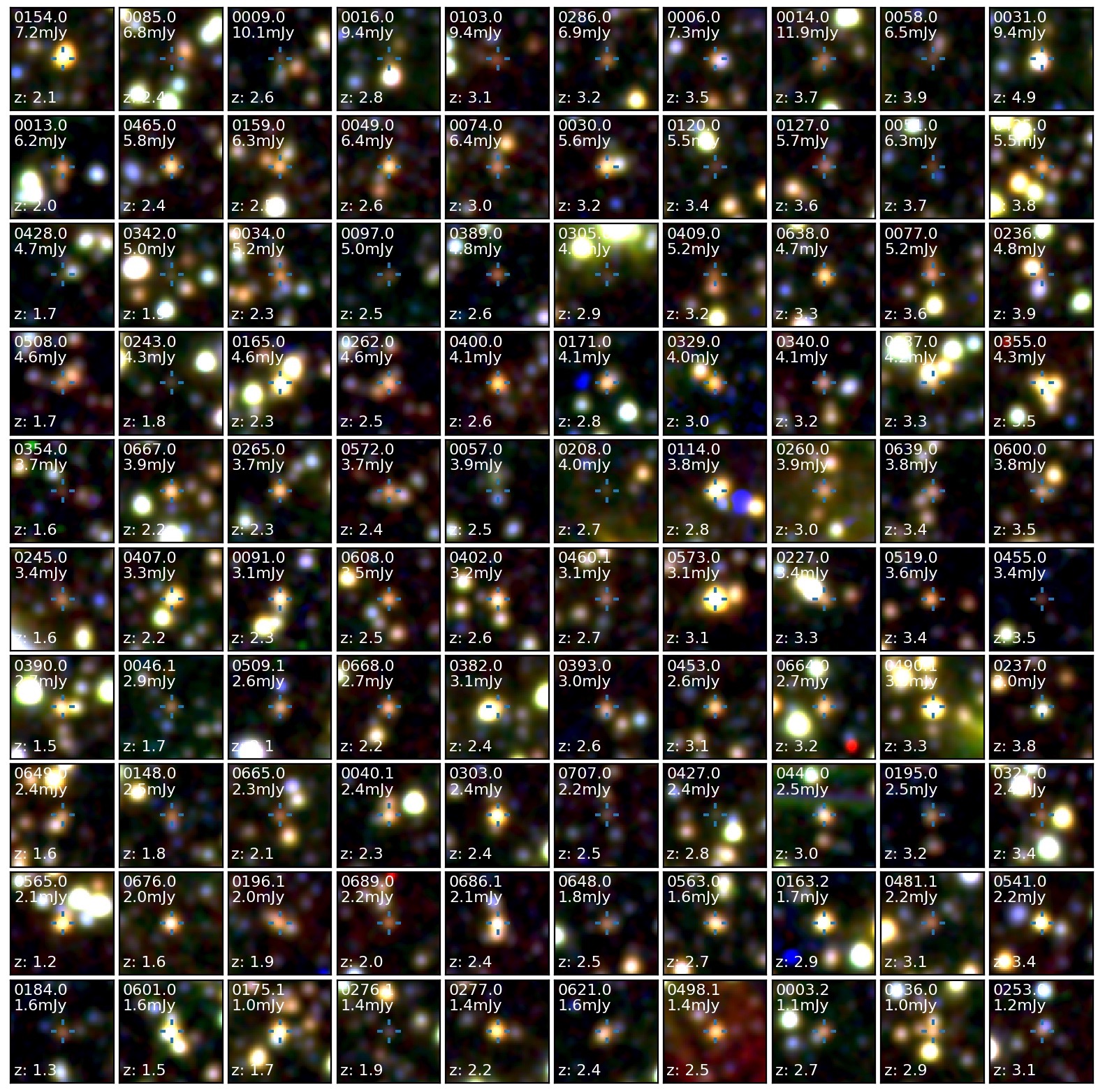}
  \caption{Examples of 100 of the AS2UDS ALMA-identified SMGs from our sample. The 25$'' \times$\,25$''$ ($\sim$\,200-kpc square at their typical redshifts) colour images are composed of $K$, IRAC 3.6$\mu$m and IRAC 4.5 $\mu$m bands with the ALMA position of the source given by the open cross. The sources are selected to be representative of the near-infrared properties of the full sample: thumbnails are ranked in deciles of flux (each row) and deciles of $z_{\mathrm{phot}}$ within each flux range (each column). We see that SMGs are in general redder than the neighbouring field galaxies. There is a weak trend for SMGs to become fainter and/or redder with redshift, but there is no clear trend of observed properties with $S_{870}$ flux density.}
  \label{fig:thumbs}
\end{figure*}

\subsubsection{\it{Spitzer} IRAC \& MIPS observations} \label{irac}

Next, we turn to the mid-infrared coverage of the UDS, in particular from
\emph{Spitzer} IRAC and MIPS observations.  At these wavelengths, the observed
3.6--8.0\,$\mu$m emission samples the rest-frame near-infrared at the
expected redshifts of the SMGs. These wavelengths are less dominated by the
youngest stellar populations, and significantly less affected by dust
than the rest-frame optical or UV.  Observations of the UDS in the mid-infrared were
taken with IRAC onboard the \emph{Spitzer} telescope as part of the
\emph{Spitzer} Legacy Program (SpUDS;\,PI: J.\ Dunlop).

We obtained reduced SpUDS images of the UDS from the \textit{Spitzer}
Science Archive.  These IRAC observations at 3.6, 4.5, 5.8 and
8.0\,$\mu$m reach 3-$\sigma$ depths of 23.5, 23.3, 22.3
and 22.4\,mag, respectively.  The astrometry of all four IRAC images
was aligned to the ALMA maps by stacking the IRAC thumbnails of the
ALMA positions of 707 AS2UDS sources and corrections in RA/Dec of
(+0.00$''$, +0.15$''$),
(+0.08$''$,+0.12$''$), (+0.08$''$,+0.00$''$) and
(+0.60$''$,$-$0.08$''$) were applied to the 3.6, 4.5, 5.8 and 8.0\,$\mu$m
images, respectively.  To measure the photometry, and minimise the
effect of blending, we extract 2$''$-diameter aperture photometry for all of
the ALMA SMGs, as well as for all 205,910 galaxies in the
UKIDSS DR11 catalogue, and calculate aperture corrections to total
magnitudes from point sources in the images.  The UKIDSS DR11 catalogue
contains aperture-corrected magnitudes measured in the  3.6- and 4.5-$\mu$m bands and we
confirm our photometry at these wavelengths by comparing the
respective magnitudes, with relative offsets of just
$\Delta$[3.6]\,/\,[3.6]$_{\mathrm{DR11}}$\,=\,0.001$_{-0.005}^{+0.007}$
and
$\Delta$[4.5]\,/\,[4.5]$_{\mathrm{DR11}}$\,=\,0.002$_{-0.003}^{+0.009}$.

\begin{table}
\caption{Photometric coverage and detection fractions for AS2UDS SMGs in representative photometric bands.}
\begin{tabular}{|p{0.12\linewidth}|p{0.14\linewidth}|p{0.14\linewidth}|p{0.14\linewidth}|p{0.2\linewidth}|}

\hline
\hline
Band & $N_{\rm covered}$ & $N_{\rm detected}$ & \%$_{\rm detected}$ &  Depth (3-$\sigma$) \\ \hline \hline

$U$ & 634   & 162  & 26  &  27.1\,AB   \\
$V$ & 590   & 330  & 56  &  27.6\,AB \\
$K$ & 634   & 526  & 83  & 25.7\,AB \\
3.6\,$\mu$m & 644 & 580$^1$ & 90$^2$ & 23.5\,AB\\
24\,$\mu$m & 628 & 304 & 48 & 60\,$\mu$Jy \\
350\,$\mu$m & 707 & 417 & 59 & 8.0\,mJy \\
1.4\,GHz & 705 & 272 & 39 &  18\,$\mu$Jy \\ \hline
\label{table:photom}
\end{tabular}
\footnotesize{ 
$^1$ Including 109 potentially contaminated sources (see \S \ref{irac})\\
 $^2$ 73\% if excluding 109 potentially contaminated sources\\
Notes: $N_{\rm covered}$ -- number of SMGs covered by imaging; $N_{\rm detected}$ -- number of SMGs detected above 3-$\sigma$;  \%$_{\rm detected}$ -- per centage of total sample detected.
}
\end{table}

Due to the relatively large PSF of the IRAC images (typically $\sim$\,2$''$ FWHM), blending with nearby sources is a
potential concern (see Fig.~\ref{fig:thumbs}).  We, therefore, identify all of the ALMA SMGs that have a second, nearby $K$-band detected, galaxy within 2.5$''$ and
calculate the possible level of contamination assuming that the flux ratio
of the ALMA SMG and its neighbour is the same in the IRAC bands as observed in the higher-resolution $K$-band images. This is conservative as the SMGs are expected to be typically redder than any contaminating field galaxies. For
any ALMA SMG, if the contamination from the nearby source is likely
to be more than 50 per cent of the total flux, the respective IRAC magnitudes are treated as 3-$\sigma$ upper limits. This transformation of detected fluxes into upper limits affects 109 sources.

From the photometry of the ALMA SMGs in the IRAC bands, we determine that 581\,/\,645 or 90 per cent of the  SMGs covered by IRAC are detected at 3.6\,$\mu$m,
or 73 per cent when we apply the
conservative blending criterion from above.  The increased fraction of the sample
that are detected in the IRAC bands, compared to $K$-band, most likely
reflects the (rest-frame) 1.6-$\mu$m stellar ``bump'' that is redshifted
to $\gtrsim$\,3\,$\mu$m for an SMG at $z\gtrsim$\,1.  We will return to
a discussion of the mid-infrared colours in \S\,\ref{photometry}.

To demonstrate the typically red colour of the SMGs (in particular
compared to the foreground field galaxy population), in Fig.~\ref{fig:thumbs} we show
colour images (composed of $K$, IRAC 3.6-$\mu$m and 4.5-$\mu$m bands)
 for 100 representative AS2UDS SMGs  ranked in terms of $S_{870}$ and photometric redshift (see \S\,\ref{redshift} for the
determination of the photometric redshifts). This figure demonstrates
that SMGs generally have  redder near-/mid-infrared colours than  neighbouring field
galaxies and also that on average higher-redshift SMGs are fainter and/or redder
in the near-infrared bands than low redshift ones for each of the ALMA
flux bins. We see no strong trends in observed properties with 870-$\mu$m flux density in any redshift bin.

Mid-infrared observations of the UDS were also taken at 24\,$\mu$m with
the Multiband Imaging Photometer (MIPS) on board \emph{Spitzer} as
part of SpUDS.  The 24-$\mu$m emission provides useful constraints on
the star formation and AGN content of bright SMGs since at the typical
redshift of our sample, the filter samples continuum emission
from heated dust grains. This spectral region also includes broad emission features associated with
polycyclic aromatic hydrocarbons (PAHs) -- the most prominent of which
appear at rest-frame 6.2, 7.7, 8.6, 11.3, and 12.7\,$\mu$m, as well as absorption by
amorphous silicates centred at 9.7 and 18\,$\mu$m \citep{2008pope,2009menendez}.  This MIPS 24-$\mu$m
imaging is also employed to provide a constraint on the
positional prior catalogue that is used to deblend the \emph{Herschel}
far-infrared maps \citep[e.g.][]{2012roseboom,2013magnelli,2014swinbank}.  We
obtained the reduced SpUDS\,/\,MIPS 24-$\mu$m image from the NASA
Infrared Astronomy Archive.  This imaging covers the entire UDS survey
area and reaches a 3-$\sigma$ (aperture corrected) limit of 60\,$\mu$Jy.
From the 24-$\mu$m image, we identify $\sim$\,35,000 sources, and
cross-matching the $>$3-$\sigma$ detections in the 24-$\mu$m catalogue
with our ALMA catalogue with a 2$''$ matching radius, we determine that 48 per cent of the SMGs are detected.  This detected
fraction is also consistent with that of other fields with similar ALMA and MIPS
coverage (e.g.\ 41 per cent in ALESS from \citealt{2014simpson}).

\subsection{Far-infrared and Radio Imaging}

\subsubsection{\it{Herschel} SPIRE \& PACS observations}
 
To measure reliable far-infrared luminosities for the ALMA SMGs, we
exploit observations using the Spectral and Photometric Imaging
Receiver (SPIRE) and the Photodetector Array Camera and Spectrometer
(PACS) on board the {\it Herschel} Space Observatory.  These observations
were taken as part of the \emph{Herschel} Multi-tiered Extragalactic
Survey (HerMES; \citealt{2012oliver}) and cover the observed wavelength range from
100--500\,$\mu$m.  These wavelengths are expected to span the dust-peak
of the SED, which (in local ULIRGs) peak
around 100\,$\mu$m, corresponding to a characteristic dust temperature
of $T_{\rm d}\simeq$\,35\,K \citep[e.g.][]{2013symeonidis,2018clements}.  At
$z \sim$\,2.5, the dust SED is expected to peak around an observed wavelength of 350\,$\mu$m
(e.g.\ see \citealt{2014casey} for a review).

Due to the coarse resolution of the \emph{Herschel}\,/\,SPIRE maps
($\sim$\,18$''$, 25$''$ and 36$\arcsec$ FWHM at 250, 350 and
500\,$\mu$m, respectively), we need to account for the effect of source
blending \citep{2012roseboom,2013magnelli}.  We, therefore, follow the same procedure as \cite{2014swinbank}.
Briefly, the ALMA SMGs, together with \emph{Spitzer}\,/\,MIPS
24-$\mu$m and 1.4-GHz radio sources, are used as positional priors in the deblending of the SPIRE maps. A Monte Carlo algorithm is used to deblend the SPIRE maps by fitting the
observed flux distribution with beam-sized components at the position
of a given source in the prior catalogue.  To avoid ``over-blending''
the method is first applied to the 250-$\mu$m data, and only sources
that are either (i) ALMA SMGs, or (ii) detected at $>$\,2-$\sigma$ at
250-$\mu$m are propagated to the prior list for the 350-$\mu$m
deblending. Similarly, only the ALMA SMGs and/or those detected at
$>$\,2-$\sigma$ at 350\,$\mu$m are used to deblend the 500-$\mu$m map.  The
uncertainties on the flux densities (and limits) are found by attempting
to recover fake sources injected into the maps (see \citealt{2014swinbank} for details), and the typical 3-$\sigma$ detection limits are
7.0, 8.0 and 10.6\,mJy at 250, 350 and 500\,$\mu$m
respectively.  The same method is applied to the PACS 100- and
160-$\mu$m imaging, with the final 3-$\sigma$ depths of 5.5\,mJy at
100\,$\mu$m and 12.1\,mJy at 160\,$\mu$m.

Given the selection of our sources at 870\,$\mu$m, the fraction of ALMA
SMGs that are detected in the PACS and/or SPIRE bands is a strong
function of 870-$\mu$m flux density, but we note that 69 per cent (486/707) of the
ALMA SMGs are detected in at least one of the PACS or SPIRE bands.
This will be important in \S\,\ref{analysis} when deriving useful constraints on the far-infrared luminosities and dust temperatures.

\begin{figure*}
  \includegraphics[width=\textwidth]{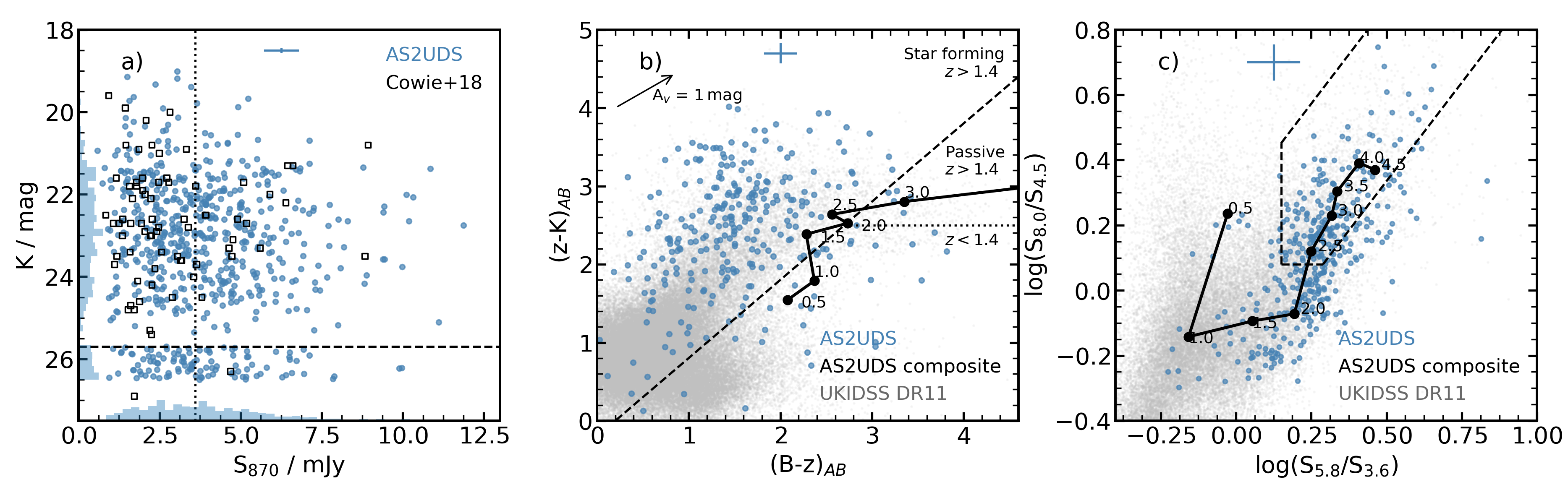}
  \caption{Distributions of observed magnitudes and colours of the SMGs from AS2UDS. 
  \textbf{(a)} $K$-band magnitude versus $S_{870}$ flux density. The dashed line shows the $K$-band 3-$\sigma$ limit of $K$\,$=$\,25.7 and the dotted line indicates the flux limit of the parent SCUBA-2 survey at $S_{870}$\,$=$\,3.6\,mJy. There are 526 $K$-band detections of SMGs and we plot the 108 limits scattered below the $K$-band limit. The  histograms show the $K$-band magnitude distribution as the ordinate and $S_{870}$ flux density distribution as the abscissa. For comparison, we also show the \protect\cite{2018cowie} sample from CDFS, which covers a similar parameter range. No strong correlation of 870-$\mu$m flux density and $K$-band magnitude is observed, but we highlight that we see a two order of magnitude range in $K$-band brightness at a fixed 870-$\mu$m flux density. 
  \textbf{(b)} $(B-z)$ versus $(z-K)$ colour-colour diagram for 290 SMGs with detections in all three bands and the $BzK$ classification regions. We stress that these are typically the brighter and bluer examples  and so are not representative of the full population. The placement of the sources on the diagram suggests that the majority (253/290) of these SMGs are high-redshift star-forming galaxies, most of which are significantly redder than the field population. The reddening vector for one magnitude of extinction in the $V$-band is plotted in the top left. 
  The solid line shows the track predicted by the composite SMG SED  track at increasing redshift (labelled). We see that the average colours of SMGs lies close to the classification boundary and so it is likely that fainter and redder SMGs would be misclassified using the $BzK$ colours. 
  \textbf{(c)} IRAC colour-colour diagram for 388 SMGs detected in all four IRAC bands. The dashed line indicates the IRAC colour criteria for AGN selection (up to a redshift of $z \sim$\,2.5) from \protect\cite{2012donley}. The solid line shows the composite SED as a function of redshift (labelled). We see
that a large fraction of SMGs have colours suggestive of  AGN, but the majority of these lie at
too high redshifts ($z\gtrsim$\,2.5) for the reliable application of this classification criterion -- with their power-law like IRAC colours resulting from the redshifting of the 1.6-$\mu$m bump longward of
the 5.8-$\mu$m passband. The field galaxies are also plotted (in grey) and it is clear that SMGs have significantly redder colours, with the bulk  of the field sample falling  off the bottom left corner of the plot. The average error is shown in the top of each panel.}
  \label{fig:obs_obs}
\end{figure*}

In terms of the field galaxies, just 3.6 per cent of the $K$-band sample have a
MIPS 24-$\mu$m counterpart, and of these only 2,396 (out of a total of
205,910 galaxies in DR11) are detected at 250\,$\mu$m, with 1,497 and
500 detected at 350\,$\mu$m and 500\,$\mu$m, respectively.  Thus the majority of the field population
are not detected in the far-infrared (in contrast to the ALMA SMGs,
where the majority of the galaxies are detected).

\subsubsection{VLA 1.4\,GHz Radio observations}

Finally, we turn to radio wavelengths.  Prior to ALMA, high-resolution
($\sim$\,1$''$) radio maps had often been employed  to identify likely
counterparts of single-dish sub-millimetre sources \citep[e.g.][]{1998ivison}.  Although the radio
emission does not benefit from the  negative $k$-correction
experienced in the sub-millimetre waveband, the lower-redshift ($z\lesssim$\,2.5)
ALMA SMGs tend to be detectable as $\mu$Jy radio sources due to the
strong correlation between the non-thermal radio and far-infrared
emission in galaxies \citep[e.g.][]{2001yun,2002ivison,2007ivison,2007vlahakis,2011biggs,2013hodge}. The standard explanation of this relationship is
that both the far-infrared emission and the majority of the radio
emission traces the 
same population of high-mass stars ($\gtrsim$\,5\,M$_\odot$).  These
stars both heat the dust (which then emits far-infrared emission) and
produce the relativistic electrons responsible for synchrotron
radiation when they explode as supernovae \citep[e.g.][]{1985helou,1992condon}.
However, the lack of a negative $k$-correction in the
radio waveband means that at higher redshifts ($z\gtrsim$\,2.5), where a large fraction of the SMGs lie,
their radio flux densities are often too faint to be detectable, for example, \cite{2013hodge} show that up to 45 per cent of ALMA SMGs in their ALESS survey are not
detected at 1.4\,GHz.

The UDS was imaged at 1.4\,GHz with the Very Large Array (VLA) using
$\sim$\,160 hours of integration.  The resulting map has an rms of
$\sigma_{\rm 1.4GHz}\simeq$\,6\,$\mu$Jy\,beam$^{-1}$ \citep[Arumugam et al.\ in prep.; for a brief summary see][]{2013simpson}. In total 6,861 radio sources are detected at
SNR$>$\,4, and 706/707 of the ALMA SMGs are covered by the map.  Matching the
ALMA and radio catalogues using a 1.6$''$ search radius ($\sim$1 per cent false-positive matches) yields 273
matches at a 3-$\sigma$ level, corresponding to a radio detection fraction of 39 per cent (see also \citealt{2018an}), which is similar to the detected
radio fraction in other comparable SMG surveys
($\sim$\,30--50 per cent; e.g.\,\citealt{2013hodge,2011biggs,2017brisbin}, although see \citealt{2011lindner}).  In \S\,\ref{redshift} we will
discuss the redshift distribution of the radio-detected versus
non-detected fractions, as well as the influence of the radio emission
on the SED modelling we perform.

\subsection{Photometric properties of SMGs in comparison to the field population} \label{photometry}

To illustrate the broad photometric properties of our SMG sample and the constraints
available on their SEDs, we list the number of SMGs detected (above 3-$\sigma$) in a range of representative optical and
infrared photometric bands in Table~\ref{table:photom}.  It
is clear that fewer detections are observed in the bluer optical
wavebands, while $\sim$\,70--80 per cent of the sample
(which are covered by the imaging) are detected in $K$ or the IRAC bands; this
drops to 56 per cent in the $V$-band. In the far-infrared, 69 per cent of the ALMA
SMGs are detected in at least one of the PACS or SPIRE bands. Thus we have good photometric coverage for the bulk of the sample longward of the near-infrared, but with more limited detection rates in the bluer optical bands.

Before we discuss the multi-wavelength SEDs, we first compare the
optical and near/mid-infrared colours of the SMGs and field galaxies
in our sample. As this study makes use of a $K$-band selected catalogue, we investigate the distribution of $K$-band magnitudes compared to the ALMA $S_{870}$ fluxes Fig.~\ref{fig:obs_obs}a.

Colour selection of galaxies can provide a simple
method to identify high-redshift galaxies.  For example,
\cite{2004daddi} suggested a criteria based on $(B-z)$ and $(z-K)$
(\emph{BzK}) with $BzK= (z-K)$\,--\,$(B-z)$ to
select star-forming galaxies at $z \simeq$\,1.4--2.5.  Although the SMGs are
likely to be more strongly dust-obscured than  typical
star-forming galaxies at these redshifts, this diagnostic still
provides a useful starting point to interpret the rest-frame
UV/optical colours, and we show the SMGs in the $(z-K)$\,--\,$(B-z)$
colour space in Fig.~\ref{fig:obs_obs}b.  
We see that compared to a field galaxy sample, 
as expected, the SMGs are significantly redder, likely due to their
higher dust obscuration and higher redshifts.   Nevertheless, for our sample of 290 AS2UDS
SMGs with detections in all three $B$, $z$ and $K$-bands, 87 per cent (253/290) of
sources lie above $BzK =-$0.2, which is the suggested limit that
separates star-forming galaxies from passive galaxies, indicating
that the majority of these $BzK$-detected (hence bluer than average) SMGs have the colours expected for a
star-forming population. However, we caution that 14 per cent of our sample of these $BzK$-detected highly dust-obscured star-forming galaxies are misclassified as ``passive''. Moreover, we note that the SMG sub-set
shown on this $BzK$ plot is strongly biased due to the large fraction that are not shown because they are undetected in the optical
bands, especially the $B$-band. To highlight this, we overlay the track
for our composite SED (see
\S~\ref{composite}), which should more accurately represent the ``typical'' SMG, as a function of increasing redshift.  This indicates that at $z\simeq$\,1.5--2.5 the
{\it average} SMG has $BzK$ colours which lie on the border of the star-forming criterion, suggesting that a significant fraction of $z \lesssim$\,2.5 SMGs would not be selected as star-forming systems based on their $BzK$ colours, even if we had extremely deep $B$-band observations.

Given that the detection rate of ALMA SMGs is much higher in the mid-infrared IRAC bands, in Fig.~\ref{fig:obs_obs}c we show the
$S_{5.8}/S_{3.6}$ versus $S_{8.0}/S_{4.5}$  colour-colour plot for 388 SMGs
that are detected in all four IRAC bands. This colour-colour space
has been used to identify high-redshift star-forming galaxies, as well
as isolate candidate AGN at $z\lesssim$\,2.5 from their power-law spectra \citep[e.g.][]{2012donley}.  In this figure, on average the IRAC-detected ALMA SMGs
are again significantly redder than the field population \citep[see also][]{2019stach}.  We overlay the track formed from the composite SED of our
sample (see \S\,\ref{composite}), which demonstrates that these
IRAC-detected SMGs are likely to lie at $z\simeq$\,2--3. Hence, although it might appear from
Fig.~\ref{fig:obs_obs}c that many of the SMGs have mid-infrared colours suggestive of an AGN (power-law like out to 8\,$\mu$m), this is simply because many of these lie at $z >$\,2.5 where sources cannot be reliably classified using this colour selection. Indeed, \cite{2019stach} estimates a likely AGN fraction in AS2UDS based on X-ray detections of just 8\,$\pm$\,2 per cent. As seen from the composite SED track, the sources in the AGN colour region are on average at higher redshifts ($z >$\,2.5), where the 1.6-$\mu$m stellar ``bump'' falls beyond the 5.8-$\mu$m band, and the \cite{2012donley} AGN criteria breaks down.

In summary, the basic photometric properties of SMGs show them to
be redder than average field galaxies across most of the UV/optical to mid-infrared regime, likely due to a combination of
their higher redshifts and higher dust obscuration. High-redshift SMGs
are also fainter than the low-redshift SMGs in the optical and near-infrared
wavebands (Fig.~\ref{fig:thumbs}), but with a large dispersion in properties at any redshift.

\section{\textsc{magphys}: testing and calibration} \label{magphys}

To constrain the physical properties of the AS2UDS SMGs we employ {\sc
magphys} \citep{2008dacunha,2015dacunha,2019battisti} -- a physically motivated
model that consistently fits rest-frame SEDs
from the optical to radio wavelengths. An energy balance
technique is used to combine the attenuation of the stellar emission
in the UV/optical and near-infrared by dust, and the reradiation of
this energy in the far-infrared.  The {\sc magphys} model includes the
energy absorbed by dust in stellar birth clouds and the diffuse ISM.
This approach provides several significant advances compared to modelling the
optical and infrared wavelengths separately \citep[e.g.][]{2014simpson,2014swinbank}, allowing more control of the covariance
between parameters and generally providing more robust constraints on the physical
parameters (e.g.\ redshifts, stellar masses, and star-formation
rates). However, we note that the modelling assumes that sub-millimetre and optical emission is coming from a region of comparable size, which is a simplification of the true system. 

Before we apply {\sc magphys} to the SMGs in our sample, we briefly
review the most important aspects of the model that are likely to affect our
conclusions and discuss a number of tests that we apply to validate
our results.  For a full description of {\sc magphys} see \cite{2008dacunha,2015dacunha} and \cite{2019battisti}.

{\sc magphys} uses stellar population models from
\cite{2003bruzual&charlot}, a Chabrier IMF
\citep{2003chabrier} and metallicities that vary uniformly from 0.2 to 2 times solar. 
Star-formation histories are modelled as continuous delayed
exponential functions \citep{2010lee} with the peak of star formation occurring in range of 0.7--13.3\,Gyr after the onset of star formation. The age is drawn randomly in the range of 0.1--10\,Gyrs. To model starbursts, {\sc magphys} also
superimposes bursts on top of the star-formation history.  These
bursts are added randomly, but with a 75 per cent probability that they
occurred within the previous 2\,Gyr.  The duration of these bursts
varies in range of 30\,--\,300\,Myr with a total mass formed in stars
varying from 0.1 to 100\,$\times$ the mass formed by the underlying
continuous model. In this way, starbursts, as well as more quiescent
galaxies, can be modelled.  We note that the star-formation rate returned from {\sc magphys}
for a given model is defined as the average of the star-formation
history over the last 100\,Myr.

The far-infrared  emission from dust in {\sc magphys} is determined
self-consistently from the dust attenuated stellar emission.  Dust
attenuation is modelled using two components following \cite{2000charlot&fall}:
a dust model for young stars that are still deeply embedded in
their birth clouds; and a dust model for the intermediate/old stars in
the diffuse ISM.  The far-infrared luminosity we report is measured by
integrating the SED in the rest-frame between 8--1000\,$\mu$m and is
calculated through the sum of the birth cloud and ISM luminosities,
which also include contributions from the polycyclic aromatic
hydrocarbons, and mid-infrared continuum from hot, warm and cold
dust in thermal equilibrium.  The dust mass is calculated using the
far-infrared radiation and a wavelength-dependent dust mass coefficient.
For a full description of how each parameter is modelled see
\cite{2015dacunha} and \cite{2019battisti}.

For our analysis, we used the updated {\sc magphys} code from
\cite{2015dacunha} and \cite{2019battisti}, which is optimised to fit SEDs of high redshift
($z>$\,1) star-forming galaxies.  This code includes modifications
such as extended prior distributions of star-formation history and
dust optical depth effects, as well as the inclusion of intergalactic
medium absorption of UV photons.  The updated version also includes
photometric redshift as a variable.

To fit the photometry of a galaxy, {\sc magphys} generates a
library of SEDs for a grid of redshifts for each star-formation history
considered.  {\sc magphys} identifies the models that best-fit the
multi-wavelength photometry by matching the model SEDs to the data
using a $\chi^2$ test and returns the respective best-fit parameters.  In this study, we focus on eight of the derived
parameters: photometric redshift ($z$); star-formation rate (SFR); stellar mass ($M_\ast$);
mass-weighted age (Age$_m$); dust temperature ($T_{\rm d}$); dust attenuation ($A_{\rm V}$);
far-infrared luminosity ($L_{\rm IR}$) and dust mass ($M_{\rm d}$).

For each parameter, {\sc magphys} returns the probability distribution
(PDF) from the best-fit model.  The derived parameters (e.g.\ photometric redshift, stellar
mass, etc) are taken as the median from the PDF, with uncertainties
reflecting the 16--84$^{\rm th}$ percentile values of this distribution (we note that
if we instead adopted the peak value from the PDF, none of the
conclusions below are significantly affected).  In a small number of
cases, the SEDs are overly constrained due to the finite sampling, and the PDFs are highly peaked, meaning the returned uncertainties are unrealistically low. In these cases, we take a conservative approach and
adopt the median uncertainty from the full sample for that
derived parameter. We flag the sources where this has occurred in the
on-line catalogue (Table A1 in Appendix \ref{appendixA}).

\begin{figure*}
  \includegraphics[width=\textwidth]{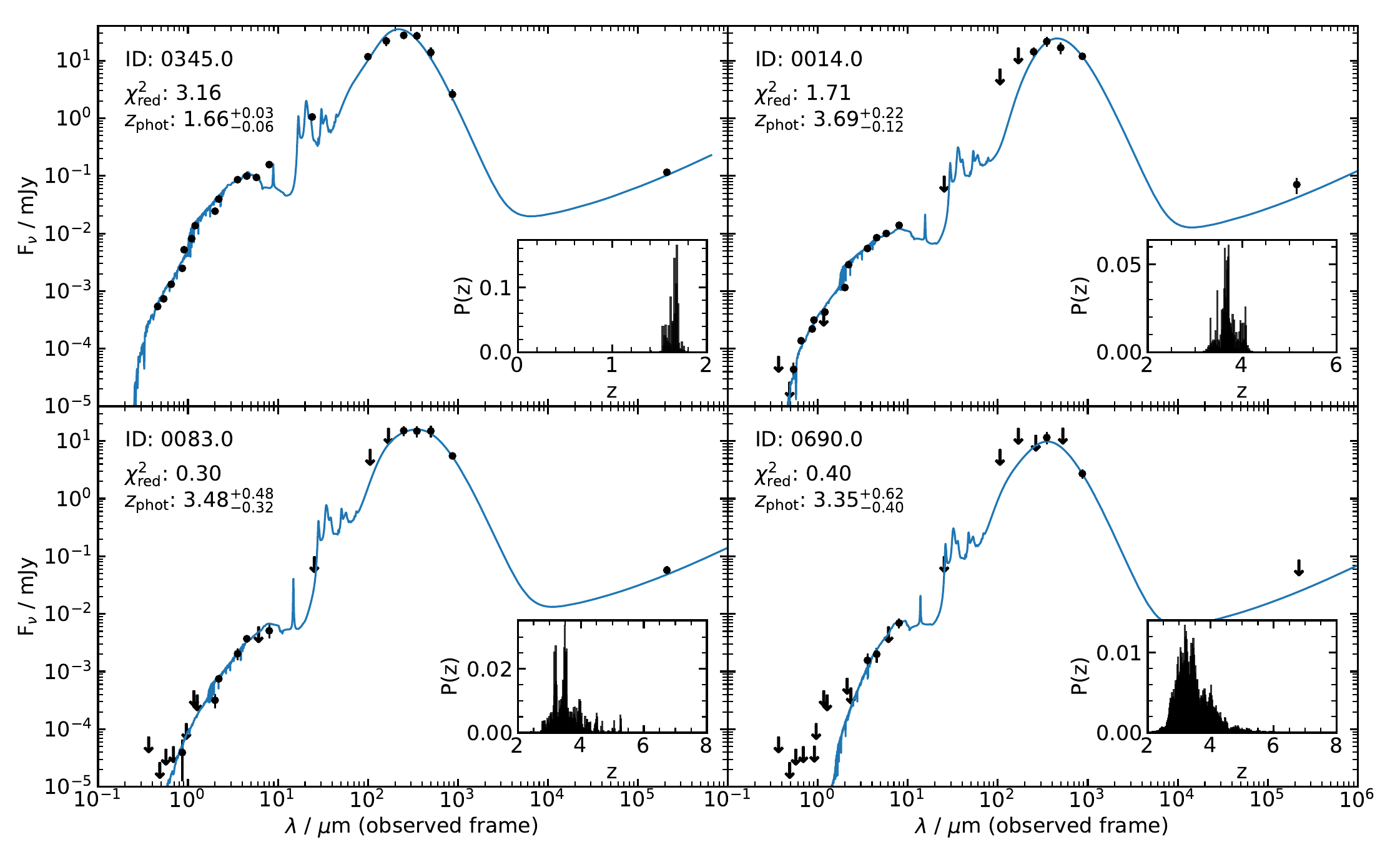}
  \caption{The observed-frame optical to radio spectral energy distributions of four example AS2UDS SMGs selected to have a decreasing number of photometric detections: 22/22, in top left; 16, in top right; 11, in bottom left; and 5 in bottom right. Limits in the optical/near-infrared wavebands ($U$-band to IRAC 8$\mu$m) were treated as 0\,$\pm$\,3$\sigma$, while those beyond 10\,$\mu$m (MIPS 24$\mu$m to Radio 1.4GHz)  are set to 1.5$\sigma\,\pm$\,1$\sigma$. These limits are indicated as  arrows. The solid line shows the predicted SED at the peak redshift of the best-fit PDF. The inset plots show the redshift probability distributions. As expected, as the number of photometric detections decreases, the redshift distribution becomes wider and the predicted photometric redshifts  becomes more uncertain. For reference, of our 707 SMGs 50 per cent have $\geq$\,11 photometric detections, while 82 per cent have $\geq$\,5 detections.}
  \label{fig:seds}
\end{figure*}

A significant fraction of the SMGs in our sample are
faint or undetected in one or more of the 22 wavebands that we employ in our analysis -- 
most frequently this is at the bluest optical wavelengths (see Table \ref{table:photom}) due to their high redshift and dusty natures. Thus, we first assess how
the flux upper limits affect the model fitting.

As a first step, in any given waveband, we treat a source as detected if it has at
least a 3-$\sigma$  detection.  For non-detections, we conservatively
adopt a flux of zero and a limit corresponding to 3-$\sigma$ in the
UV-to-mid-infrared bands (i.e.\ up to 8\,$\mu$m).  This is motivated by a
stacking analysis of ALMA SMGs in ALESS where the individually
optically faint or undetected SMGs yielded no or only weak detections
in the stacks \citep[e.g.][]{2014simpson}.  In the far-infrared, most
of the ``non-detections'' occur in the \emph{Herschel} maps, which are
confusion-noise dominated.  Stacking analysis of SMGs at 250--500\,$\mu$m has
demonstrated that the flux densities of ALMA SMGs at these wavelengths
are often just marginally below the confusion noise \citep[e.g.][]{2014simpson}. To this end, for non-detected sources in the infrared (beyond 10\,$\mu$m), we
adopt a flux density of 1.5\,$\pm$\,1.0$\sigma$. Other choices of limits were tested (e.g.\ 0\,$\pm$\,1$\sigma$ for all wavebands, 0\,$\pm$\,1$\sigma$ for optical/near-infrared and 1.5\,$\pm$\,1.0$\sigma$ for infrared) with no significant changes found for any of the derived physical parameters. 

We run {\sc magphys} on all 707 ALMA SMGs in our sample, and in
Fig.~\ref{fig:seds} we show the observed photometry and best-fit {\sc
magphys} model for four representative examples. All SED fits are shown on-line (Fig. A1 in Appendix \ref{appendixA}).
These examples are selected to span
the range in the number of photometric detections included in the SEDs: from sources that are
detected in all of the available 22 photometric bands (37 per cent of sources have coverage in 22--16 bands), 16 bands ( 28 per cent have coverage in 16--11 bands), 11 bands (20 per cent have coverage in 11--5 bands) and down to 5 bands (15 per cent have coverage in 5 or less bands).
We also plot the resulting photometric redshift PDF for each of
these SMGs. This demonstrates that when the SED is well-constrained
(e.g.\ the galaxy is detected in a large fraction of the photometric bands), the range of possible
photometric redshifts is narrow, e.g.\, with a median 16--84$^{\rm th}$ percentile
range of $\Delta z$\,$=$\,0.20 for SMGs with detections in all 22 bands. 
However, as the number of detection decreases, this range broadens.  
For our full sample of SMGs, the median
number of bands that are detected is 12, which yields a median 16--84$^{\rm th}$ percentile
redshift range on any given SMG of $\Delta z$\,$=$\,0.50. For reference, the median uncertainty for
the 18 per cent of SMGs that are detected in $\leq$\,6 bands is $\Delta z$\,$=$\,0.86. Note also that in some cases the reduced $\chi^2$ decreases as
the number of detections decreases. This does not necessarily
indicate a better fit, but rather often reflects the large
uncertainties in non-detected wavebands.

Finally, before testing the accuracy of the photometric redshifts, we ensure that the energy balance technique is appropriate and the far-infrared photometry is not affecting the redshift prediction significantly. We run {\sc magphys} on SMGs with $K$-band detections including only photometry up to 8\,$\mu$m and compare the predicted photometric redshifts to the values derived using the full UV-to-radio photometry. We find that the scatter of photometric redshifts is within the error range as the median is ($z_{\rm full} - z_{\rm \leq 8\mu m}$)/($z^{84th}_{\rm full} - z^{16th}_{\rm full}$)\,=\,0.11 with 68$^{\rm th}$ percentile range of -1.0--0.95. Thus, coupling far-infrared information into the estimation of photometric redshifts is not introducing any significant biases.

\begin{figure}
\centering
  \includegraphics[width=\columnwidth]{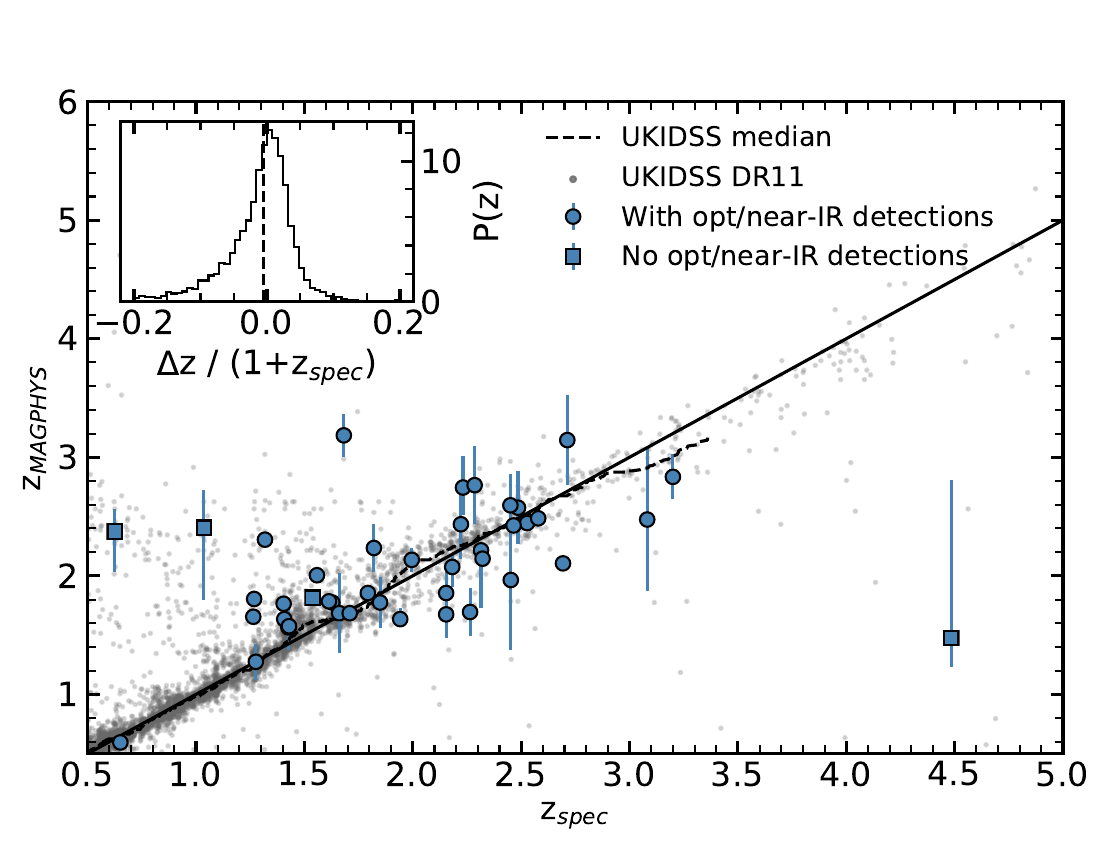}
  \caption{Comparison of {\sc magphys} photometric redshifts versus spectroscopic redshifts. The 44 AS2UDS SMGs with spectroscopic redshifts are plotted, 
as well as field sample of 6,719 $K$-detected UDS galaxies with spectroscopic redshifts. The dashed line shows the running median for the field galaxies, which tracks the spectroscopic redshifts closely up to $z\sim$\,3.5.  For the SMGs, we identify the four that lack detections in the optical bands. The inset panel shows the fractional offset of photometric redshifts from spectroscopic values for the field sample. The median offset is $(z_{\rm spec}-z_{\rm phot})/(1+z_{\rm spec})=-$0.005$\pm$0.003 with a dispersion of 0.13.} 
  \label{fig:mag_v_spec}
\end{figure}

\subsection{Testing against spectroscopic redshifts}

Before discussing the redshift distribution of our SMGs, we first
confirm the reliability of {\sc magphys} to measure photometric redshifts, and critically their uncertainties, \citep[see also][]{2019battisti}
by comparing the photometric and spectroscopic redshifts for both
the SMGs and the field galaxies in the UDS.

First, we run {\sc magphys} on all 6,719 $K$-band detected galaxies in the UKIDSS DR11 catalogue   
that have archival spectroscopic redshifts, and that have no photometric contamination flags (\citealt{2008smail}; Hartley et al.\ in prep.; Almaini et al.\ in prep.).
This includes 44 of the SMGs from our sample (including new
spectroscopic redshifts from KMOS observations; Birkin et al.\ in prep.). We note that it is possible, and indeed probable, that given the wide variety of sources from which these
redshifts were taken and the faintness of many of the target galaxies, that some of these spectroscopic
redshifts are incorrect.  As a result, we concentrate on the quality of the agreement achieved
for the bulk of the sample, giving less emphasis to outliers.  We also note that, given
the heterogeneous sample selection, the sample contains a mix of
populations, which is likely to include an increasing fraction of AGN-hosts at higher redshifts, the SEDs for which are not reproduced by the current version of {\sc magphys}.

We further isolate a sub-sample of all field galaxies with no photometric contamination flags above $z $\,$=$\,2 and include 500 galaxies with spectroscopic redshifts below $z $\,$=$\,2 to form a 
field sample biased towards higher-redshift/fainter sources 
that is more representative of the distribution of high-redshift SMGs. {\sc magphys} run on this sub-sample yields a
median offset between the spectroscopic- and photometric-redshifts of
$\Delta z$/\,$(1+z_{\rm
spec})\,$\,$=$\,0.004\,$\pm$\,0.001, although with larger systematic
offsets at redshift above $z\simeq$\,2.5 ($\Delta z$/\,$(1+z_{\rm
spec})\,$\,$=$\,0.040$\pm$0.003). At these
redshifts, the photometric redshift has sensitivity to the intergalactic medium (IGM) opacity
as the Lyman break (rest-frame 912--1215\AA) pass
through the observed $B$-band for sources that are bright enough to be detectable.  
Adjusting the IGM absorption coefficient in the SED model can reduce this systematic $\Delta z$
offset \citep[e.g.][]{2011wardlow}. The IGM effective absorption optical depth of each model is drawn from a Gaussian distribution centred at the mean value given in \cite{1995madau}, with a standard deviation of 0.5. We, therefore, rerun {\sc magphys}
for the spectroscopic sample with IGM absorption coefficients between
0.2--1.0 of each drawn model value.  From
this test, we find that tuning the IGM coefficient to 0.5 of the initially drawn value minimises the systematic offset between the
spectroscopic and photometric redshifts above $z\sim$\,2, whilst
maintaining the closest match at lower redshift, thus we adopt it in any subsequent analysis.
In Fig.~\ref{fig:mag_v_spec} we show the comparison of the spectroscopic
and photometric redshifts for the field galaxies and SMGs. We see that for the SMGs the three most extreme outliers are optically undetected, leading to uncertain estimation of their redshifts. The fourth outlier is a secondary ALMA source within a single SCUBA-2 map, where the optical photometry may have been mismatched. Over the
full redshift range, the offsets between the spectroscopic and
photometric redshifts for all 6,719 field galaxies is $\Delta z$/\,$(1+z_{\rm
spec})\,$\,$= -$0.005\,$\pm$\,0.003, and $\Delta z$/\,$(1+z_{\rm
spec})\,$\,$= -$0.02\,$\pm$\,0.03, with a 1$\sigma$ scatter of $\Delta z$/\,$(1+z_{\rm
spec})\,$\,$= $0.13, if
we just consider the 44 SMGs. The photometric redshift accuracy we obtain is comparable to that found for SMGs in the COSMOS field by \cite{2019battisti}.

We check what effect the error on the photometric redshift has on our inferred physical properties by running {\sc magphys} on the AS2UDS sub-sample of 44 SMGs with spectroscopic redshifts at their fixed spectroscopic redshifts.  We investigate whether the
change in the derived value of the property at the spectroscopic redshift and the photometric redshift is encompassed by the quoted errors (at the photometric redshift and including the covariance due to the uncertainty in this value) by calculating the fractional difference, $X_{\rm spec}/X_{\rm phot}$, where $X$ is any given parameter. The change for all the predicted parameters was, on average, less than $\lesssim$\,15 per cent, which is less than the typical errors. Therefore we confirm that the error uncertainty effect on any given parameter is captured in its error range and is not affecting final parameter distribution.

\subsection{Modelling EAGLE galaxies with {\sc magphys} -- a comparison of simulated and {\sc magphys} derived properties}
\label{eagle}

As well as empirically testing the reliability of the predicted
photometric redshifts from {\sc magphys}, we also wish  
to test how well the other {\sc magphys}-derived parameters are expected
to track the corresponding physical quantities.  This is more challenging, as
we lack knowledge of the ``true'' quantities (e.g.\ stellar mass or star-formation rate) 
for observed galaxies in our
field and so we have to adopt a different approach.  We, therefore, 
take advantage of the simulated galaxies from the Evolution and Assembly of GaLaxies and their
Environments  \citep[EAGLE,][]{2015schaye,2015crain} galaxy formation model to test how well {\sc
magphys} recovers the intrinsic properties of realistic model galaxies.

The EAGLE model is a smoothed-particle hydrodynamical simulation that
incorporates processes such as accretion, radiative cooling,
photo-ionisation heating, star formation, stellar mass loss, stellar
feedback, mergers and feedback from black holes. The full description
of the simulation as a whole can be found in \cite{2015schaye} and the
calibration strategy is described in \cite{2015crain}.  The most
recent post-processing analysis of the model galaxies in EAGLE
includes dust reprocessing using the {\sc skirt} radiative transfer code \citep{2011baes,2015camps&baes}.
This yields predicted SEDs of model galaxies covering the rest-frame UV-to-radio
wavelengths \citep[e.g.][]{2018camps,2019mcalpine}, and is
calibrated against far-infrared observations from the \emph{Herschel}
Reference Survey \citep{2010boselli}.  Our primary goal here is to run
{\sc magphys} on the model photometry of EAGLE galaxies and so test whether the
uncertainties on the derived quantities from {\sc magphys} encompass the 
known physical properties of the model galaxies. This will provide us with a
threshold that we can use to test the significance of any trends we
observe in our real data in $\S$~\ref{analysis}. We stress that {\sc magphys}
makes very different assumptions about the star-formation histories and dust properties
of galaxies than are assumed in EAGLE and {\sc skirt} and so this should
provide a fair test of the robustness of the derived parameters from {\sc magphys}
for galaxies with complex star-formation histories and mixes of dust and stars.

To select a sample of galaxies from the EAGLE model we use the
largest volume in the simulation set -- Ref-L0100N1504, which is a
100\,cMpc on-a-side periodic box (total volume 10$^6$\,cMpc$^3$).
However, we note that the volume of even the largest published EAGLE
simulation contains only a modest number of 
high-redshift galaxies with star-formation rates (or predicted 870-$\mu$m
flux densities) comparable to those seen in AS2UDS \citep{2019mcalpine}.
As a result, to match the observations as closely
as possible, but also provide a statistical sample for our comparison,
we select all 9,431 galaxies from EAGLE with
SFR\,>\,10\,M$_\odot$\,yr$^{-1}$ and $z>$\,0.25, but also isolate the 100
most strongly star-forming galaxies in the redshift range
$z$\,$=$\,1.8--3.4 (the 16--84$^{\rm th}$ percentile redshift range of our
survey).  To be consistent with the observations, for each model
galaxy we extract the predicted photometry in the same photometric bands as our observations and run {\sc magphys} to predict their physical properties.   

We show the comparison of intrinsic EAGLE properties
versus derived {\sc magphys} properties for these 9,431 galaxies on-line (Fig. A2 in Appendix~\ref{appendixA}). We concentrate our comparison on the stellar mass, star-formation rate,
mass-weighted age, dust temperature and dust mass, since these are the
quantities we will focus on in \S~\ref{analysis}.  We note that there are
systematic differences in the derived quantities from {\sc magphys} compared
to the expected values from EAGLE,
although in all cases {\sc magphys} provide remarkably linear
correlations with the intrinsic values (see Fig. A2).  The largest
difference is in the
stellar mass, where {\sc magphys} predicts a stellar mass that is
0.46\,$\pm$\,0.10\,dex lower than the ``true'' stellar mass in EAGLE, consistent with previous studies of systematic uncertainty in SMG masses \citep[e.g.][]{2011hainline}.
This difference is likely to be attributed to variations in the
adopted star-formation histories,  dust model and geometry
between {\sc magphys} and those in the radiative transfer code {\sc skirt}.  Accounting
for these differences is  beyond the scope of this work, and
indeed, more critical for our analysis is the scatter around the
line of best fit, since we can use this to further estimate the minimum
uncertainty on a given parameter in our data (even if the PDF suggests
the parameter is more highly constrained).

The stellar and dust masses have a scatter of 30 per cent and 10 per cent around the best fit, respectively.  The star-formation rates have
a scatter of 15 per cent around the best fit, and the scatter in the ages is
50 per cent. The scatter in dust temperature is 9 per cent, and we note that dust temperatures are estimated using very different methods in the simulations and from the observations.
Finally, we also use the quartile range of the scatter as a proxy to
assess the significance of any trends we observe in \S~\ref{analysis} (i.e.\ we adopt a significance limit that any trend in
these derived quantities seen in the SMGs must be greater than the
quartile range of the scatter in Fig. A2).  For the quantities in
Fig.~A1, these correspond to ratios of the $R$\,$=$\,75th/25th quartile
values of $R(T_{\rm d})\simeq $\,1.2, $R({\rm Age_{\rm m}})\simeq $\,4.2, $R(M_{\rm d})\simeq $\,2.7, $R(M_{\ast})\simeq $\,3.7 and $R({\rm SFR})\simeq $\,2.6. 

\subsection{Comparing observed and {\sc magphys}-derived quantities} \label{der_props}

\begin{figure*}
\centering
  \includegraphics[width=1\textwidth]{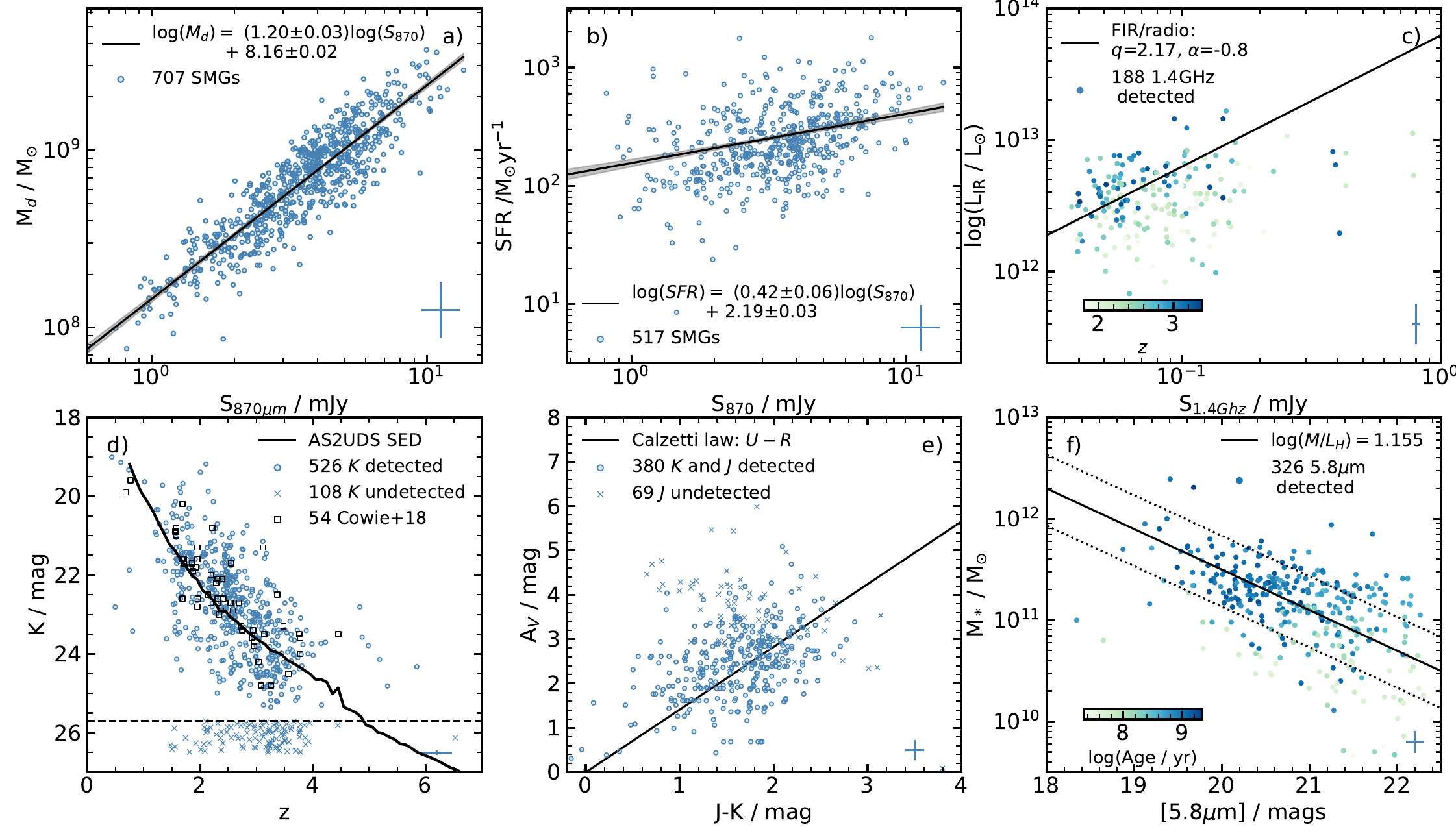}
  \caption{Observed photometry versus predicted physical parameters for the AS2UDS SMGs. Panels (a) and (d) include the full sample, while the other panels are for SMGs with $z$\,$=$\,1.8--3.4. In all panels, circles indicate sources that are detected.
  The typical errors are shown in the bottom right of each panel. 
\textbf{(a)}: Dust mass versus ALMA 870$\mu$m flux for all 707 SMGs. The best-fit line has a slope of 1.20$\pm$0.03
and the shaded region indicates the $\pm$1$\sigma$ error range. The strong positive correlation between the observed 870-$\mu$m flux and dust mass indicates that the 870-$\mu$m emission has the most sensitivity to cold dust mass. 
\textbf{(b)}: Star-formation rate versus ALMA 870-$\mu$m flux for 517 AS2UDS sources. We plot the best-fit line with a gradient of 0.42$\pm$0.06 and $\pm$1$\sigma$ errors shown as the shaded region. A positive correlation is observed, as expected for dusty SMGs, where the emission from young/hot starts is re-emitted at far-infrared wavelengths. 
\textbf{(c)}: Far-infrared luminosity versus 1.4-\,GHz flux for the SMGs. The solid line shows the FIR--radio correlation with $q_{\rm IR}$\,$=$\,2.17 at the median redshift of the AS2UDS sample. The radio-detected population are roughly consistent with the trends expected from the FIR--radio correlation with the scatter being mainly driven by redshift variation.
\textbf{(d)}: $K$-band magnitude versus photometric redshift for 526 $K$-band detected SMGs. The 108 SMGs with no $K$-band detection are plotted below the $K$-band aperture-corrected magnitude limit of $K$\,$=$\,25.7. We see a negative correlation due to the positive $k$-correction in the $K$-band. The fact that $K$-band undetected SMGs have redshifts down to $z\sim$\,1.5--2.5 highlights that some of the sources may be very obscured. The expected variation with redshift for the composite SED from our SMG sample is shown as a solid line. 
\textbf{(e)}: $V$-band dust attenuation versus  $(J-K)$ colour. The solid line shows the predicted reddening from the Calzetti reddening law. As expected, the rest-frame $(U-R)$ colour (observed $(J-K)$ at the median redshift of AS2UDS) follows the predicted reddening law well, indicating that SMGs with redder colours are likely to be more dust-obscured. 
\textbf{(f)}: Stellar mass versus  IRAC 5.8-$\mu$m magnitude, coloured by estimated age. The solid line shows the track of the mass inferred from the median $H$-band mass-to-light ratio at the median redshift. The dashed lines indicate $H$-band mass-to-light ratios of $\log(M/L_{\rm H} [M_\odot/L_\odot])=$\,2.5 and $\log(M/L_{\rm H} [M_\odot/L_\odot])=$\,0.5. Rest-frame $H$-band (corresponding to $\sim$\,5.8\,$\mu$m at the median redshift of the AS2UDS SMGs) correlates well with the predicted stellar mass. The scatter is mainly due to covariance of the mass with the mass-weighted age, as shown by the age trend at a given 5.8-$\mu$m magnitude.}
  \label{fig:obs_pred}
\end{figure*}

Before we discuss any of the physical parameters for the SMG population and their evolution, we compare the derived quantities returned from {\sc
magphys} with those observables which they are empirically expected to
correlate with (e.g.\ the dust mass is expected to correlate broadly
with 870-$\mu$m flux density).

In Fig.~\ref{fig:obs_pred} we plot the derived quantities returned from {\sc
magphys} against observed properties for the SMGs. For some quantities, we restrict the
sample to the redshift range $z $\,$=$\,1.8--3.4 (which represents the
16--84$^{\rm th}$ percentile) to reduce the degeneracies with redshift.
We first focus on those quantities that are most sensitive to the far-infrared 
part of the SED and see how these
correlate with the far-infrared photometry.  The main source of sub-millimetre
radiation is the thermal continuum from dust grains -- the rest-frame
UV/optical radiation from young/hot stars is absorbed by dust and re-emitted
at far-infrared wavelengths.  Hence observed 870-$\mu$m flux density
should trace both the dust mass and star-formation rate \citep[e.g.][]{2002blain,2014scoville}. In
Fig.~\ref{fig:obs_pred} a) we, therefore, plot the 870-$\mu$m flux density
versus estimated dust mass and star-formation rate.  As this shows there is a strong correlation between
870-$\mu$m flux density and dust mass ($M_{\rm d}$), which follows
 $\log_{10}(M_{\rm d}[M_\odot])\,=\,(1.20\,\pm\,0.03)\,\times \log_{10}(S_{\rm
870}[{\rm mJy}])\,+\,8.16\,\pm\,0.02$. 
This tight correlation suggests that, as expected, the 870-$\mu$m flux density tracks the cold dust mass \citep{2014scoville,2018liang}.
The trend of 870-$\mu$m flux density with star-formation rate
is also clear in Fig.~\ref{fig:obs_pred} b). Fitting to the SMGs,
the correlation between 870\,$\mu$m flux density and star-formation rate
has the form $\log_{10}(\rm{SFR}[{\rm M_\odot yr^{-1}}])\,=\,(0.41\,\pm\,0.05)\log_{10}(S_{\rm
870}[{\rm mJy}])\,+\,2.19\,\pm\,0.03$. The trend observed with star-formation rate is weaker than that of dust mass and has more dispersion, thus constraints from shorter rest-frame far-infrared wavelengths are needed to reliably measure the star-formation rate.

The predicted star-formation rates and far-infrared luminosities from {\sc magphys} closely follow the \cite{1998kennicutt} relation with an offset of $\rm{SFR}/\rm{SFR}_{\rm K98}(\rm{L}_{\rm FIR}) $\,$=$\,0.87$\pm$0.01 (where $\rm{SFR}_{\rm K98}(\rm{L}_{\rm FIR})$ is the predicted Kennicutt relation). In addition, the total far-infrared luminosity should correlate with the {\it observed} radio luminosity although this is used in the SED fitting) due to the far-infrared--radio
correlation \citep{1971vanderkruit,1973vanderkruit}.  As discussed in
\S~\ref{obs}, the radio luminosity is expected to be dominated by synchrotron radiation from
relativistic electrons that have been accelerated in supernovae
remnants \citep{1975harwit&pacini}. The far-infrared and radio luminosities
are correlated since the supernovae remnants arise from the same
population of massive stars that heat and ionise the H{\sc ii}
regions, which in turn heats the obscuring dust. In Fig.~\ref{fig:obs_pred} c) we, therefore, plot the 
{\sc magphys} far-infrared
luminosity (integrated between 8--1000\,$\mu$m) as a function of the
observed 1.4-GHz flux density, again restricting the sample to a
redshift range of $z $\,$=$\,1.8--3.4 (to reduce the effects of the geometrical dimming).  We overlay the far-infrared/radio correlation from \cite{2010ivison} for the
median redshift of our sample SMGs ($z $\,$=$\,2.61) with $q_{\rm IR}$\,$=$\,2.17 \citep{2010magnelli} and $\alpha =-$0.8 \citep{2010ivison}, appropriate
for high redshift, strongly star-forming galaxies \citep{2010magnelli}, where $q_{\rm IR}$ is the logarithmic ratio of bolometric infrared and monochromatic radio flux and $\alpha$ is the radio spectral index. This shows a rough correlation between the predicted far-infrared luminosities and the observed radio luminosities, which is consistent in form and normalisation with that
derived for the AS2UDS sample. The scatter is mainly due to variations in redshift. A more detailed analysis of the far-infrared - radio correlation in AS2UDS is given in Algera et al (in prep.).

Next, we turn to the optical and near-infrared wavelengths.  The
observed optical/near-infrared emission at $z\sim$\,2 corresponds to
rest-frame far-UV to $R$ band, which traces the stellar-dominated SED around the
Balmer (3646{\AA}) and 4000{\AA} breaks -- the former is more
prominent in star-forming galaxies, while the latter is more prominent
in older, quiescent galaxies, giving an indication of the galaxy's recent star-formation history. To
test how the derived quantities
correlate with basic observables, in Fig.~\ref{fig:obs_pred} we plot stellar mass, optical
extinction and redshift as a function of observed magnitudes and colours of the SMGs.

First, we note that the observed $K$-band magnitude increases with increasing
redshift, as a result of positive $k$-correction \citep{2004smail}.  As a
guide, we, therefore, overlay the average $K$-band magnitude expected as a function of
redshift based on the composite SMG SED from our sample (see
$\S$~\ref{composite}).  We also overlay the
ALMA-detected  SMGs in the CDFS from \citet{2018cowie}, which show a similar
trend. We note that there are 108 SMGs in our sample that are
undetected in the  $K$-band ($K>$\,25.7).  The {\sc
magphys}-derived redshifts for this sub-sample lie in the range $z $\,$=$\,1.5--6.5
with a median of $z $\,$=$\,3.0\,$\pm$\,0.1.  We will discuss this population further in
\S~\ref{analysis}.

Next, we assess the $V$-band dust attenuation, $A_{V}$.  The optical
extinction returned from {\sc magphys} reflects the stellar
luminosity-weighted average across the source. At $z\sim$\,2, the extinction is expected to correlate with the rest-frame optical
colours.  In Fig.~\ref{fig:obs_pred} e) we, therefore, plot the $A_{V}$ versus $(J-K)$ colour (which corresponds approximately to
rest-frame $(U-R)$ colour at these redshifts and so is indicative of the optical SED slope).
We also overlay in Fig.~\ref{fig:obs_pred} e) a track
representing the expected rest-frame $(U-R)$ colours (corresponding to observed $(J-K)$ at the median redshift of AS2UDS) based on the Calzetti reddening law \citep{2000calzetti}.
This reproduces the trend we see and suggests that our
estimates of $A_V$ for the SMGs from {\sc magphys} are reliable.
Reassuringly, the majority of the 181
SMGs with no detection in either $J$- or $K$-band have a higher $A_V$,
indicating that it is likely that their higher dust obscuration is responsible
for their non-detection.

Finally, we turn to the stellar mass. It is expected that
the dominant stellar population by mass in these galaxies arises from the lower mass stars, which can be better traced from the rest-frame $H$-band
luminosity.  At $z\sim$\,2, this corresponds to the mid-infrared,
around $\sim$6$\mu$m and so in Fig. \ref{fig:obs_pred} f) we plot the {\sc
magphys}-derived stellar mass as a function of the
observed-frame IRAC 5.8-$\mu$m magnitude. As expected, brighter 5.8-$\mu$m
magnitudes correspond to higher stellar masses, and for SMGs in the
range $z\sim$\,1.8--3.4 we derive a correlation with
$\log M_\ast =(-0.25\pm0.03)\,S_{5.8}+(16.4\pm0.6)$. 
We also overlay the prediction of mass for a median $H$-band mass-to-light ratio (1.155) 
for our sample SMGs and find that it follows the observed properties well. The correlation in Fig.~\ref{fig:obs_pred} f) shows a
scatter of 0.05\,dex at fixed 5.8\,$\mu$m magnitude on
average. 
This scatter is due to variations in the star-formation history and dust extinction, but is also correlated with the predicted
mass-weighted age of the stellar population in the sense that for a
given observed 5.8-$\mu$m magnitude, the younger the inferred
age of the galaxy the lower the stellar mass. We note  that independent tests of the reliability of the {\sc magphys} predictions
for the reddening
and stellar masses using the simulated {\sc eagle} galaxies also provide mutual support for the reliability of the other parameter,
given the strong covariance expected between these two quantities in any SED fit (see Fig. A2 in Appendix \ref{appendixA}).

\begin{figure}
\centering
  \includegraphics[width=\columnwidth]{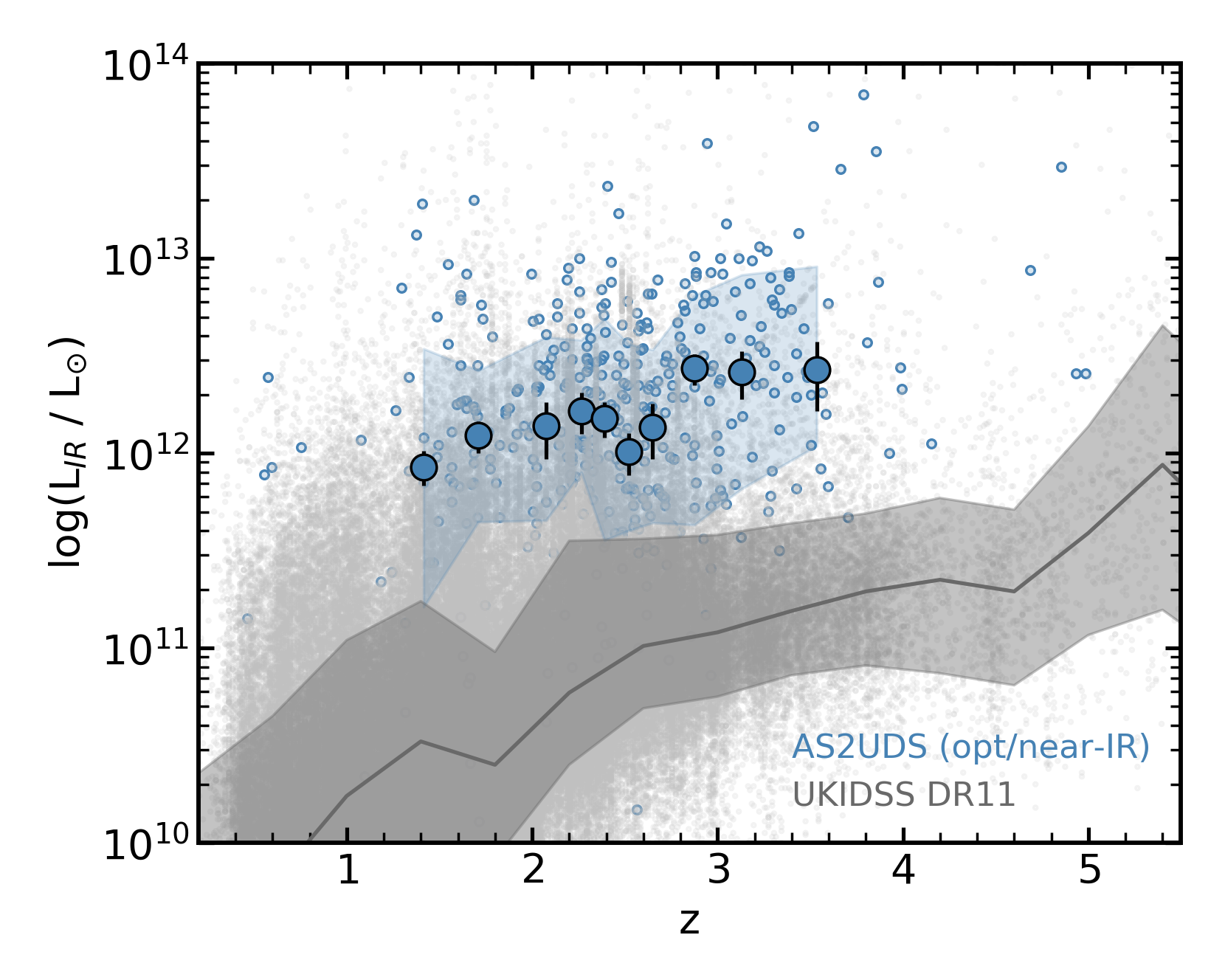}
   \caption{Predicted far-infrared luminosity as a function of redshift for a $K$-band selected field galaxy sample with the most reliable photometry based on a {\sc magphys} analysis of just the photometry shortward of the IRAC 8.0-$\mu$m band. The solid line shows the running median and the shaded region indicates the inter-quartile range. We
also plot the far-infrared luminosity derived for the  511 $K$-band detected SMGs, similarly limiting the model fit to photometry shortward of 8.0\,$\mu$m. We indicate with filled circles the binned medians of samples of 50 SMGs ranked in redshift and the blue region shows the interquartile range. It is clear that even when {\sc magphys} only has information on the optical/near-infrared SED, it still predicts AS2UDS SMGs to be significantly more far-infrared luminous than a typical field galaxy.} 
\label{fig:lz_field}
\end{figure}

\subsubsection{Predicting the far-infrared properties of the field
galaxies in UDS} \label{test_fir}

As we discussed in $\S$~\ref{photometry}, we derive the physical properties of
205,910 $K$-band selected galaxies in the UDS field from the UKIDSS DR11
catalogue by applying {\sc magphys} to their optical and near-infrared photometry (up to IRAC
8.0\,$\mu$m) in an analogous way to our SMG sample.  We will use this sample for a range of tests, but here we explicitly test the dust attenuation laws (and the
degeneracies between age and reddening), by determining whether far-infrared luminosity can be predicted just using the optical/near-infrared part of the SED.

In Fig.~\ref{fig:lz_field} we plot predicted
far-infrared luminosity versus redshift for the $K$-band selected field galaxy sample.  
We also plot those
SMGs that are $K$-band detected and where we have similarly derived the predicted far-infrared luminosities
based on {\sc magphys} modelling of {\it just} their optical/near-infrared photometry up to 8.0\,$\mu$m. 
Remarkably,
on average {\sc magphys} is able to identify the SMGs as dusty and
highly star-forming and thus far-infrared luminous
using only the information shortward of $\sim$\,2\,$\mu$m in the rest-frame.  
Indeed, for the
$K$-band detected $S_{870} >$\,3.6\,mJy ALMA SMGs, the mean ratio of
far-infrared luminosity from the $\leq $\,8-$\mu$m fit to that
from the full-SED including far-infrared/sub-millimetre photometry is
$L_{\rm IR}^{\leq 8\mu\rm m}/L_{\rm IR}^{\rm full}$\,$=$\,1.1$\pm$0.1. 

However, it is clear from Fig.~\ref{fig:lz_field} 
that {\sc magphys} also
predicts a population of $\sim$\,2,000
galaxies at $z\sim$\,1.5--4, which are claimed to be far-infrared luminous, but which
are not detected in the SCUBA-2 850-$\mu$m survey.
We suspect that many of these faux-SMGs may be either sources with AGN contributions to their optical/near-infrared SEDs or hotter dust sources, missed by our 850-$\mu$m selection. Hence, while this test does confirm
that the dust modelling and energy balance in {\sc magphys} provides
robust constrains on the far-infrared emission, it can only be
used reliably if far-infrared photometric constraints are available,
otherwise, the false-positive rate is high. 

\begin{figure*}
  \includegraphics[width=\textwidth]{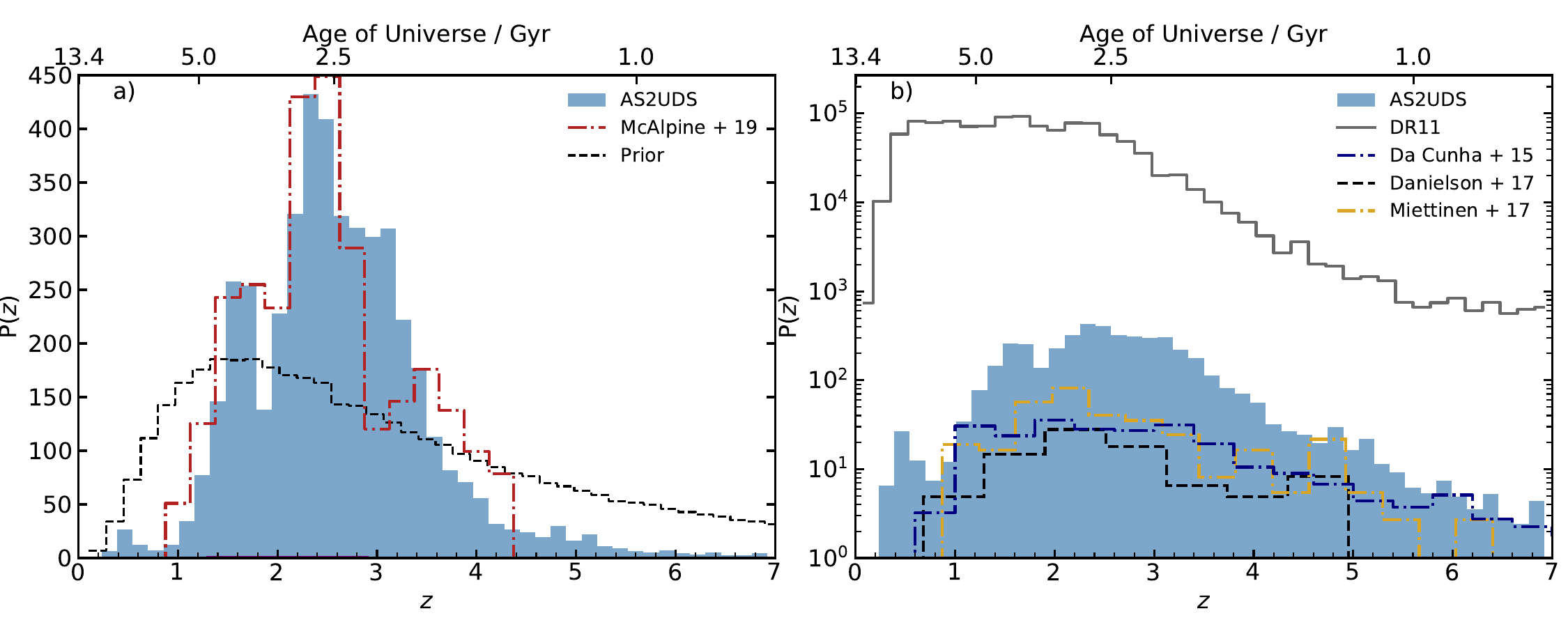}
\caption{\textbf{(a)} The redshift distribution from summeld likelihood distributions for our complete sample of 707 AS2UDS SMGs with a median of $z\,$\,$=$\,2.61$\pm$0.08 (68th per centile range of $z\,$\,$=$\,1.8--3.4 and 6 per cent at $z >$4). The dashed line indicates the prior distribution. For comparison, we also overlay theoretical predictions for SMG type galaxies from \protect\cite{2019mcalpine}, who find a median redshift of $z$\,$=$\,2.4$\,\pm\,$0.1.
\textbf{(b)} Comparison of the AS2UDS redshift distribution to the equivalent distribution for the  99 ALESS SMGs from \protect\cite{2015dacunha}, 52 spectroscopically identified ALESS SMGs from \protect\cite{2017danielson} and 124 spectroscopically identified SMGs from \protect\cite{2017brisbin}. We also include a comparison to the field galaxies from our {\sc magphys}-derived distribution for 205,910 $K$-band selected UKIDSS UDS sources with a median redshift of $z =$\,1.75\,$\pm$\,0.03. 
The distributions are normalised by their survey area.}
  \label{fig:red_dist}
\end{figure*}

This comparison of derived parameters from {\sc magphys} modelling of the complete SEDs of SMGs, compared
to the results when restricted to {\it only} fitting photometry shortward of 8\,$\mu$m, indicates
a poor recovery of those parameters that are most sensitive to details of the dust SED, such as dust temperature
or dust mass.  However, it also suggests little change for this $K$-detected sub-set of the SMG population
in the derived median: 
photometric redshifts, $(z^{\rm full}-z^{\leq 8\mu\rm m})/(1+z^{\rm full})$\,$=$\,0.008$\pm$0.004 (with 1$\sigma$ dispersion of 0.13); 
dust reddening, $(A_V^{\rm full}-A_V^{\leq 8\mu\rm m})/A_v^{\rm full}$\,$=$\,0.01$\pm$0.02 (with 1$\sigma$ dispersion of 0.30); 
or stellar mass, $(M_\ast^{\rm full}-M_\ast^{\leq 8\mu\rm m})/M_\ast^{\rm full}=-$0.02$\pm$0.01 (with 1$\sigma$ dispersion of 0.68); and
a modest bias towards younger ages when including the $>$\,10\,$\mu$m photometry: 
 $({\rm Age}_m^{\rm full}-{\rm Age}_m^{\leq 8\mu\rm m})/{\rm Age}_m^{\rm full}=-$0.25$\pm$0.05 (with 1$\sigma$ dispersion of 1.85).

\section{Analysis and Results} \label{analysis}

\subsection{Redshift distribution} \label{redshift}

The redshift distribution of SMGs can provide stringent constraints on
galaxy formation models, and indeed, in some instances has forced changes in the
way rapidly star-forming galaxies are modelled \citep[e.g.][]{2005baugh}.  The early measurements of the redshift distribution of
SMGs were hampered  by incompleteness and errors in the identification of counterparts for
single-dish sources \citep{2005chapman,2006pope,2011wardlow}, although the
results favoured
a median redshift of $z \simeq$\,2.3.  More recent studies have overcome
some of the  weaknesses of the early work,  both by unambiguously 
 identifying the SMGs using sub-/millimetre interferometry with
ALMA, and also by using a variety of methods to account for incompleteness
in the estimation of
redshift for the $\sim$10--20 per cent of SMGs that are too faint
in the optical/near-infrared to locate multi-wavelength counterparts \citep{2014simpson, 2015dacunha,2017danielson,2017miettinen,2018cowie}.

These studies  suggest a slightly higher median redshift, $z\simeq$\,2.6 \citep[e.g.][]{2014simpson}, for
the SMG population at mJy-flux density limits. 
However, exploiting these samples to go beyond just a crude redshift distribution to 
investigate evolution in the properties of SMGs with
redshift, have been hampered by the 
modest sample sizes available ($\lesssim $\,100 SMGs),  which weakens our ability to statistically identify 
 trends in the data (e.g.\  with 870\,$\mu$m flux density,
star-formation rate or mass). Here, our sample of 707 ALMA-identified
SMGs, combined with the {\sc magphys} analysis of their 
multi-wavelength properties from deep ancillary data, provides both, {\it complete}
redshift information and the large sample size necessary to 
simultaneously sub-divide the sample on the basis of, e.g.\,mass and star-formation rate to search for evolutionary trends \citep[e.g.][]{2019mcalpine}.

We begin by deriving the redshift distribution of our SMG sample. We
note that redshift prior in {\sc magphys} has a broad peak at $z\sim$\,1.5 (see  Fig.~\ref{fig:red_dist}a), thus we have also tested the influence of the prior on the photometric redshifts by running {\sc magphys} on all of the SMGs with a flat prior distribution (from $z$\,$=$\,0--7). For the SMGs, the resulting change in the redshift
distribution is negligible, with $\Delta z$\,$=$\,0.100\,$\pm$\,0.007. Hence, we conclude that the prior does not have a significant effect on
our estimate of the photometric redshift distribution.

With the reliability of the {\sc magphys} photometric redshifts confirmed in \S\,\ref{magphys}, we derive a photometric
redshift distribution for the SMGs and show this in Fig.~\ref{fig:red_dist}a.  To capture the uncertainties in the redshifts (and the range of quality
reflected in their PDFs) 
we stack the individual likelihood redshift distributions of all of the SMGs. For the complete sample of 707 870\,$\mu$m selected
SMGs, we determine a median redshift of $z $\,$=$\,2.61\,$\pm$\,0.08. The quoted error combines the
systematic uncertainty derived  from 
comparison of the {\sc magphys} redshifts to those
for the 6,719 $K$-band
galaxies with spectroscopic redshifts in the UDS and the bootstrap error on the redshift
distribution.  The photometric redshift distribution is 
strongly peaked, with a 16--84$^{\rm th}$ percentile range of $z$\,$=$\,1.8--3.4 and just $\sim$\,6 per cent of SMGs at $z>$\,4, while we find only five examples of SMGs at $z<$\,1 even though this redshift range encompasses 57 per cent of the age of the Universe -- underlining the identification of SMGs as a high-redshift population.  Moreover, it is possible that some of these $z<$\,1 systems are incorrectly identifications resulting from galaxy-galaxy lensing \citep[e.g.][]{2017simpson,2017danielson}. In Fig.~\ref{fig:red_dist}a, we also overlay the predicted redshift
distribution for SMGs with $S_{850} \geq$\,1\,mJy from the EAGLE
simulation  \citep{2019mcalpine}.  The median redshift for the EAGLE
 SMGs is $z$\,$=$\,2.4\,$\pm$\,0.1, with a sharp decrease above
$z\sim$\,2.5, driven in part by an increasing dust temperature in sources at higher redshifts. Therefore, this model distribution is a reasonable match to our observations.

In Fig.~\ref{fig:red_dist}b we, next, compare our sample to the earlier study of 99 SMGs
from ALESS \citep{2015dacunha}.  This sample has a single-dish 870-$\mu$m flux density
limit of $S_{870} \geq$\,3.5\,mJy, similar to our survey
and the photometric redshifts were also derived using {\sc magphys}. \cite{2015dacunha}  estimate a median redshift of
$z$\,$=$\,2.7$\,\pm\,$0.1 for their sample, comparable to what we find, although the
ALESS SMGs appear to have a 
shallower decline in number density beyond $z\gtrsim $\,3.5--4, compared
to AS2UDS. In Fig.~\ref{fig:red_dist}b we also compare to the 1.1-mm selected sample of 124
SMGs in COSMOS from \citet{2017miettinen}, who have also used {\sc magphys} to derive
their properties. \citet{2017miettinen} estimated a median redshift for
their sample, which has a  median equivalent 870\,$\mu$m flux density of 4.2\,$\pm$\,0.2\,mJy (adopting $S_{870}$\,/\,$S_{1100}\sim$\,2.7, \citealt{2015ikarashi}), and a median redshift of $z$\,$=$\,2.30\,$\pm$\,0.13, marginally lower than our measurement. 
The significance of this difference is only $\sim$\,2-$\sigma$, before considering cosmic variance or differences in the
initial waveband selection, and so we conclude that the distributions are consistent.

We next compare our distribution to those from spectroscopic SMG samples. \citet{2017danielson} provides
spectroscopic redshifts for 52 ALMA-identified SMGs from ALESS with
$S_{870}>$\,2\,mJy.  This sample has a median redshift of
$z$\,$=$\,2.4\,$\pm$\,0.1 (see Fig.~\ref{fig:red_dist}b), which is also similar to the median of the redshift
distribution from  the spectroscopic survey of radio-identified SMGs
in \cite{2005chapman}.  Both of these results are slightly lower than the median
we derive, most likely due to a combination of selection effects:
both the optical/near-infrared brightness of the counterparts (which aids
spectroscopic identifications) and in the case of \cite{2005chapman},
radio biases.  To assess the former bias, we note that the majority
of spectroscopic SMGs in \cite{2017danielson} have $K\lesssim$\,23.5.
Cutting our sample at $K\leq$\,23.5 yields a median redshift of
$z$\,$=$\,2.44\,$\pm$\,0.08, in much better agreement to their result.
Similarly, to demonstrate the potential influence of the radio identifications, if we limit our sample in AS2UDS to the 273 radio-detected SMGs
then we obtain a median redshift of $z$\,$=$\,2.5$\pm$0.1, which is within
the uncertainty of the result from \cite{2005chapman}.  

In addition, we have also run {\sc magphys} on all 205,910 $K$-band selected
galaxies in the field with no contamination flags to allow us to compare the properties of the ALMA SMGs directly to the less active field population in a consistent manner. The redshift distribution of the field sample is also
shown in Fig.~\ref{fig:red_dist}b, where we derive a median redshift of $z $\,$=$\,1.75$\pm$0.08. 

\begin{figure*}
  \centering
  \includegraphics[width=\textwidth]{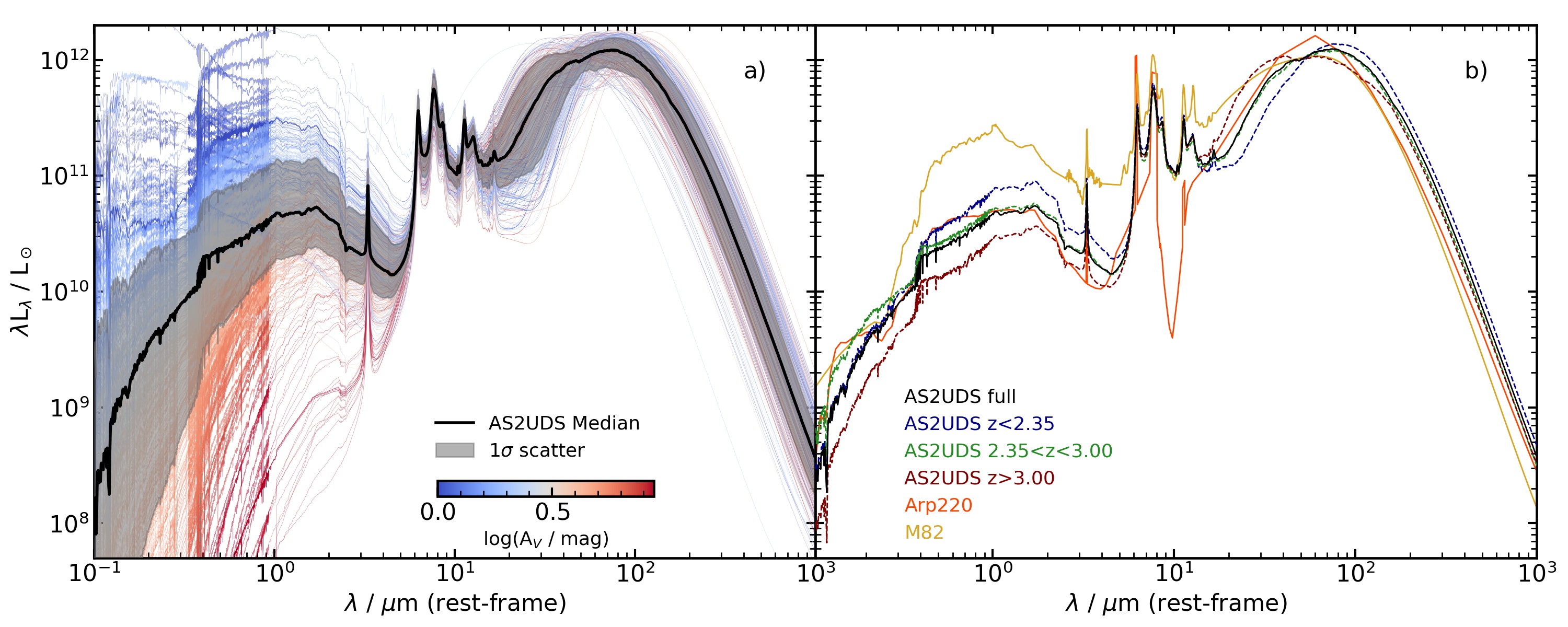}
\caption{\textbf{(a)} The best-fit  rest-frame SEDs of all 707 AS2UDS SMGs 
normalised to $L_{\rm IR}$\,$=$\,2\,$\times$\,10$^{12}$\,L$_\odot$.
The individual SEDs are coloured
  by their $V$-band dust attenuation.
The dispersion
  in the flux between the SEDs in
the  far-infrared and sub-millimetre is $\sim$\,2--3$\times$, but
  this increases below
$\sim $\,5$\mu$m to several orders of magnitude at rest-frame
wavelengths of $\lesssim $\,1\,$\mu$m. The thick
  solid line shows the median composite SED from this sample and the shaded region
  indicates the 16--84$^{\rm th}$ percentile region. We see that AS2UDS SMGs
have a wide variety of  colours and luminosities even in the
rest-frame optical, thus it is
very hard to construct a sample of
 star-forming galaxies which is complete for
even the most obscured examples based on selection in the  
observed optical or even near-infrared
wavelengths. \textbf{(b)} A comparison of AS2UDS composite to local
galaxies. We plot  the  composite of the full sample and
the SEDs for sub-samples split on redshift (dashed lines) into roughly equal-sized 
sub-sets: $z<$\,2.35,
$z$\,$=$\,2.35--3.00 and $z>$\,3.00. 
We see that high-redshift SMG's composite  SED is more dust-obscured and
peaks at shorter wavelength (i.e.\ hotter dust temperatures) than the lower-redshift
composites. For comparison, we also plot the SEDs of the local galaxies M82
and Arp\,220. The full AS2UDS composite appears to be much more dust-obscured than M82, while it resembles Arp\,220 quite closely at
optical and near/mid-infrared wavelengths. However, in the
far-infrared Arp\,220's SED  peaks at shorter wavelengths (e.g.\ hotter
characteristic dust temperature) than the majority of the SMGs at
$z<$\,3.  Thus Arp\,220 is a poor far-infrared template for typical SMGs,
but can provide an appropriate match to the typically hotter 
sources seen at higher redshift ($z >$\,3).}
\label{fig:comp_sed}
\end{figure*}

One major benefit of the use of {\sc magphys} in our analysis is the inclusion of the far-infrared and radio photometry in the SED modelling and the photometric
redshift determination. Hence, we are able to investigate the redshift
distribution of optical/near-infrared-faint and -bright SMGs using redshifts
derived in a consistent manner.  
The photometric redshift distribution
for the SMGs which are undetected in the $K$-band (17$\pm$1 per cent), with $K>$\,25.7 has a median redshift of
$z$\,$=$\,3.0\,$\pm$\,0.1, but exhibits a broad 
distribution with a 16--84$^{\rm th}$ percentile range of $z$\,$=$\,2.0--3.8 (see  Fig.~\ref{fig:obs_pred} d)).
Thus, {\sc magphys} predicts that the $K$-blank SMGs
are typically at higher redshifts than the $K\leq$\,25.7 sub-set (which have $z$\,$=$\,2.55\,$\pm$\,0.08),
although there is considerable overlap between the two redshift distributions and we 
particularly highlight that around $\sim$\,25 per cent of the near-infrared-blank SMGs lie at relatively low redshifts, $z\leq $\,2.5.
{\sc magphys} indicates that the main reason for the difference in the $K$-band brightness of these two sub-samples is
dust reddening: the $K$-detected SMGs have  optical reddening of $A_V$\,$=$\,2.61\,$\pm $\,0.05, but the
$K$-blank SMGs exhibit much higher reddenings, $A_V$\,$=$\,5.33\,$\pm $\,0.18 (and  $A_V$\,$=$\,6.0\,$\pm $\,0.2 for
those $K$-blank SMGs at $z<$\,2.5). Thus while higher redshifts is an explanation for the $K$-band faintness of many of these SMGs, that is not the case for all.
As both sub-samples have similar dust mass values and far-infrared luminosities, the difference in the dust attenuation cannot be attributed to higher dust content in the $K$-band undetected SMGs. Instead those $K$-band undetected SMGs at $z\lesssim$\,3 must differ physically in the geometry of their dust and stars -- either they have different viewing angles (disk-like systems viewed edge-on would result in higher dust attenuation) or these are more compact sources with higher dust column. In fact, from the sub-sample of AS2UDS SMGs with 870-$\mu$m sizes from \cite{2019gullberg}, $K$-band faint sources have smaller  sizes of $R_e$\,$=$\,1.60$\pm$0.10\,kpc, compared to those detected in the $K$-band,  $R_e$\,$=$\,1.98$\pm$0.10\,kpc \citep{2019gullberg}. This suggests that the relative distribution of stars and dust may be the main factor in their near-infrared faintness.

\subsection{SMG spectral energy distributions} \label{composite}

In this section, we analyse the SEDs of the
707 ALMA SMGs in our sample in order to quantify the variation in 
SEDs within the SMG population, and to compare the overall properties of
the SMGs to other populations, including local galaxies.

In Fig.~\ref{fig:comp_sed}a we plot the rest-frame SEDs of all the SMGs in our sample.  We normalise the
SEDs by their far-infrared luminosity (8--1000$\mu$m) to roughly the
median of our sample, $L_{\rm IR}$\,$=$\,2\,$\times$\,10$^{12}$\,L$_\odot$. Each of
the SEDs is colour-coded by the source's estimated $V$-band dust
attenuation ($A_V$),
which indicates that the galaxies with the reddest UV/optical SEDs
are also the most highly obscured. Therefore, we derive a composite SED for our whole population by measuring the median value at each
wavelength, and overlay this on to the individual spectra in
Fig.~\ref{fig:comp_sed}a. We also indicate the variation in the dispersion between the SEDs of SMGs as a function of wavelength.  
This  highlights the wide variation in the rest-frame
UV/optical luminosities for a far-infrared selected sample. In
the wavelength range 0.1--5\,$\mu$m (rest-frame UV/optical--near-infrared),
the dispersion is $\sim$\,1--2\,dex, with the full range of the
population spanning five orders of magnitude. It should be stressed
that this variety is for a population which has far-infrared luminosities in excess of 10$^{12}$\,L$_\odot$ and typical stellar masses in the range
$M_\ast\sim$\,10$^{10-11}$\,M$_\odot$.  This highlights the
difficulty in constructing complete samples of star-forming galaxies
in the optical/near-/mid-infrared and how even near-infrared surveys are unable to
identify fully mass-complete samples of strongly star-forming galaxies. 

To search for evolution in  the SEDs of SMGs, we split our sample
into three redshift ranges containing roughly equal numbers
of sources: $z<$\,2.35, $z$\,$=$\,2.35--3.00 and
$z>$\,3.00, with median redshifts of $z$\,$=$\,1.86$\pm$0.05, 
$z$\,$=$\,2.58$\pm$0.02 and $z$\,$=$\,3.35$\pm$0.04 respectively. We 
determine the median SED of each sample and overlay these in 
Fig.~\ref{fig:comp_sed}b. At $\lesssim$\,5$\mu$m we see a factor of $\sim$\,3--4$\times$
variation in brightness of the composite SEDs between the different redshift 
ranges -- with the lower redshift samples being consistently brighter in the
rest-frame optical/near-infrared than those at
higher redshifts (we see the same trend when we limit our sample to the luminosity-selected SMGs, see \S~\ref{luminosity}). Looking at the derived median far-infrared luminosities,
stellar masses and dust reddening for the three sub-sets (see \S~\ref{properties}), this variation
appears to be due primarily to higher far-infrared luminosities and dust temperatures
at higher redshifts, along with slightly higher reddening and
slightly lower stellar masses.  There are few observational
constraints on the shape of the SED at rest-frame $\sim$\,10$\mu$m and
perhaps, as a result, the three sub-sets show similar mid-infrared
luminosities. At longer wavelengths, there is one notable difference
between
the SEDs, with the higher-redshift sub-sets peaking at
progressively
shorter wavelengths, indicating hotter characteristic dust
temperatures (a similar trend was indicated \cite{2015dacunha}, although sample size did not allow for confirmation), which are further discussed in \S~\ref{dustTemp}. 

\begin{figure*}
  \includegraphics[width=1\textwidth]{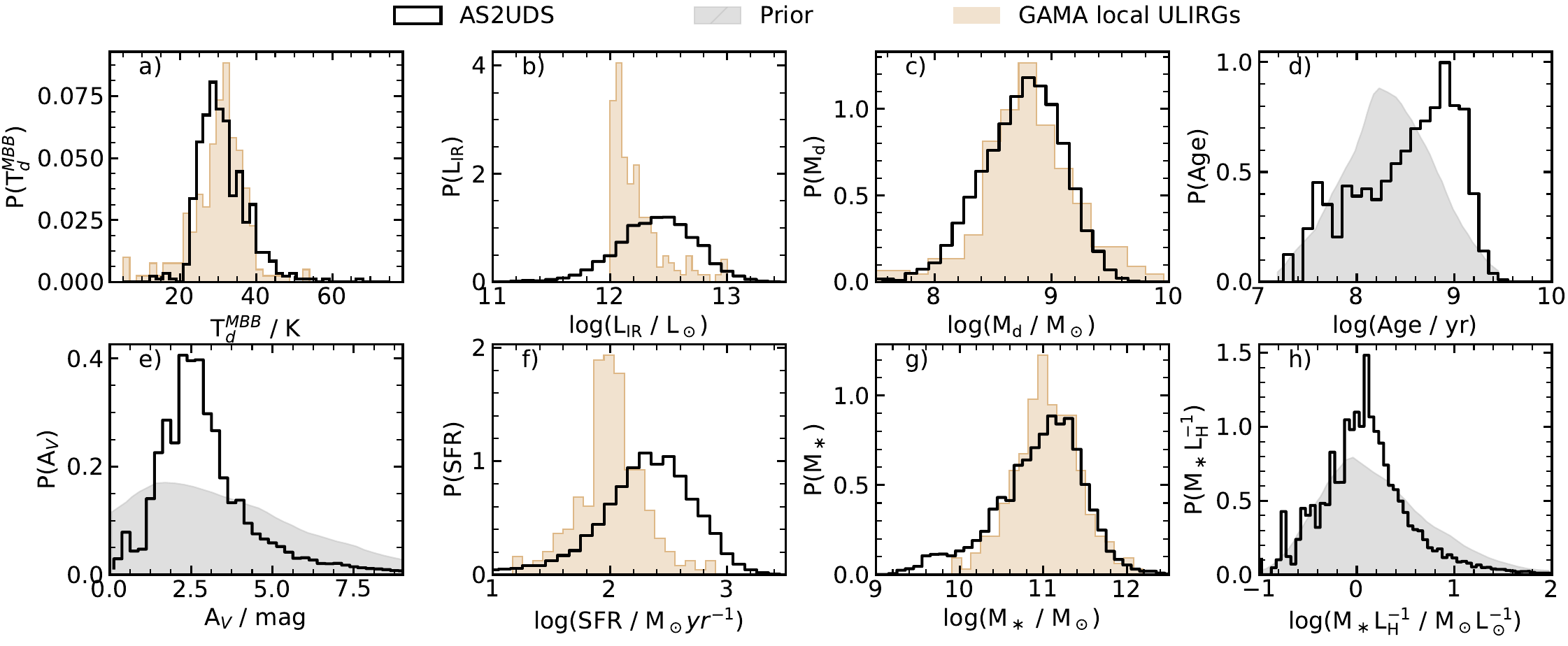}
  \caption{Distributions of the physical properties of the AS2UDS SMG
    population predicted by \textsc{magphys} or derived directly from the SEDs. To demonstrate that the posterior likelihood distributions are not affected by the model priors, we overlay them where applicable. For comparison the available properties from a sample of local ULIRGs from the GAMA survey \protect\citep{2018driver}, selected to have $L_{\rm IR} > $\,10$^{12}$\,L$_\odot$ and $z <$\,1, is
    also shown.
    In each panel, we plot the median stacked likelihood distribution from
    combining the  PDFs of the individual SMGs. 
    From top left the distributions show: (a) Optically thin modified blackbody
    temperature ($T_{\rm d}$) for those galaxies that are detected in at least one SPIRE band;
    (b) Far-infrared luminosity ($L_{\rm IR}$); (c) Stellar mass ($M_\ast$); (d) Mass-weighted
    age, (e) $V$-band dust attenuation ($A_V$); (f) Star-formation rate, g) Dust
    mass ($M_{\rm d}$), (h) restframe $H$-band mass-to-light ratio ($M/L_{H}$). 
    We see broad similarities between the properties of
    the SMGs and the local ULIRGs, with the exception that the SMGs (which have much
higher volume densities than the comparably luminous galaxies at $z<$\,1)
    are typically more luminous in the far-infrared and have
    correspondingly higher star-formation rates.}
\label{fig:distribution}
\end{figure*}

For comparison to our SMG composites,  we also show the SEDs
of  the local
starburst galaxies M82 and Arp\,220 (scaled to the same far-infrared
luminosity) in Fig.~\ref{fig:comp_sed}b. The full-sample AS2UDS SED
(and all three sub-sets) differs significantly
from M82, which is much brighter in the optical/infrared relative to the far-infrared than
a typical SMG.  
The full SED of
the SMGs is better matched to Arp\,220 in the rest-frame optical/near-infrared.
In the mid-infrared, Arp\,220 has a strong silicate absorption feature
at 9.8$\mu$m which falls in a poorly sampled part of our SED, where
the  predicted SED is dependent upon the
details of the model in {\sc magphys} (as this wavelength is only
sampled at $z<$\,1 by our MIPS coverage where we have few SMGs).  
However, the
limited 
mid-infrared spectroscopy available for SMGs indicates that
most do not show 
silicate absorption as strong as seen in Arp\,220
\citep{2008farrah,2009menendez}.
While in the far-infrared, the SED of Arp\,220 peaks at a shorter wavelength
($\lambda_{\rm peak}\sim$\,60$\mu$m) than the full SMG SED, which peaks at
$\lambda\sim $\,70--80$\mu$m, implying a hotter characteristic dust
temperature in Arp\,220. The far-infrared SED of Arp\,220 is better matched to the higher redshift bins with $z>$\,2.5 and the ratio $L_{\rm opt}$/$L_{\rm FIR}$ of Arp\,220 is similar to $z\sim$\,2.5 SMGs. 
Hence,  Arp\,220 template may be an appropriate 
template for SMG dust SED-fitting in the high-redshift regime ($z$\,>\,2.5), but is not well matched to
the typical SMGs below $z\sim$\,2.5.

\subsection{Physical properties} \label{properties}

The composite SEDs of  our SMGs shown in Fig.~\ref{fig:comp_sed} indicate potential
differences between the  properties of low- and high-redshift SMGs, suggesting
 evolutionary changes within the population (or the influence of sample selection).
In the following, we investigate the physical properties of SMGs and the variation within the population, to search for evolutionary trends.

To quantify the typical properties of the SMGs we begin by constructing
the stacked likelihood distributions of far-infrared luminosity
($L_{\rm{IR}}$), dust mass ($M_{\rm d}$), age, $V$-band dust attenuation ($A_V$), star-formation rate, stellar mass ($M_\ast$), and rest-frame $H$-band
mass-to-light ratio ($M/L_H$), and show these in
Fig.~\ref{fig:distribution}. We also include a histogram of the characteristic dust temperature from the modified blackbody fits ($T_{\rm d}^{\rm MBB}$), which is further explained in $\S$\ref{dustTemp}. By stacking the likelihood
distributions we include the
uncertainties (and covariance) between the derived values, including
the uncertainties in the photometric redshifts.  Where applicable, in
Fig.~\ref{fig:distribution} we also overlay the {\sc magphys} prior in
order to illustrate their potential influence on our derived distributions. We
note that the reliability of some of these derived quantities have been demonstrated by their correlation
with the observables as discussed in \S~\ref{der_props}, see also
Fig.~\ref{fig:obs_pred}.  

Before we discuss these derived quantities, we identify a
comparison sample of local ULIRGs with which we can compare these
distributions and average properties.  For this purpose we select 
ULIRGs from analysis of the GAMA survey undertaken
by \cite{2018driver}.
They used {\sc magphys} to model
the multi-wavelength 
photometry of this sample from rest-frame UV--radio wavelengths, including 
both, {\it GALEX} far-UV and {\it Herschel}/SPIRE far-infrared photometry, which broadly matches
the rest-frame wavelength coverage of the AS2UDS SMGs.  This similarity in the
multi-wavelength coverage and the use of the same SED modelling code
will minimise systematic uncertainties in any comparison 
of the properties of these local ULIRGs with high-redshift SMGs.  
The GAMA local ULIRG sample we use comprises 353 galaxies 
which have spectroscopic redshifts
of $z<$\,1 (with a median of $z=$0.59), are brighter than
$r\leq$\,19.8 (roughly equivalent to $H\sim$24 at z$\sim$2.5), have at least one PACS or SPIRE detection and have far-infrared luminosities $L_{\rm  IR}>$\,10$^{12}$\,L$_\odot$. For comparison, we plot the distributions of the available parameters for the local ULIRGs in Fig. \ref{fig:distribution}.
We similarly compare to previously published results on
two high-redshift ULIRG samples from  the  {\sc magphys} analyses of 
the 870\,$\mu$m ALMA sample in ALESS \citep{2015dacunha} and a 
comparably sized 1.1-mm selected SMG sample in COSMOS studied with 
ALMA by \cite{2017miettinen}.

\begin{figure*}
\centering
  \includegraphics[width=\textwidth]{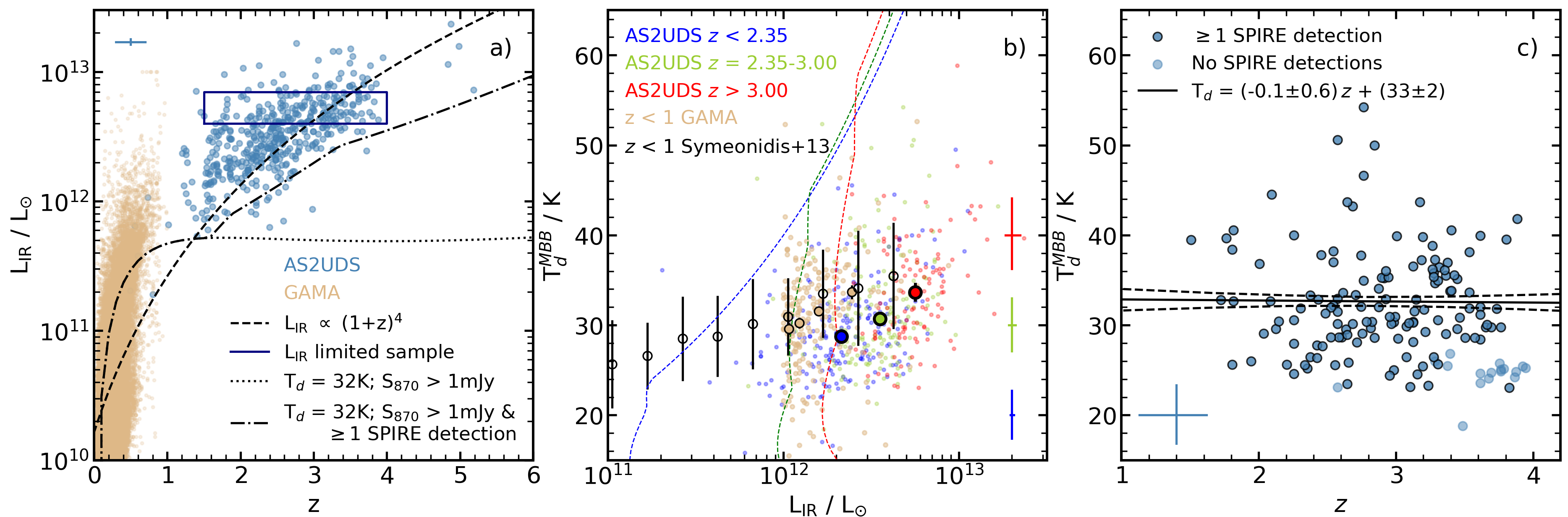}
  \caption{\textbf{(a)} The far-infrared luminosity of the AS2UDS SMGs as a function of redshift for those sources which have at least one SPIRE detection. The dashed line shows the luminosity evolution according to $L_{\rm IR}\propto (1+z)^4$. The dotted line denotes the selection function of a $S_{870}\gtrsim$\,1\,mJy SMG for a modified blackbody dust SED with the median dust temperature of the sample, $T_{\rm d}\,$\,$=$\,32K. While the dot-dashed line shows the selection including the requirement that the dust SED includes at least one detection above the flux limits of the available SPIRE observations at 250, 350 or 500\,$\mu$m.  We see that the latter model provides a reasonable description of the variation of the lower boundary in $L_{\rm IR}$ with redshift which we observe. The rectangle shows the limits of the unbiased, luminosity-limited sub-set we use below to test evolutionary trends.  We also show the low redshift ($z < $\,1) sample of far-infrared detected galaxies from the GAMA survey \protect\citep{2018driver} and note that we compare to the $\geq$10$^{12}$L$_\odot$ sub-set of these. 
\textbf{(b)} The temperature--luminosity relation of those AS2UDS SMGs with at least one SPIRE detection, split into three redshift bins: $z<$\,2.35, $z$\,$=$\,2.35--3.00 and $z>$\,3.00. We highlight the median values of each of the redshift sub-samples with their associated uncertainties. The selection function of AS2UDS sources with at least one SPIRE detection and $S_{870} \geq$\,1 mJy at redshifts of $z$\,$=$\,1, 2.35 and 3.00 are plotted.  We see an apparent evolution of the far-infrared luminosities and dust temperatures of the SMGs which is roughly parallel to the trend seen within each redshift slice and does not
appear to be influenced by the selection limits. For comparison, we plot the median values from local $z $\,$=$\,0--1 SPIRE-selected LIRGs and ULIRGs from \protect\citep{2013symeonidis} and the distribution with the median values derived for the ULIRGs from the GAMA survey.  These show that the 870-$\mu$m selected SMGs are between $\sim$\,4--7\,K cooler than comparably far-infrared luminous galaxies at $z<$\,1. 
\textbf{(c)} The variation in dust temperature with redshift for our complete luminosity-limited SMG sample, which lies within the rectangle plotted in panel a) ($L_{\rm IR} $\,$=$\,(4--7)\,$\times 10^{12}$\,L$_\odot$ and $z $\,$=$\,1.5--4.0). No evolution is seen in the dust temperature at fixed luminosity across this redshift range.}
\label{fig:ldust_td}
  \end{figure*}

\subsubsection{Far-infrared luminosity} \label{luminosity}

To investigate the dust properties of SMGs, we first determine their
far-infrared luminosities, which is derived by integrating the
rest-frame SED between 8--1000\,$\mu$m.  For our sample, the median
far-infrared luminosity is $L_{\rm
  IR}$\,$=$\,(2.88\,$\pm$\,0.09)\,$\times$\,10$^{12}$\,L$_\odot$, with
a 16--84$^{\rm th}$ per centile range of $L_{\rm
  IR}$\,$=$\,(1.5--5.4)\,$\times$\,10$^{12}$\,L$_\odot$. 
The vast majority of our sample are classed as ULIRGs with
$L_{\rm  IR}$\,$=$\,1--10\,$\times$\,10$^{12}$\,L$_\odot$, with
53 (7 per cent) being LIRGs with $L_{\rm  IR}<$\,1\,$\times$\,10$^{11}$\,L$_\odot$ typically at $z\sim$\,1.8,
and 14 (2 per cent) are HyLIRGs with $L_{\rm  IR}>$\,1\,$\times$\,10$^{13}$\,L$_\odot$ lying at $z\sim$\,3.5. 
Comparison to local ULIRGs of Fig. \ref{fig:distribution}b shows that local ULIRGs have considerably lower far-infrared luminosities with a median of  $L_{\rm
  IR}$\,$=$\,(1.41\,$\pm$\,0.03)\,$\times$\,10$^{12}$\,L$_\odot$ and a 16--84$^{\rm th}$ percentile range of $L_{\rm IR}$\,$=$\,(1.1--2.4)\,$\times$\,10$^{12}$\,L$_\odot$.

Restricting  the sample at the original SCUBA-2 single-dish flux density limit
of $S_{850}>$\,3.6\,mJy, yields 364 SMGs with a median of $L_{\rm
  IR}$\,$=$\,(3.80\,$\pm$\,0.14)\,$\times$\,10$^{12}$\,L$_\odot$.
    In the following analysis, we also make use of a sub-set of our sample which are detected in at least one of the {\it Herschel} SPIRE bands, as these sources have more reliable measurements of their dust temperatures and hence their far-infrared luminosities. There are 475 SMGs in this SPIRE-detected subset with a median $L_{\rm
  IR}$\,$=$\,(3.39$\pm$0.14)\,$\times$\,10$^{12}$\,L$_\odot$ and a 68$^{\rm th}$ percentile range of $L_{\rm
  IR}$\,$=$\,(1.7--5.9)\,$\times$\,10$^{12}$\,L$_\odot$ and lying at a median
  redshift of $z$\,$=$\,2.48\,$\pm$\,0.08 (68$^{\rm th}$ percentile range of $z$\,$=$\,1.8--3.2).

The median far-infrared luminosity of our SMGs is comparable with that
derived  for the sample in ALESS $L_{\rm IR
}$\,$=$\,(3.7\,$\pm$\,0.1)\,$\times$\,10$^{12}$\,L$_\odot$ for a similar
870$\mu$m flux density limit \citep{2015dacunha}, and also comparable
to the median far-infrared luminosity of the 1.1-mm selected SMG sample
from \citet{2017miettinen}\footnote{Note that the errors on Miettinen's values are
the 16--84$^{\rm th}$ percentile ranges, rather than the uncertainty in the median value.}  who derive a median of $L_{\rm
  IR}$\,$=$\,(4.0$\pm$\,0.3)\,$\times$\,10$^{12}$\,L$_\odot$ for a sample
with an equivalent 870\,$\mu$m flux density range of 1.5--20\,mJy (adopting $S_{870}/S_{1100}\sim$\,1.8, equivalent to
a $\nu^{-2.5}$ spectral index, based on the average flux ratio of AS2UDS SMGs with published 1.1-mm photometry from ALMA in \citealt{2017ikarashi}).

To illustrate the evolution in our sample, we plot the variation of far-infrared luminosity with redshift for the AS2UDS SMGs in Fig.~\ref{fig:ldust_td}a. We include in this plot only those SMGs which have at least one SPIRE detection. The SMGs show a trend in redshift for the brightest luminosities which is roughly reproduced by evolution of the form $L_{\rm IR}\propto(1+z)^\gamma$, with $\gamma\sim$\,4, consistent with the behaviour previously claimed
for luminous dusty galaxies at $z<$\,2 \cite[e.g.][]{2011bethermin}.  
We also need to consider the influence of our sample selection on this trend and so we also show
in Fig.~\ref{fig:ldust_td}a the far-infrared luminosity of a source
with a dust SED modelled by a modified blackbody with a temperature of $T_{\rm d}$\,$=$\,32\,K (the median for this sample) and an 870\,$\mu$m flux density $S_{870}=$\,1\,mJy, which is the typical completeness level of our ALMA maps. We see that due to the negative $k$-correction the resulting far-infrared luminosity limit is almost
constant out to $z \sim$\,6.   
In addition, we overlay a selection function for the same $T_{\rm d}$\,$=$\,32\,K model with the additional constraint that the SED must be detected in at least one SPIRE band at 250, 350 or 500\,$\mu$m, which is the requirement placed on the sub-set of the AS2UDS sample we are plotting.   We see that this selection  results in an increasing 
far-infrared luminosity limit at higher redshifts, which reproduces the behaviour we see in our sample.
Hence, the apparent deficit in Fig.~\ref{fig:ldust_td}a of lower luminosity sources (with $L_{IR}\lesssim$\,2--3\,$\times$\,10$^{12}$\,L$_\odot$) at $z\gtrsim$\,2.5--3, can be accounted for
by the sample selection. While the SPIRE-detected subset of our SMG sample is biased towards more luminous sources at higher redshifts, we retain this selection because these have more robust estimates of their far-infrared properties.  However, to control for the resulting bias in far-infrared
luminosity with redshift, and so assess evolutionary trends, we will also exploit our large sample to construct an {\it unbiased} sample of SMGs at $z$\,$=$\,1.5--4, selected to lie in a narrow range of far-infrared luminosity ($L_{\rm IR}$\,$=$\,4--7\,$\times$\,10$^{12}$\,L$_\odot$), where our sample is
complete with respect to the SPIRE detection limits (this selection is shown by the rectangle plotted in Fig.~\ref{fig:ldust_td}a).

\subsubsection{Characteristic dust temperature}\label{dustTemp}

Although {\sc magphys} can estimate a characteristic dust temperature, it is derived from a complex calculation involving five free parameters which describe the temperature and luminosity contributions from the warm (birth cloud) and cold (diffuse inter-stellar medium) components. The far-infrared SEDs of our sources are covered by at most six photometric bands, thus we choose to adopt a simpler, more conservative approach and fit modified blackbody functions to the available {\it Herschel} PACS and SPIRE, and ALMA 870-$\mu$m photometry. This approach also has an added advantage that the dust SEDs of the comparison samples can be fitted
in a very similar way, allowing for more reliable comparison, free from systematic uncertainties resulting from the fitting procedures. We use a modified blackbody function of the form: 
\begin{equation}
    S_{\nu_{\mathrm{obs}}} \propto (1 - e^{-\tau_{\mathrm{rest}}}) \times B(\nu_{\mathrm{rest}}, T),
    \label{eq:bb}
\end{equation}

where $B(\nu_{\mathrm{rest}}, T)$ is the Planck function, $\tau_{\mathrm{rest}}$ is the frequency-dependent optical depth of the dust of the form $\tau_{\mathrm{rest}}$\,$=$\,$\Big(\frac{\nu_{\rm rest}}{\nu_0}\Big)^\beta$, $\nu_0$ is the frequency at which optical depth is equal to one and $\beta$ is the dust emissivity index. We adopt $\beta$\,$=$\,1.8 as used in previous SMG studies and consistent with the finding for local star-forming galaxies \citep{2011planck,2013clemens,2013smith}. Making the  assumption that the region from which the dust emission originates is optically thin, thus $\nu_0 \gg \nu_{\rm rest}$, Eq.~\ref{eq:bb} simplifies to: 
\begin{equation}
    S_{\nu_{\mathrm{obs}}} \propto \nu_{\mathrm{rest}}^\beta \times B(\nu_{\mathrm{rest}}, T),
\end{equation}

The dust temperature derived using the optically-thin approximation does not represent the true temperature of the dust emission regions, as \cite{2013riechers,2017simpson} and others have shown that the emission from SMGs is, on average, optically thick at $\lambda_0 \lesssim$\,75\,$\mu$m (we explore this further in $\S$~\ref{dust_gas}). Thus, this estimate is only a simplified model which we will refer to as the characteristic dust temperature. The best-fit temperature is acquired by fitting this modified blackbody function using a Markov Chain Monte Carlo sampler (see \citealt{2017simpson}).  

The resulting characteristic temperature distribution for our SPIRE-detected SMGs is shown in Fig.~\ref{fig:distribution}a. Comparing the dust temperatures  for the SMGs from  the modified blackbody fits with the predicted characteristic dust temperature from {\sc magphys}, we find a typical fractional difference of $({T^{\rm MBB}_{\rm d}-T^{\rm MAGPHYS}_{\rm d}})/T^{\rm MBB}_{\rm d} =-$0.28$\pm$0.01.
The median characteristic dust temperature for our ALMA SMGs with at least one SPIRE detection is $T^{\rm MBB}_{\rm d}$\,$=$\,30.4\,$\pm$\,0.3\,K with a 68$^{\rm th}$ percentile range of $T^{\rm MBB}_{\rm d}$\,$=$\,25.7--37.3\,K, this is shown in Fig.~\ref{fig:distribution}a. For comparison, the same method to derive characteristic dust temperature was applied to the local ULIRGs sample. The median temperature of the local ULIRGs sample is slightly higher but within error range to SMGs, with a median characteristic dust temperature of $T^{\rm MBB}_{\rm d}$\,$=$\,31.1\,$\pm$\,0.4\,K. However, we stress that the typical far-infrared luminosity of the GAMA ULIRGs is a factor of 2--3\,$\times$ lower than the SMGs and, as we discuss below, when we compare $L_{\rm IR}$-matched samples then the local ULIRGs are on average hotter than the SMGs.

Due to the similarities in their physical properties (e.g. stellar mass and dust mass, see Fig.~\ref{fig:distribution}), SMGs have been proposed to be analogues of the local ULIRGs. Indeed, as seen in Fig.~\ref{fig:comp_sed}b the SED for at least one archetypal ULIRG, Arp\,220, shares some similarities with the higher-redshift SMGs. Local ULIRGs exhibit a dust temperature--luminosity relation, so we now investigate the correlation between far-infrared luminosity and characteristic dust temperature for the AS2UDS SMGs in Fig.~\ref{fig:ldust_td}b.
We find a positive correlation between far-infrared luminosity and dust temperature for the AS2UDS SMGs similar to previous SMG studies \citep[e.g][]{2012magnelli,2013symeonidis,2014swinbank}.  Moreover, we see a correlation between luminosity and temperature within each of the three
redshift sub-sets and a similar trend between the medians of the three sub-sets. We also show in Fig.~\ref{fig:ldust_td}b the  selection functions for three redshifts which
illustrate the selection  of our 870$\mu$m observations  as a function of redshift, far-infrared
luminosity and dust temperature. Comparing these selection boundaries to our 
SPIRE-detected samples indicates that they should not be strongly influencing the correlations we observe. Indeed, when we look at the variation
of $T_{\rm d}$ with $L_{\rm IR}$ for our unbiased luminosity-limited sub-sample, we find a similarly strong  $L_{\rm IR}$--$T_{\rm d}$ trend, $\Delta T_{\rm d} \sim$\,12\,$\Delta \log_{10}(L_{\rm IR})$.   

Fig.~\ref{fig:ldust_td}b also shows the  $L_{\rm IR}$--$T_{\rm d}$   distribution for 
$z $\,$=$\,0--1 SPIRE-selected LIRGs and ULIRGs from \cite{2013symeonidis} and the   $z < $\,1 ULIRGs from the GAMA survey.  These show a significant offset in dust temperature at a fixed luminosity relative to the SMGs: $\Delta T_{\rm d}=$\,3\,$\pm$\,1\,K for
samples with $L_{\rm IR}\sim$\,2--4\,$\times$\,10$^{12}$\,L$_\odot$. This comparison ought not to be influenced by the
selection limits on our SMG sample, although we have not modelled those for the local samples. We note, that the temperature difference between the samples is comparable with the uncertainty derived from {\sc eagle} comparison, however, this is a systematic offset at all luminosity bins. So we tentatively conclude that at a fixed luminosity the AS2UDS SMGs
appear to show cooler median dust temperatures than the local samples, possibly due to more compact dust distribution in local ULIRGs \citep{2009iono,2014wilson}. 

\begin{figure*}
\centering
  \includegraphics[width=\textwidth]{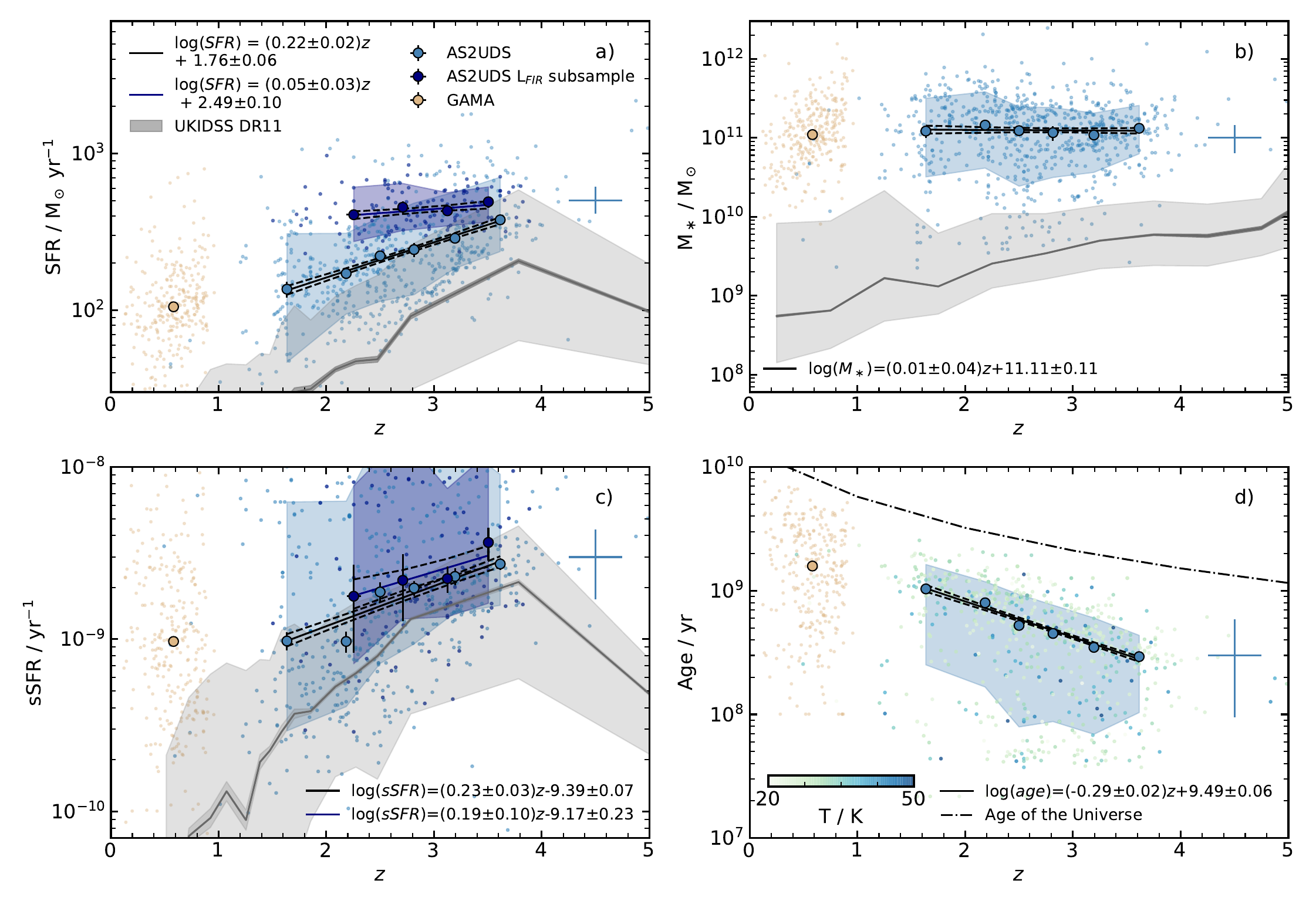}
  \caption{The evolution of the {\sc magphys} stellar parameters with redshift. In each panel, the large circles show the median in bins with equal number of sources. The sample of $z<$\,1 ULIRGs from the GAMA survey and their median are similarly shown. The solid line shows the best-fit to the the binned medians and the dashed lines show the associated uncertainty. The median error on any individual source is shown on the right side of each panel. The 16--84$^{\rm th}$ percentile range of the given property are shown as blue shaded regions.
  In panels (a), (b) and (c) we plot the median trend derived from the UKIDSS DR11 field population with the error on the median in the grey shaded region and their associated 16--84$^{\rm th}$ percentile range indicated as the light shaded region. 
\textbf{(a)} Star-formation rate versus redshift for the SPIRE-detected SMG sample and the unbiased luminosity-selected sub-sample. As with the variation of $L_{\rm IR}$ with redshift shown earlier, there is a highly significant increase in median SFR with redshift within our sample, however, when the sample is limited to the unbiased luminosity-selected sub-sample this trend disappears. In comparison to a $K$-selected sample we again see that typical SMGs are significantly above the median trend seen in ``normal'' star-forming field galaxies at all redshifts.  
\textbf{(b)} Stellar mass versus redshift for the AS2UDS SMGs. We see no strong variation in the estimated stellar mass of the SMGs with redshift, with this extending down to the $z<$\,1 ULIRGs. In comparison to $K$-band selected sample, SMGs have significantly higher stellar masses at all redshifts.
\textbf{(c)} Specific star-formation rate versus redshift, which shows a moderate (5.5$\sigma$) trend of $d({\log_{10}(\rm sSFR}))/dz$\,$=$\,0.23$\pm$0.03(. This trend weakens as we limit the sample to the unbiased luminosity-selected sub-set. SMGs lie above the median of a $K$-band selected sample out to $z\sim$\,3--4. 
\textbf{(d)} Mass-weighted age versus redshift. The median derived age for the SMGs is (4.6$\pm$0.2)\,$\times$\,10$^8$\,yr and the best-fit line has a gradient of $d({\log_{10}\rm (Age_{\rm m}}))/dz=-$0.29$\pm$0.02.  The AS2UDS points are coloured by dust temperature, showing that the strength of this trend could be partly due to the model assigning younger ages to galaxies with higher dust temperatures (and far-infrared luminosities), which are typically found at higher redshifts. The dashed line shows the maximum formation redshift allowed by {\sc magphys}, which corresponds to a cosmological lookback time of 13.4\,Gyrs at $z\sim$0. 
}
\label{fig:pars_z}
\end{figure*}

The median values at each redshift slice in Fig.~\ref{fig:ldust_td}b may suggest a trend of characteristic dust temperature with redshift. Thus, we select the unbiased luminosity-limited sample sources and plot the variation of their dust temperature with redshift in Fig.~\ref{fig:ldust_td}c. No evolution in the dust temperature at a fixed luminosity is seen in this redshift range.

\subsubsection{Star-formation rate}

Far-infrared luminosity traces dust-obscured star formation, thus it is possible to infer star-formation rates using the conversion from $L_{\rm IR}$ given in \cite{1998kennicutt}. Models in {\sc magphys}, however, allow  dust heating by old stellar populations and thus the model also estimates the star-formation rate in the optical regime after accounting for dust attenuation. Even though a wide range of model star-formation histories were included, we find a good correlation between the far-infrared and {\sc magphys} derived star-formation rates for the SPIRE-detected sub-set, with a dispersion of $\sim$\,25 per cent estimated from the 16--84$^{\rm th}$ percentile range. We determine a median star-formation rate of SFR\,$=$\,290$\pm$14\,M$_\odot$\,yr$^{-1}$ with a 68$^{\rm th}$ percentile range of SFR\,$=$\,124--578\,M$_\odot$\,yr$^{-1}$ (based on the SPIRE-detected sub-sample) which is consistent with \cite{2015dacunha} who found SFR\,$=$\,280$\pm$70\,M$_\odot$\,yr$^{-1}$ for the ALESS sample. In comparison to the local ULIRG sample from GAMA, the typically higher far-infrared luminosities of our SMGs suggest higher star-formation rates, which is indeed the case, with the former having a median star-formation rate of 108$\pm$4\,M$_\odot$\,yr$^{-1}$ (see Fig.~\ref{fig:distribution}f). 

We investigate the evolution of star-formation rate in the SMGs with redshift in Fig.~\ref{fig:pars_z}a. We also include the local ULIRGs sample, and as noted earlier, we observe that local ULIRGs typically have lower star-formation rates than seen in the SMGs in our sample. The best-fit line with a gradient of $d(\log_{10}({\rm SFR}))/dz$\,$=$\,0.22$\pm$0.02 indicates a significant 11-$\sigma$ trend. However, as seen in \S\ref{luminosity}, our selection affects the trends seen with redshift. When we limit our sample to the unbiased luminosity-selected sample, we observe no significant star-formation rate evolution with $d(\log_{10}({\rm SFR}))/dz$\,$=$\,0.05$\pm$0.03, as seen in Fig. \ref{fig:pars_z}a. We compare the star-formation rates of SMGs at different redshifts with the UKIDSS DR11 field sample. For this comparison we select field galaxies which have stellar masses above the 16$^{\rm th}$ percentile value of the AS2UDS sample ($M_\ast >$\,3.5\,$\times$\,10$^{10}$\,M$_\odot$). The shaded regions shows the 16--84$^{\rm th}$ percentile ranges of the SMGs and the field sample in their respective colours. As seen in Fig.~\ref{fig:pars_z}a, the typical SMGs in our sample have significantly higher star-formation rate than a mass-selected sample at all redshifts probed.  

\subsubsection{Stellar Emission Properties}

Next, we look at the physical properties inferred from the stellar emission which typically dominates
the rest-frame UV/optical/near-infrared region of the SED of galaxies. We investigate the derived stellar masses as it is one of the most fundamental properties of SMGs. Robust stellar masses can provide tests of the evolutionary links between the SMGs and field galaxies, such as determining the fraction of massive galaxies which may have evolved through an SMG-like phase.

The median stellar mass of the full AS2UDS sample is $M_\ast $\,$=$\,(12.6\,$\pm$\,0.5)\,$\times$\,$10^{10}$\,M$_\odot$ and we see no strong variation in this with redshift, as shown in Fig.~\ref{fig:pars_z}b.
Our median mass is in good agreement with the 1.1-mm selected sample from \cite{2017miettinen} who find median a stellar mass of $M_\ast$\,$=$\,12$^{+19}_{-9}\times10^{10}$\,M$_\odot$ and also see no evolution with redshift in their sample.
However, our derived mass is higher than the $M_\ast$\,$=$\,(8.9$\pm$0.1)\,$\times$\,10$^{10}$\,M$_\odot$  found by \cite{2015dacunha}. Limiting both samples to the same 870-$\mu$m flux limit doesn't eliminate this disagreement, but we note that due to the broad distribution of $P(M_\ast$) there is a wide range of acceptable stellar masses for our sample, the 16--84$^{\rm th}$ percentile range for AS2UDS being 5.9\,$\times$\,10$^{10}$ to 22\,$\times$\,10$^{10}$\,M$_\odot$ (see Fig.\ref{fig:distribution}). This difference may, therefore, be due to either sampling statistics or cosmic variance.

When comparing with local ULIRGs (see Fig.~\ref{fig:distribution}g), we see no significant differences in the distributions of stellar mass, even though the Universe is roughly three times older at the epoch of the GAMA population than it was at the era when the SMGs peak. However, the $r=$19.2\,mag limit of GAMA is $\sim$1.5\,mag brighter than our equivalent $H$-band limit (at $z\sim$\,2.5), thus GAMA may be biased to higher stellar masses.

Next, we investigate the attenuation of stellar emission at UV to near-infrared wavelengths, which, in {\sc magphys}, is estimated using a two-component model of \cite{2000charlot&fall}. The two components model the effective attenuation in the $V$-band from dust in both stellar birth clouds and in the diffuse ISM.
The median $V$-band dust attenuation derived for the AS2UDS sample is $A_V$\,$=$\,2.89$\pm$0.04\,mags with a 68$^{\rm th}$ percentile range of $A_V$\,$=$\,1.89--4.24\,mags. The posterior likelihood distribution is significantly more peaked than the prior (see Fig.~\ref{fig:distribution}e).
Moreover,  the prediction, shown in Fig.~\ref{fig:lz_field}, that the vast majority of SMGs are
indeed far-infrared luminous based solely on the {\sc magphys} modelling of the $U$--8.0\,$\mu$m SEDs provides strong support that the derived $A_V$ have diagnostic power as this parameter is the main driver of that prediction. As expected, we find that optically brighter SMGs (those with more than three detections in the optical/near-infrared bands) have $A_V$\,$=$\,2.5$\pm$0.2\,mags, while the optically fainter examples (fewer than four detections) have $A_V$\,$=$\,3.6$\pm$0.2\,mags. We note that this estimate of reddening is an angle averaged dust attenuation, which is measured by the classical definition -- comparing the intrinsic and obscured $V$-band magnitudes. Even though they are lower limits on the true total extinction (as they give lower weight to more extinct emission), they are still significant and underline the difficulty of constructing robust mass-limited samples of high redshift galaxies in the face of the significant dust obscuration found in some of the most active and massive examples.   

We compare our sample to the 99 ALESS SMGs from \cite{2015dacunha} 
which yields a median $A_V$\,$=$\,1.9$\pm$0.2 mags (restricting this analysis to the sub-set of 52 of these SMGs with spectroscopic redshifts from \citealt{2017danielson} doesn't change this estimate).  This is significantly lower than the median reddening derived for the AS2UDS galaxies, although both distributions span a similar range in $A_V$. This difference does not seem to relate to differences in the 870\,$\mu$m flux density limits, redshift distributions or stellar masses of the two samples.  Instead, it appears to reflect a population of highly obscured, $A_V\gtrsim$\,3.5, SMGs at $z\lesssim$\,3.0, which are seen in AS2UDS, but are absent in ALESS. 

Finally, to measure the overall age of a given source, {\sc magphys} outputs a mass-weighted age, which depends strongly on the form of the star-formation history. We find a median age for our SMGs of Age$_{\rm m} =$\,(4.6$\pm$0.2)\,$\times$\,10$^8$\,yr. We note that the posterior likelihood distribution differs significantly from the prior in Fig.~\ref{fig:distribution}d, suggesting the model is varying this parameter when fitting the SED. We return to compare the mass-weighted ages of the SMGs to other estimates of age from the derived physical properties in \S \ref{ages}.

\subsubsection{Comparison with the ``main sequence''}

We wish to relate the SMG population to the more numerous and less active and massive galaxies seen in the field population.  One tool to do this is to assess the distribution of this population on the stellar mass versus star-formation rate plane, in particular, the relative position of the SMGs compared to the broad
relation between star-formation rate and stellar mass exhibited by star-forming galaxies (the so-called ``main sequence'', \citet{2007daddi}) -- as
assessed through their relative specific star-formation rates (${\rm sSFR} = {\rm SFR}/M_\ast$). Specific star-formation rates significantly above the median trend of the field population have been argued to be a signature of starburst activity, potentially resulting from galaxy-galaxy mergers and interactions which enhance the star-formation rates of galaxies and so increase their sSFR  \citep{2012magnelli}. Alternatively, it may be possible for galaxies to achieve high star-formation rates without such triggers, merely through significant gas accretion -- enabling high star-formation rate systems to inhabit the high-mass end of the sequence of normal star-forming galaxies \citep{2010dave}. Alternatively, samples of highly star-forming galaxies could represent a heterogeneous mix of these two classes of systems, encompassing both physical processes  \cite[e.g.][]{2011hayward,2015narayanan,2019mcalpine}.

\begin{figure*}
\centering
  \includegraphics[width=\textwidth]{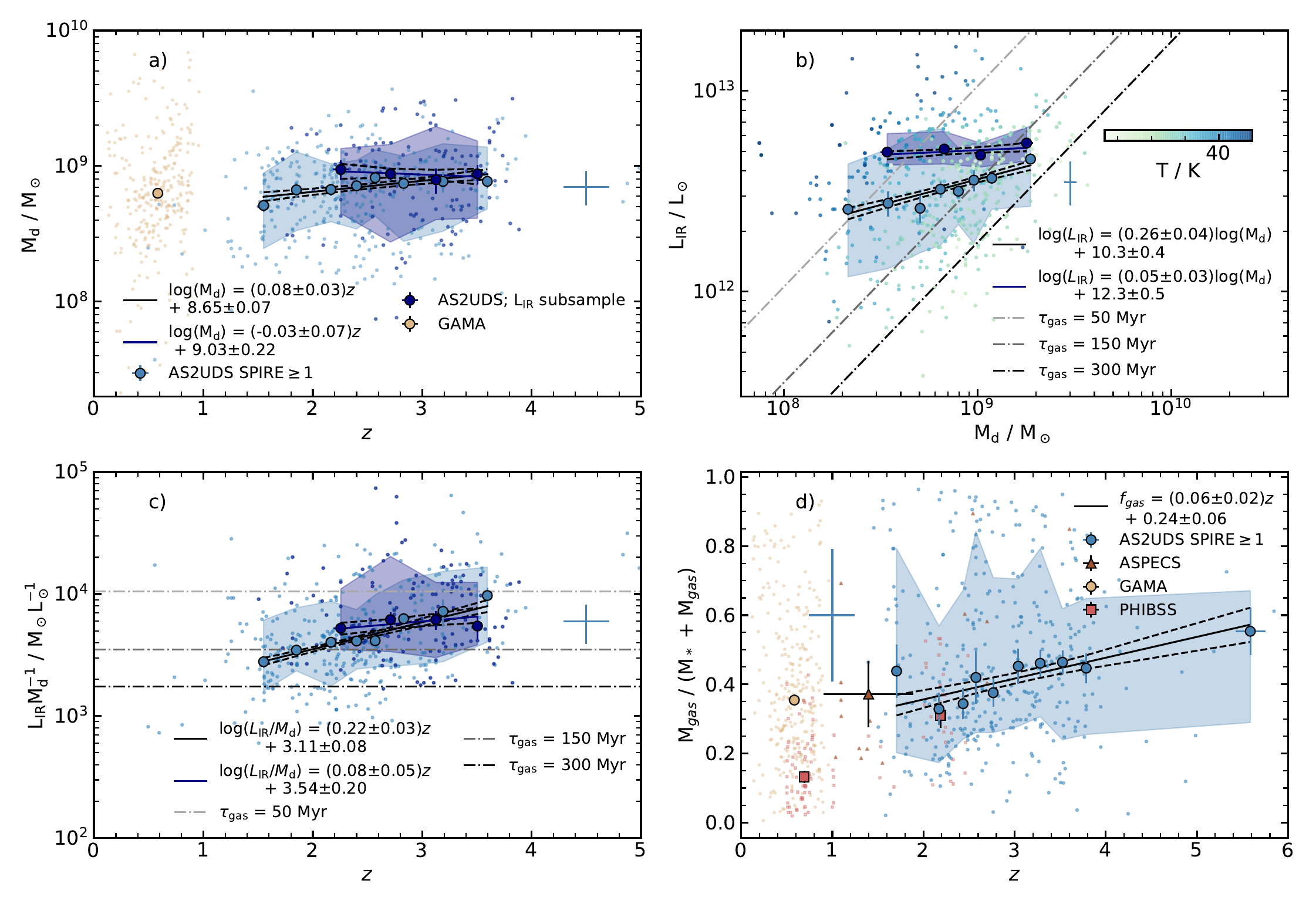}
\caption{The redshift evolution of the {\sc magphys} predicted dust properties for those AS2UDS SMGs detected in at least one SPIRE band and the unbiased luminosity-selected sub-sample of SMGs. In each panel the large circles show the binned median in bins with equal numbers of sources,  the solid line shows the best-fit line to the binned data and the dashed lines show the associated uncertainty. The 16--84$^{\rm th}$ percentile range of the given property are shown as blue shaded regions. The sample of $z<$\,1 ULIRGs from the GAMA survey and their median are similarly shown. The median error on any individual AS2UDS source is shown in each panel.
\textbf{(a)} Dust mass as a function of redshift. We see that the 870-$\mu$m selection of our joint S2CLS/ALMA survey yields a uniform selection in dust mass with redshift over the full redshift range probed by our study, with no evolution seen in the median dust mass with redshift. 
\textbf{(b)} Dust mass versus far-infrared luminosity. Our sample spans an order-of-magnitude range in both $L_{\rm IR}$ and $M_{\rm d}$ with a weak positive correlation with a
slope of 0.26$\pm$0.04. We note that the dispersion is driven in part by variations in dust temperature, whereby SMGs that have lower temperatures are observed to have higher dust masses for a given far-infrared luminosity. We also indicate lines of constant gas depletion. 
\textbf{(c)} The ratio of far-infrared luminosity to dust mass, a proxy for star-formation efficiency (or the inverse of gas depletion timescale), versus redshift. We again indicate lines of constant gas depletion. We see a strong increase in star-formation efficiency with increasing redshift within our SPIRE-detected sample with a gradient of 0.22\,$\pm$\,0.03. However, this trend weakens significantly if we restrict our analysis to the unbiased, luminosity-limited sub-set suggesting that it is driven primarily by the variation in $L_{\rm IR}$ or star-formation rate with redshift. 
\textbf{(d)} Gas fraction as a function of redshift. We derive a median gas mass fraction of $f_{\rm gas}$\,$=$\,0.41\,$\pm$\,0.02 with a 68$^{\rm th}$ percentile range of $f_{\rm gas}$\,$=$\,0.24--0.72 and we see modest evolution of this quantity with redshift, with a gradient of 0.06$\pm$0.02. For comparison we overlay results from the ASPECS blind CO-survey  from \protect\cite{2019aravena} and the
CO-detected typical star-forming galaxies at $z \sim$\,1--3 from \protect\cite{2018tacconi}.
}
\label{fig:FIRpars_z}
\end{figure*}

In Fig.~\ref{fig:pars_z}c we look at the variation with redshift in the sSFR estimates for the SMGs compared to the less-active field population (derived in the same manner with robustness tested in \S~\ref{test_fir}). We observe a modest positive correlation between specific star-formation rate and redshift for the SMGs.  As shown in Fig.~\ref{fig:pars_z}a, we find no evolution of stellar mass with redshift for our
SMG sample, thus this trend is set by the variation in the star-formation rate. However, as we showed before, the latter is due to selection effects on far-infrared luminosity. Indeed, when we limit our sample to the unbiased luminosity-selected sample, we find no significant trend in sSFR with redshift, with the SMG population on average spanning an order of magnitude in sSFR at all redshifts.

We compare the SMGs to the distribution from our
{\sc magphys} analysis of the $K$-band selected sample of galaxies in the UDS field, which we take to represent the ``main sequence'' (consistent with \citealt{2015tasca}). For a consistent comparison, we select field galaxies that have stellar masses above the 16$^{\rm th}$ percentile of the AS2UDS sample ($M_\ast \gtrsim$\,3.5\,$\times$\,10$^{10}$\,M$_\odot$). Fig.~\ref{fig:pars_z}c demonstrates that field population has lower median sSFR values at all redshifts, but the difference between two populations decreases with redshift and SMGs lie close to the evolved ``main sequence'' at $z \gtrsim$\,3.5, at which point the number density of SMGs in our sample is declining rapidly (see also \citealt{2015dacunha}). This suggests that the bulk of the SMG population we detect brighter than $S_{870}\sim$\,1\,mJy have specific star-formation rates which put them above the ``main sequence'' at their respective redshifts. 
Indeed, using the sources in our 16--84$^{\th}$ percentile range of $z$\,$=$\,1.8--3.4, we find that 82\,$\pm$\,4 per cent lie above the ``main sequence'' defined by the $K$-band selected sample, with 34$\pm$3 per cent lying more than a factor of four above it (the arbitrary definition of a ``starburst''). 

\subsubsection{Dust and gas masses} \label{dust_gas}

We now investigate the properties of dust and gas in SMGs. Dust mass estimates, together with properties calculated from stellar emission, allow us to assess how efficient SMGs are at forming stars from gas, which in turn can provide a constraint on the lifespan of the sub-millimetre luminous phase in these systems.

We derive a median dust mass for  the full AS2UDS sample  of $M_{\rm d}$\,$=$\,(6.8$\pm$0.3)\,$\times$\,10$^8$\,M$_\odot$ with a 68$^{\rm th}$ percentile range of $M_{\rm d}$\,$=$\,(3.0--12.6)\,$\times$\,10$^8$\,M$_\odot$, which broadly agrees with the median
estimated for the ALESS sample: $M_{\rm d}$\,$=$\,(5.6$\pm$0.1)\,$\times$\,10$^8\,$\,M$_\odot$ \citep{2015dacunha}. Similarly, \cite{2017miettinen} estimate a median dust mass of $M_{\rm d}$\,$=$\,10$^{+6}_{-5}\times$\,10$^8$\,M$_\odot$ for their 1.1-mm selected SMG sample, which again is similar to our measurement. It is expected that dust mass will be closely correlated with sub-/millimetre flux density, hence this agreement may simply reflect the roughly similar effective flux density limits of the single-dish surveys followed-up in these three ALMA studies.   

Indeed, in Fig.~\ref{fig:FIRpars_z}a we see a relatively tight lower boundary to the distribution (for the $S_{870}\geq$\,3.6\,mJy sample this corresponds
to $M_{\rm d}$\,$\geq$\,5\,$\times$\,10$^{8}$\,M$_\odot$), confirming that 870\,$\mu$m flux density selection provides an
approximately uniform dust mass selection across a wide redshift range.  The ratio of dust mass to 870\,$\mu$m flux density gives a simple conversion between the observable and the intrinsic property of $\log_{10}(M_{\rm d}[M_\odot])\,=\,(1.20\,\pm\,0.03)\,\times \log_{10}(S_{\rm
870}[{\rm mJy}])\,+\,8.16\,\pm\,0.02$, with a 1-$\sigma$ dispersion of $\sim$\,30 per cent, within the error derived from {\sc eagle} comparison.
We also see only a moderate increase in dust mass with redshift in our sample, corresponding to $\sim$\,30 per cent across the redshift range $z$\,$=$\,1.8--3.4.  
This is qualitatively consistent with the variation in median redshift with $S_{870}$ flux density found by \cite{2019stach} -- who demonstrated
that SMGs from AS2UDS which are brighter at 870\,$\mu$m on-average lie at higher redshifts. However, this trend weakens when we only consider the unbiased luminosity-selected sub-sample (see Fig.~\ref{fig:FIRpars_z}a).

We note that if the gas-to-dust ratio of this strongly star-forming population does not vary significantly over this redshift range, then our 870\,$\mu$m selection will correspond to a similarly uniform selection in terms of molecular gas mass. The conversion factor from dust mass to molecular gas mass has been derived for a small sample of high-redshift SMGs with CO(1--0) observations, yielding a gas-to-dust ratio, $\delta _{\rm gdr}$, of $\sim$\,100 \citep{2014swinbank} similar to that estimated for Arp\,220, which we adopt for our study. We note that the gas-to-dust ratio is expected to vary as a function of metallicity and hence potentially stellar mass and redshift.  However, the lack of reliable gas-phase metallicities for SMGs means we choose to adopt a fixed ratio in our analysis.

We see an order of magnitude range in both $M_{\rm d}$ and $L_{\rm IR}$ across our sample in Fig.~\ref{fig:FIRpars_z}b with a weak correlation between these two parameters, although there is a clear variation across the distribution in terms of dust temperature. We also overlay onto the figure lines corresponding to constant gas depletion (or star-formation efficiency), which we estimate assuming half of the gas is available to form stars (the other half being expelled) \citep{2002Pettini}: 

\begin{equation}
    \tau_{\rm dep}\sim \frac{0.5 \times M_{\rm gas}}{\rm SFR}
\end{equation}
     
We see that the population spans a range of 
a factor of six in gas depletion timescales, from 50 to 300\,Myr with a median of 146$\pm$5\,Myr.  Hence, the estimated length of the SMG phase, assuming the sources are typically seen half-way through this period, is $\sim$\,300\,Myr.
We also observe that the SMGs with the hottest characteristic dust temperatures show the shortest gas depletion timescales (or equivalently the highest star-formation efficiencies).   

As a corollary to the $L_{\rm IR}$--$M_{\rm d}$ plane, we also plot in Fig.~\ref{fig:FIRpars_z}c the ratio of $L_{\rm IR}$ (as a simple observable linked to star-formation rate) and $M_{\rm d}$ (as a proxy for gas mass) in our sample as a function of redshift. This ratio reflects the expected gas depletion timescale and we see that it declines by a factor of $\sim$\,2 between $z$\,$=$\,1.5 and $z$\,$=$\,4.0 from $\sim$\,200\,Myrs to $\sim$\,50\,Myrs (\citealt{2015dacunha} analysis of the ALESS sample shows very similar behaviour). However, as seen in Fig.~\ref{fig:FIRpars_z}c, when we restrict our analysis to the unbiased, luminosity-limited sub-set this trend weakens suggesting that it is driven primarily by the incompleteness in $L_{\rm IR}$, and thus star-formation rate, with redshift -- rather than a fundamental variation in the gas depletion timescale with redshift.

We can also compare the estimates of the dust and stellar masses for the SMGs. 
For our full SMG sample we derive a median dust to stellar mass ratio of $M_{\rm d}/M_\ast$\,$=$\,(5.4$\pm$0.2)$\times$10$^{-3}$ with a 16--84$^{\rm th}$ range of 0.0028--0.0131 and little change with redshift, while for the GAMA ULIRGs we estimate $M_{\rm d}/M_\ast$\,$=$\,(11$\pm$2)$\times$10$^{-3}$.
At the upper end of our observed range these values are above the expected yields for dust from SNe and AGB stars \citep{2017calura} unless the IMF is biased to high-mass stars \citep{2005baugh,2018zhang} or that much of the dust grain growth in these systems occurs in the ISM \citep{2009draine,2020burgarella}, which might be a result of the high ISM densities found in the SMGs \citep{2011swinbank,2017simpson,2018zhukovska}. 

We also derive the gas fraction, the ratio of the molecular gas mass to the total baryonic mass of the galaxy:

\begin{equation}
f_{\rm gas} = \frac{M_{\rm gas}}{M_{\rm gas}+M_\ast}
\end{equation}

and show its variation with redshift in Fig.~\ref{fig:FIRpars_z}d. We derive a median gas mass fraction of $f_{\rm gas}$\,$=$\,0.41$\pm$0.02 with a 68$^{\rm th}$ percentile range of $f_{\rm gas}$\,$=$\,0.24--0.72, comparable to the median gas fraction of $f_{\rm gas}=$\,0.38$\pm$0.03 for the local ULIRGs from the GAMA survey estimated in an identical manner. We see modest evolution of $f_{\rm gas}$ with redshift, with a gradient of $df_{\rm gas}/dz$\,$=$\,0.06$\pm$0.02. For comparison we overlay the ASPECS blind CO-selected sample from \cite{2019aravena} and the CO-detected ``main sequence'' galaxies from the PHIBSS compilation of star-forming galaxies by \cite{2018tacconi}. Gas fraction of SMGs and ``main sequence'' galaxies at z$\gtrsim$2 appears to be similar.

Finally, we estimate the dust obscuration to the source of far-infrared emission averaged along the line-of-sight. In order to estimate this dust obscuration, we first estimate the column density of hydrogen atoms for the 154 sources that have 870$\mu$m sizes (Gullberg et al.\,2019, further discussed in $\S$~\ref{sizes}) assuming $\delta _{\rm gdr}$\,=\,100 to convert dust mass to gas mass, and find a median of $N_{\rm H} = (1.66\,\pm \,0.14)\,\times \,10^{24}$\,cm$^{-2}$. Dust attenuation is related to hydrogen column density by, $N_{\rm H} = 2.21\,\times \,10^{21}\,A_{V}$, and thus we find a median line-of-sight dust attenuation of $A_{V} =$\,750$\pm$60 mags. The result is within the error range of \cite{2017simpson} (who found a median $A_V $\,$=$\,540$^{+80}_{-40}$\,mags) when samples are selected in the same manner, having detections in all three SPIRE bands, resulting in a median $A_{V} =$\,700$\pm$90 mags.

As the hydrogen column and dust attenuation results are suggesting dusty, highly obscured systems, we estimate the wavelength at which the optical depth, $\tau$, becomes optically thin, $\lambda_0$. We, first, derive the brightness temperature of the SMGs with 870-$\mu$m sizes using: 

\begin{equation} \label{eq:bright}
B_{\nu_{\rm rest}}(T_B) = 0.5S_{\nu_{\rm rest}}(1+z)^3/\Omega_{\nu_{\rm obs}},
\end{equation}

where $B_{\nu_{\rm rest}}$ is the full Planck function and the solid angle subtended by the source is $\Omega_{\nu_{\rm obs}} = \pi R_\nu^2/D_{\rm A}^2$, where $R_\nu$ is the emission region size and $D_{\rm A}$ is the angular diameter distance. The factor of 0.5 is included as we are considering the emission within the half-light radius. 
Using Eq.~\ref{eq:bright} we estimate a median brightness temperature of $T_B =$21\,$\pm$\,1\,K, with a 16--84$^{\rm th}$ per centile range of 16--28\,K. 
The brightness temperature can be related to the true dust temperature and optical depth by: 

\begin{equation} \label{eq:opt_dep}
\frac{1}{e^{h\nu/kT_{\rm B}}-1}=\frac{1-e^{-\tau_\nu}}{e^{h\nu/kT_{\rm D}}-1}.
\end{equation}

As in $\S$~\ref{dustTemp}, we used fixed dust emissivity index of $\beta =$\,1.8 in the calculation of optical depth. We make the assumption that the emission region at 250, 350 and 500\,$\mu$m is the same size as that measured at 870\,$\mu$m. We note that, for a given source, the observed size of the emission varies with optical depth as it increases with frequency. Thus, our assumption overestimates the flux density within the 870$\mu$m half-light radius for 250, 350 and 500$\mu$m emission. Therefore, our estimated dust temperature and optical depth are the lower limits of the true values.

We use a sub-sample of 64 sources that have 870\,$\mu$m sizes and detections in all three SPIRE bands to solve for both, dust temperature and optical depth, using Monte Carlo Markov Chain method. We estimate a median true dust temperature of T$_{\rm D}=$40\,$\pm$\,2\,K and a median optical depth of unity, $\lambda_0=$106\,$\pm$\,6\,$\mu$m and we note that both of these quantities are the lower limit estimates. 
The wavelength estimate is comparable to the results from \cite{2017simpson} who found $\lambda_0 \gtrsim$75\,$\mu$m for a small sub-sample of 14 UDS SMGs.

\begin{figure*}
  \includegraphics[width=\textwidth]{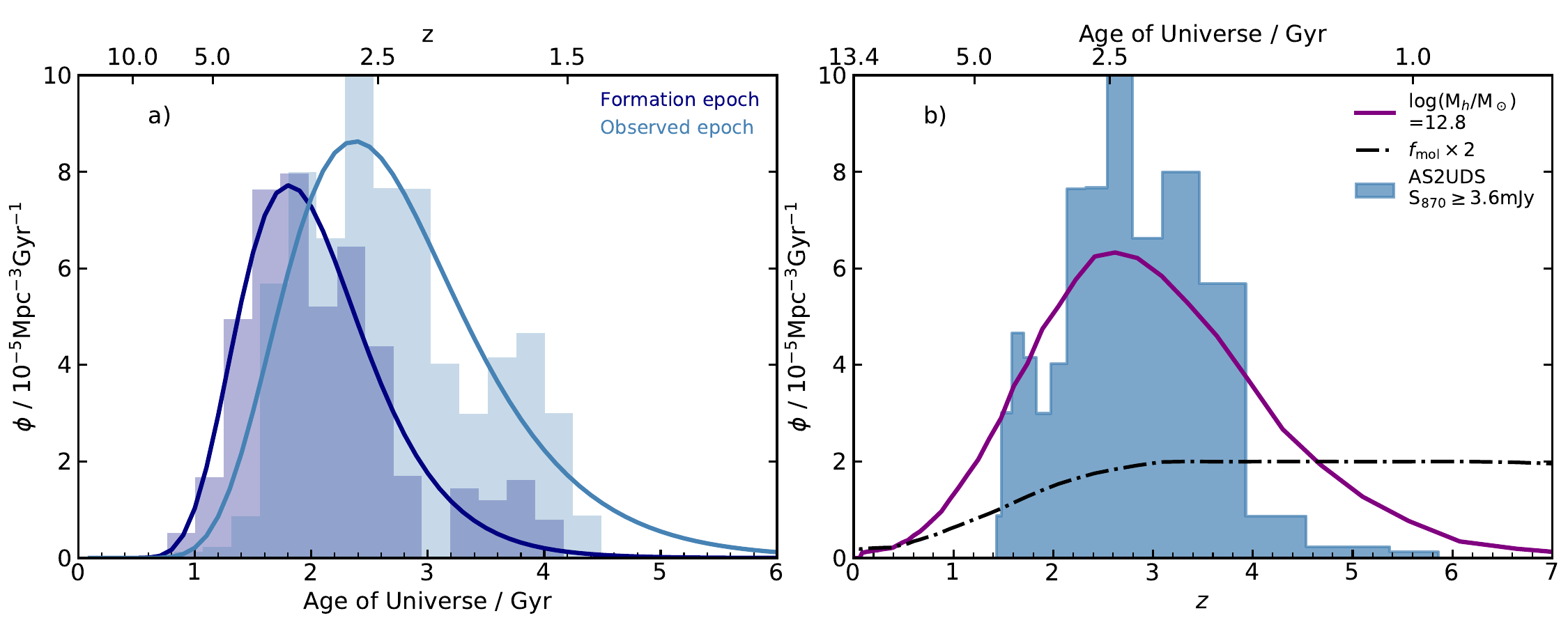}
  \caption{\textbf{(a)} The distribution of cosmic ages for the observed epochs of the AS2UDS SMGs and the inferred formation epochs for these galaxies (corrected for
their estimated ages) corrected for incompleteness following \protect\citep{2017geach}. The solid lines show log-normal fits to the respective distributions. 
We see that the observed age distribution peaks at $\sim $\,2.4 Gyr, while the inferred formation-age distribution peaks at $\sim$\,1.8 Gyr ($z\sim$\,3.5), with both well described by log-normal distributions.
  \textbf{(b)} The stacked likelihood redshift probability distribution of the  sample of 364 AS2UDS SMGs that have $S_{870}\geq$\,3.6\,mJy corrected for incompleteness following \protect\citep{2017geach}. We overlay a simple analytic model assuming that SMGs reside in haloes whose mass crosses a characteristic threshold of $\sim$6\,$\times$\,10$^{12}$\,M$_\odot$ and accounting for the evolution of molecular gas fraction with redshift ($f_{\rm mol}$, scaled by a factor of 2$\times$, is shown as dashed line) in the halos. The model follows our distribution well as shown by the solid line. This normalisation represents the duty cycle correction for the SMGs.}
\label{fig:age_reds}
\end{figure*}

\section{Discussion} \label{discussion}

Having analysed the physical properties of SMGs and their evolution, in this section we focus on combining these results to investigate three main aspects in detail: formation and evolution of SMGs, their relationship to the wider population of massive galaxies and insights into the distribution of the star-forming regions within this population.

\subsection{Evolution and lifetimes} \label{discuss_age}

It is expected to be challenging to reliably constrain the age of the stellar populations in SMGs due to their high obscuration and the influence of the intense recent star formation on the SED, as well as the degeneracies between age and other parameters such as redshift. Nevertheless, the analysis of the model SEDs of simulated strongly star-forming galaxies from EAGLE (described in $\S$\ref{eagle}) suggests that there is some diagnostic information in the derived ages from {\sc magphys}, as there is a positive linear correlation between these and the mass-weighted stellar ages in the model (see Fig. A2). 
In Fig. A2 (see Appendix~\ref{appendixA}) we see that the median scatter around the best fit line is
$\|({\rm Age}_{\rm MAGPHYS}-{\rm Age}_{\rm best-fit})\|/{\rm Age}_{\rm MAGPHYS}\,=\,0.52$ for the sample. 
In comparison, the median fractional error on ages in the AS2UDS sample is 0.54. These errors are comparable, thus the systematic error is encompassed in the error returned from {\sc magphys}. 
As such mass-weighted ages may be used to infer the typical formation epoch of the SMGs and to assess the evolution of properties of this population.  

\subsubsection{Mass-weighted ages} \label{ages}

We first compare the mass-weighted ages to estimates of ages of SMGs derived from other physical properties.
We derive a median mass-weighted age of our sample of Age$_{\rm m} =$\,(0.46$\pm$0.02)\,Gyr. We note that the posterior likelihood distribution differs significantly from the prior in Fig.~\ref{fig:distribution}d, suggesting the model is varying this parameter when fitting the SED.

We have two other methods to estimate ``ages'' for the SMGs.  Firstly, we can take the derived stellar mass and the current star-formation rate and ask how long it would take to form the observed mass?  This age parameter, $M_\ast/{\rm SFR}$, has a median ratio of $M_\ast/{\rm SFR}$\,$=$\,0.51$\pm$0.03\,Gyr, and correlates very closely with Age$_{\rm m}$ for ages $\lesssim$\,0.7\,Gyrs (corresponding to the bulk of the population at $z\gtrsim$\,2--3). $\sim$\,25 per cent of SMGs, mostly at $z\lesssim$\,2 have  $M_\ast/{\rm SFR}$ higher than Age$_{\rm m}$ indicating either a declining star-formation rate or significant previous stellar mass in these galaxies.

The second estimate we can obtain comes from the expected lifetime of the current star-formation event, given by the ratio of the estimated gas mass to the star-formation rate:
$M_{\rm gas}/{\rm SFR}$. This characteristic lifetime is also twice the gas depletion timescale as we assume to be observing the SMGs half way through the burst. This was estimated in \S\,\ref{dust_gas} as 146$\pm$5\,Myrs, yielding a lifetime of 292$\pm$10\,Myrs. This shows a weak correlation with Age$_{\rm m}$ with significant dispersion and around $\sim$\,20 per cent of the SMGs (covering the whole redshift range) have gas depletion times which are longer than their mass-weighted ages, suggesting that the current star-formation event may represent the first major star-formation episode.  However, for the bulk of the population, it appears that there is a pre-existing (older) stellar population in these systems.

In Fig.~\ref{fig:pars_z}d we plot mass-weighted age as a function of redshift. We show the limit provided the age of the Universe at a given redshift.  The best-fit trend to the Age$_{\rm m}$--$z$ plot suggests a statistically significant evolutionary trend of age with redshift with a gradient of $d{\rm Age_{\rm m}}/dz=-$0.29$\pm$0.02, so that higher-redshift SMGs are systematically younger. However, we caution that this may be a consequence of the code fitting younger ages to hotter dust components which are more prevalent at higher redshifts. For comparison, we overlay the local ULIRGs sample from the GAMA survey. We see that the median age from the local ULIRG sample agrees with the trend we observe at higher redshift, with these galaxies having overall older mass-weighted ages (as expected). 

To assess the influence of the current star-formation activity on the evolution of the SMGs we determine when the current star-formation is likely to cease. Again, using our estimate a median gas depletion timescale of $\tau_{\rm dep}\,=\,146\pm$5\,Myr with a 68$^{\rm th}$ percentile range of $\tau_{\rm dep}\,=$\,53-321\,Myr for the SMGs at $z =$\,1.8--3.4, this indicates that the star-formation activity in this population is expected to cease by $z\sim$\,2.5, soon after their peak at $z =$\,2.6. The stellar populations in these systems would then evolve to become red and quiescent by $z\sim$\,2, in the absence of subsequent gas accretion and star formation. Similarly, assuming that,  on average, we see the SMGs half way through their most active phase, we can adopt this depletion timescale as the likely age of the SMG-phase at the point we observe the SMG. Comparing this estimate to 
the median mass-weighted age of the $z =$\,1.8--3.4 SMG sub-sample, 490$\pm$20\,Myr (68$^{\rm th}$ percentile range of 97--960\,Myr), suggests that the bulk of the population had some pre-existing stellar population before the onset of the current star-formation event. We can also consider the mass produced in the last $\sim$\,150\,Myr (when the SMG-phase started) assuming a constant star-formation rate. We find a median fraction of $M_{150\rm Myr}/M_\ast\,\sim$\,0.3. This means that, for an average SMG, $\sim$\,30 per cent of the current stellar mass was formed in the last 150\,Myr, and by the end of the SMG-phase these systems would have roughly doubled their pre-existing stellar masses.

\subsubsection{Lifetimes of SMGs} \label{lifetime}

As seen in Fig. \ref{fig:red_dist}, the redshift distribution of our {\it complete} sample of 707 ALMA-identified AS2UDS SMGs  has a median redshift of
$z$\,$=$\,2.61$\pm$\,0.08, with a 68$^{\rm th}$ percentile range of $z$\,$=$\,1.8--3.4. The rapid decline in the number density of SMGs we see at both $z\lesssim$\,2 and $z\gtrsim$\,3.5 is striking. We stress that by virtue of employing full-SED modelling with {\sc magphys}, the  redshift distribution in Fig.~\ref{fig:red_dist} comprises the summed PDFs of all of the SMGs in our sample, not
just the biased sub-set which are detectable in the optical/near-infrared  \citep[e.g.][]{2014simpson}
and without having to employ a heterogeneous mix of redshift estimators \citep[e.g.][]{2017brisbin,2018cowie}.   We find a highly-peaked redshift distribution, 
which drops rapidly at higher  redshifts, 
with  $\sim$\,30 per cent of the SMGs lying at $z>$\,3, and just $\sim$\,6 per cent at $z>$\,4.   Equally, we find only five examples of SMGs at $z<$\,1, some of which may be unidentified weakly amplified galaxy-galaxy lenses \citep[e.g.][]{2017simpson}.

For the subsequent analysis, we use only the 364 ALMA SMGs with $S_{870}\geq$\,3.6\,mJy. This matches the flux density limit of the parent S2CLS survey \citep{2017geach}, which covers an area of 0.96\,deg$^{2}$ and so allows us to estimate the appropriate volume densities from the sample.  We also correct our estimates for the incompleteness in SCUBA-2 850$\mu$m sample in the UDS field \citep{2017geach}.
In Fig.~\ref{fig:age_reds}a we recast our redshift distribution to illustrate the variation in volume density ($\phi$) of bright SMGs with cosmic time.  
Fig.~\ref{fig:age_reds} shows that the volume density of AS2UDS SMGs peaks around
$\sim$\,2.4\,Gyr after the Big Bang with a 16--84$^{\rm th}$ per centile range of
1.8--4.5\,Gyr.  The distribution is log--normal, with a mean of
$\mu$\,$=$\,0.97\,$\pm$\,0.03\,Gyr, standard deviation of $\sigma$\,$=$\,0.32\,$\pm$\,0.04\,Gyr and a normalisation of $c$\,$=$\,(1.7\,$\pm$\,0.2)\,$\times$\,10$^{-4}$\,Mpc$^{-3}$Gyr$^{-1}$.  We also combine the redshifts with the mass-weighted ages of each SMG (see
$\S$\,\ref{ages}) to predict  the distribution of formation ages of
the SMGs.  This distribution also follows a log--normal shape with a
median cosmic time at formation of $\sim$\,1.8\,Gyr after the Big Bang, and log--normal parameterisation of $\mu$\,$=$\,0.68\,$\pm$\,0.03\,Gyr, $\sigma$\,$=$\,0.30\,$\pm$\,0.03\,Gyr and $c$\,$=$\,(1.08\,$\pm$\,0.08)\,$\times$\,10$^{-4}$\,Mpc$^{-3}$Gyr$^{-1}$. Fig.\ref{fig:age_reds}a shows that the SMGs begin to form in large numbers at a cosmic time of $\sim$\,1.8\,Gyr, corresponding to $z\sim$\,3.5. This confirms that the rapid rise in number density we see
in the redshift distribution at $z\lesssim$\,3.5 is being driven by the onset of this population. 

\subsubsection{Formation of SMGs} \label{formation}

Previous measurements of the spatial clustering of SMGs imply dark matter halo masses for SMGs of $M_{\rm h}$\,$\sim$10$^{13}$\,M$_\odot$ \citep{2012hickox,2017wilkinson}. More crucially, \cite{2012hickox} suggested that the SMG redshift distribution is related to the growth rate of cosmological structures. The basis of this model is the concept of a {\it critical threshold mass} for halos, which has been developed to interpret the clustering evolution of QSOs \citep[e.g.][]{2003overzier,2006farrah}. To investigate this further, we use the Millennium Simulation \citep{2005springel} to determine the growth rate of dark matter halos as a function of redshift. Using the dark matter merger trees from this 500\,$h^{-1}$\,Mpc$^3$ simulation, we measure the volume density of dark matter halos at each redshift that pass through mass thresholds of 
$M_{\rm h}$\,=\,10$^{11}$--10$^{15}$\,M$_\odot$ in steps of 0.05\,dex. To account for the evolution of the molecular mass fraction within halos, 
we convolve these volume densities with the molecular gas fraction evolution \citep[e.g.][]{2011lagos} and derive the redshift at which these distributions peak. By matching the distributions predicted by this simple model to our observed redshift, we estimate a
``critical-mass'' for haloes of bright SMGs with $S_{870}\gtrsim$\,3.6\,mJy of $\log (M_{\rm h})$\,$\sim$\,12.8\,M$_\odot$. In Fig.~\ref{fig:age_reds}b we plot volume density of bright SMGs in our sample, limiting the SMGs to those brighter than $S_{\rm 870\mu m}$\,=\,3.6\,mJy (which represents the flux density limit of the parent survey) and overlay the redshift distribution of these dark matter halos for a critical mass of
$\log (M_{\rm h})$\,$\sim$\,12.8\,M$_\odot$.  

In this model the rapid decrease in the number density of SMGs at $z\lesssim$\,1.5--2 is explained by the
decline in the molecular gas fraction in the halos \citep{2011geach,2011lagos,2018tacconi}, as well as the decrease in the number of dark matter haloes that transit above the mass threshold as the Universe expands. Fig.~\ref{fig:age_reds}b shows
that the shape of the redshift distribution of SMGs appears to be reasonably well described by this combination the cosmological growth of structure and the evolution of the molecular gas fraction in galaxies.

\begin{figure*}
  \includegraphics[width=\textwidth]{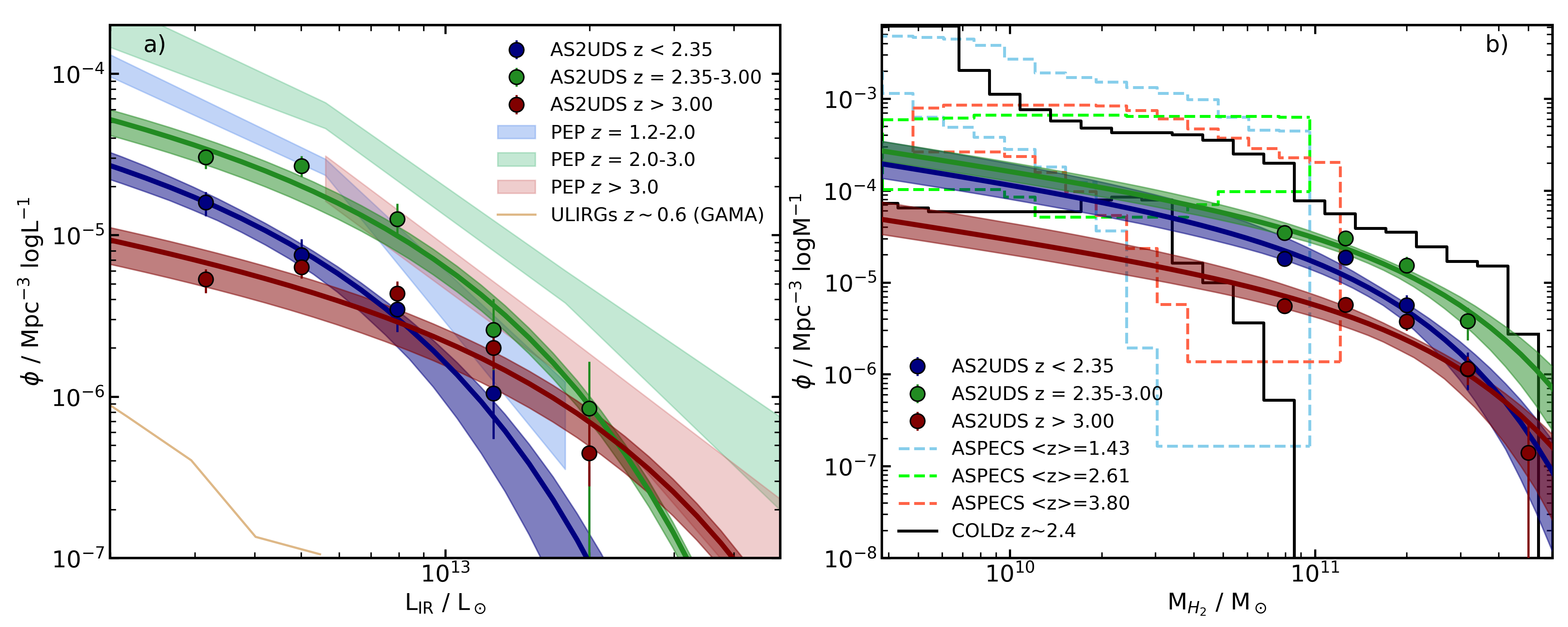}
  \caption{The evolution of the far-infrared luminosity function (left) and molecular gas mass  function (right) from the flux-limited sample of $S_{870}\geq$\,3.6\,mJy AS2UDS SMGs corrected for incompleteness. These are both plotted for three independent redshift bins
    with similar number of galaxies: $z<$\,2.35, $z$\,$=$\,2.35--3.0, $z>$\,3.0. Each of the bins is fitted with a Schechter function which is shown as the solid line in the respective colour. 
      1-$\sigma$ errors were obtained by resampling the luminosity or gas mass and redshift PDFs and the 1-$\sigma$ fitting error is shown as the shaded area. 
     {\bf (a)} We compare the AS2UDS far-infrared luminosity function to the PEP survey 100- and 160$-\mu$m selected samples from \protect\cite{2013gruppioni}. We also compare to local sample of ULIRGs ($z\sim$\,0.6) from the GAMA survey. This demonstrates the roughly two orders of magnitude increase in space density of ULIRGs between $z\sim$\,0 and $z\sim$\,2--3, with the space density peaking at $z$\,$=$\,2.35--3.00 and then declining at higher redshifts.
     {\bf (b)} We compare the AS2UDS gas mass function to results from the ASPECS blind mid-$J$ CO survey \citep{2019decarli} for three corresponding redshift ranges, where the ASPECS gas masses have been converted to the equivalent scale as our {\sc magphys} estimates. We see good agreement between the mass functions from the two surveys across the three redshift slices at higher mass end. We also compare to results from the COLDz blind low-$J$ CO survey  \protect\citep{2019riechers}. We see good agreement between the z$\sim$2.4 CO (1--0) sources and $z$\,$=$\,2.35--3.00 SMGs luminosity functions.}
\label{fig:lum_func}
\end{figure*}

The halo mass of $\sim$6$\times$10$^{12}$M$_\odot$, estimated from the SMG redshift distribution, is comparable to the clustering results for SMGs (\citealt{2012hickox,2016chen,2017wilkinson}; Stach et al.\, in prep.), which suggest that they occupy halos of $\sim10^{13}M_\odot$ at $z >$\,2.5. This halo mass is also similar to that estimated from clustering
studies for $L^\ast$ QSOs at $z\sim$\,1--2 \citep{2009ross}, supporting the evolutionary association between SMG and QSOs 
suggested by \cite{2012hickox} and others. Cosmological models of halo growth indicate that  a dark matter halo mass of $M_{\rm h}\,\sim\,6\times10^{12}\,M_\odot$ at $z\sim$\,2.6, corresponds to a median descendent halo mass at $z\sim$\,0 of $\gtrsim\,10^{13}\,M_\odot$, which is consistent with the 2--4\,$L^\ast$ ellipticals at the present day \citep{2011zehavi}.  Moreover, the characteristic halo mass we estimate agrees well with the theoretical prediction of the maximum halo mass where gas can cool and collapse within a dynamical time \cite{1978white&rees} and is thus also the halo mass associated with the highest star-formation efficiency \citep{2001gerhard,2013behroozi}. 

This may suggest that SMGs represent efficient collapse occurring in the most massive,  gas-rich halos which can host such activity. This simple model provides a natural explanation for them representing the highest star-formation rate sources over the history of the Universe, as well as for the details of their redshift distribution (Fig. \ref{fig:age_reds}). Moreover, it offers a description of why their massive galaxy descendants at $z\sim$\,0 have the highest stellar baryonic to halo mass ratios of any collapsed systems \citep{2001gerhard}.

\subsection {Evolution of the far-infrared luminosity and gas mass functions} \label{LFevolution}

Having determined the redshifts, far-infrared luminosities and dust masses for our SMG sample, we can exploit the fact that our survey is derived from a uniformly-selected sample of 850-$\mu$m
SCUBA-2  sources across a degree-scale field \citep{2017geach} to determine the luminosity and gas mass functions of SMGs and their evolution. We, therefore, use the sub-set of 364 ALMA SMGs brighter than the flux density limit of the  SCUBA-2 catalogue, $S_{870} \geq$\,3.6\,mJy, and correct for incompleteness ($\sim$20\% at $S_{870}$\,=\,3.6\,mJy, \citealt[][]{2017geach}) to obtain an 870-$\mu$m selected sample across the full UDS field. We note that $\sim$\,74 per cent of these SMGs are detected in at least one SPIRE band and hence have robust far-infrared luminosities.

\subsubsection{Far-infrared luminosity function} \label{fir_lum}

We calculate the  far-infrared luminosity
function for the 870-$\mu$m selected AS2UDS sample within the accessible volume using $\phi(L_{\rm IR})\Delta L_{\rm IR}$ =
$\Sigma (1/V_i)$, where $\phi(L)\Delta L$ is the number density of
sources with luminosities between $L$ and $L$+$\Delta L$ and $V_i$ is
the co-moving volume within which the $i$-th source would be detected
in a given luminosity bin. We split the sample of the 364 AS2UDS SMGs
brighter than $S_{870}$\,$=$\,3.6\,mJy into three redshift bins with similar number of galaxies in each: $z<$\,2.35, $z$\,$=$\,2.35--3.00 and $z >$\,3.00. The resulting luminosity functions are shown in Fig.~\ref{fig:lum_func}a.  Errors are estimated using a bootstrap method by re-sampling the photometric redshift and luminosity probability distribution functions. We fit the  luminosity functions using Schechter functions of the form, $\phi = (\phi^\ast/L^\ast)(L/L^\ast)^\alpha e^{-L/L^\ast}$, where  $\phi^\ast$ is the normalisation density, $L^\ast$ is characteristic luminosity and $\alpha$ is the power-law slope at low luminosities \citep{1976schechter}. \cite{2013clemens} derive $\alpha =-$1.3 for their  {\it Planck} detections of a local
volume-limited galaxy sample, while \cite{2011dunne} derive $\alpha =-$1.2$^{+0.4}_{-0.6}$ for a SPIRE-selected sample out to $z\sim$\,0.5, while other
studies have yielded values ranging $\alpha $\,$=$\,$-$1.0 to $-$1.7  \citep{2005vlahakis,2011dunne}. 
As our sample covers only a relatively narrow range in far-infrared luminosity at each redshift we are unable to constrain $\alpha$ directly and so instead we choose to fix it to $\alpha=-$1.5. The Schechter fits to each redshift slice are shown in Fig.~\ref{fig:lum_func}.

To demonstrate the evolution of the ULIRG population across our survey volume, we also plot in Fig.~\ref{fig:lum_func}a an estimate of the local far-infrared luminosity function from the sample derived from the GAMA survey from \cite{2018driver} at $z\sim$\,0.6. 
Examining the evolution in the luminosity function within our survey in Fig.~\ref{fig:lum_func}a, we see that the space density increases from the $z<$\,2.35  to peak in the $z$\,$=$\,2.35--3.00 bin (median redshift $z\sim$\,2.6) and then declines at $z>$\,3.00. Compared to local ULIRGs, we conclude that the AS2UDS SMGs have a space density that is a factor of $\sim$\,100\,$\times$ higher, similar to the findings for the smaller ALESS sample from \cite{2014swinbank}. In comparison to other estimates of the high-redshift far-infrared luminosity function, we find that our measurements for this rest-frame 200--300\,$\mu$m-selected samples lie below those from the PEP survey from \cite{2013gruppioni}, which is based on 100- and 160-$\mu$m selected samples. This is due to the fact that our 870-$\mu$m selection is sensitive to cooler sources, with  $T_d\lesssim$\,50--60\,K, out to $z\sim$\,4, thus we are incomplete for the hottest sources  (such as in \citealt{2013gruppioni}, see also \citealt{2011symeonidis,2019gruppioni}).

\begin{figure*}
  \includegraphics[width=\textwidth]{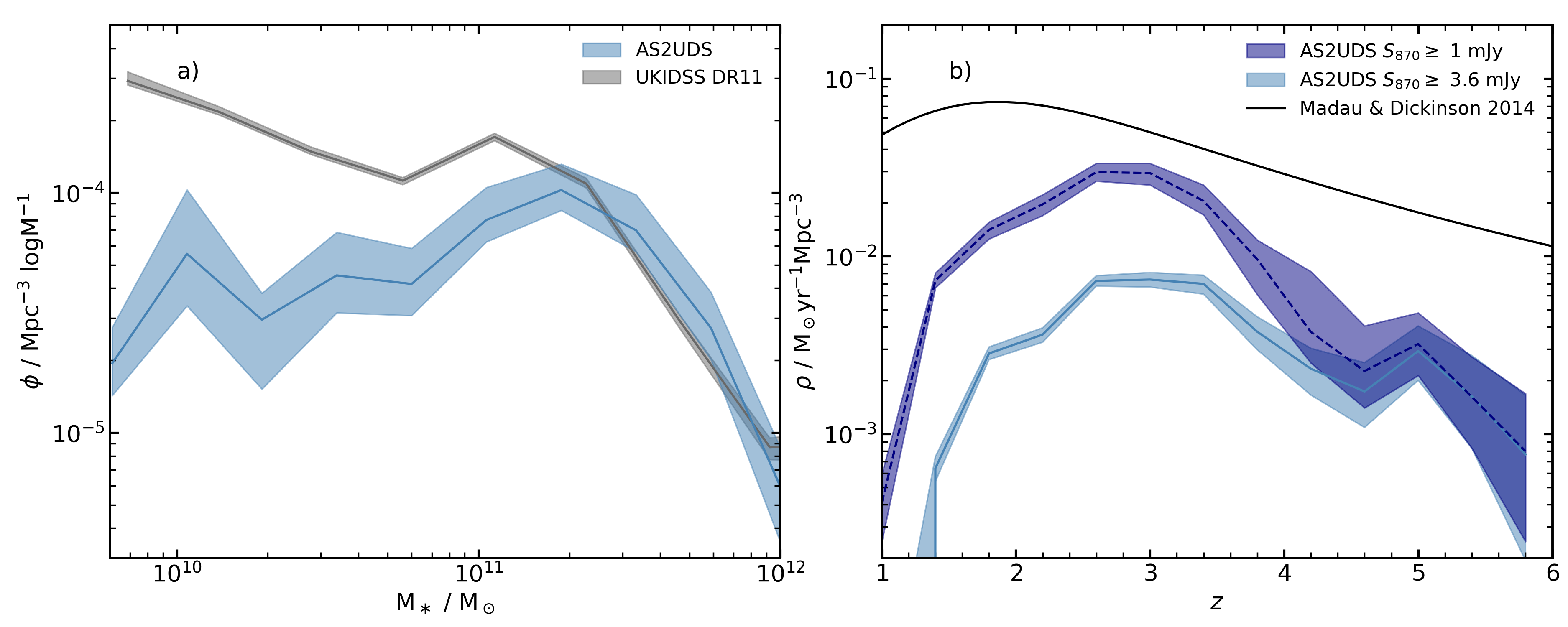}
  \caption{\textbf{(a)} The stellar mass function for $S_{870} \geq$\,3.6\,mJy SMGs at $z $\,$=$\,1.8--3.4. The blue line shows the expected number density of galaxies which had passed through an SMG-like phase, estimated from the distribution of the 
AS2UDS SMGs and corrected for duty cycle by a factor of 5.1$\times$ (see \S\,\ref{mass_func}), with the respective 1-$\sigma$ error shown as the blue shaded region. The grey line shows the stellar mass function of $K$-band selected galaxies in the UDS field with the 1-$\sigma$ error as the grey shaded region. We see that the SMGs make an increasing contribution to the total mass density distribution at higher stellar masses. The corrected volume density of AS2UDS SMGs corresponds to $\sim$\,30 per cent of the total number density of galaxies above $M_\ast $\,$=$\,3\,$\times$\,10$^{10}$\,M${\odot}$, but this fraction increases rapidly so  that $\sim$\,100 per cent of galaxies at $M_\ast \gtrsim$\,3\,$\times$\,10$^{11}$\,M$_{\odot}$ are expected to have passed through an SMG-phase.
\textbf{(b)} The co-moving cosmic star-formation density as a function of redshift. We show the contribution of AS2UDS sources for  SMGs brighter than $S_{870} =$\,3.6\,mJy 
and also brighter than $S_{870} =$\,1\,mJy, where we correct the numbers
of fainter sources using the 850-$\mu$m number counts \citep{2017geach} and 1.13-mm number counts \citep{2018hatsukade}, adopting $S_{870}$/$S_{1130}= $1.8. The shaded regions represent the 1-$\sigma$ error which has been calculated by re-sampling the redshift probability distribution while taking into account the star-formation--redshift correlation.
This shows that the contribution from SMGs peak at higher redshift ($z\sim$\,3) than the total star-formation rate density \protect\citep{2014madau} at which epoch SMGs with $S_{870} \geq$\,3.6\,mJy contribute $\sim$\,15 per cent to the total SFRD, and $\sim$\,60 per cent if we integrate down to the $S_{870} \geq$\,1\,mJy.}
\label{fig:mass_func}
\end{figure*}

\subsubsection{Gas mass function}
In an equivalent manner as in \S\,\ref{fir_lum}, we have estimated the  gas mass function for the SMG population and its variation with redshift in three broad redshift ranges, illustrated in Fig.\,\ref{fig:lum_func}b.  

Here, we compare estimates of the gas mass function derived from the ASPECS blind mid-$J$ CO survey from \citet{2019decarli} to the space densities for our gas mass functions in Fig.\,\ref{fig:mass_func}b. We note that for this comparison we have converted
the ASPECS gas masses, which are based on a conversion from CO luminosity to molecular gas mass adopting $\alpha_{\rm CO}$\,$=$\,3.6, to agree with the gas masses derived from {\sc magphys} dust masses with a gas-to-dust ratio of 100.  \cite{2016decarli_gas} show that this translates to a reduction in their estimated gas masses of a factor of 5.3$\pm$0.8\,$\times$ and so
we apply this conversion to compare to our {\sc magphys}-derived estimates.

Our estimates of the gas mass function and those from ASPECS agree at the high gas mass end for all three redshifts
(we show the corresponding 1-$\sigma$ confidence level measurements at $z$\,$= $\,1.4, $z $\,$=$\,2.6 and $z $\,$=$\,3.8 from ASPECS), with the wide-field AS2UDS estimates adding information at the high gas mass end of the distribution which is missing from ASPECS owing to its modest survey volume.  We see that the extrapolated low-mass space densities from AS2UDS, based on our Schechter
function fits with a low-mass slope of $\alpha$\,$=$\,$-$1.5, are broadly in agreement with the ASPECS samples down to masses of $\sim$10$^{10}$\,M$_\odot$, but fall below at lower masses, however, we note that these differences could be accounted for by the uncertainty in our adopted $\alpha$ value. We also compare our results to the gas mass function derived from the COLDz blind low-$J$ CO survey of \cite{2019riechers} in Fig.\,\ref{fig:mass_func}b (converting from CO luminosity to molecular gas mass in an equivalent manner to ASPECS). The z$\sim$2.4 CO (1--0) sources and $z$\,$= $\,2.35--3.00 SMGs agree very well across the whole gas mass range.
Thus, broadly, the evolution of the gas mass function from the combined AS2UDS\,+\,CO-selected samples appears to be best characterised by an increasing space density of galaxies at a fixed gas mass from $z\sim$\,3.5 down to $z\sim$\,1.5,
with a hint that we may be seeing the space density of massive gas-rich systems beginning to decline at $z<$\,2.5.  
 
\subsection{Stellar mass function} \label{mass_func}

We next investigate what fraction of massive galaxies may have experienced a high star-formation rate phase, which would correspond to an SMG, and hence whether SMGs are a phase that all massive galaxies go through. For this comparison, we estimate the number density of massive galaxies using our {\sc magphys} analysis of the $K$-band sample in the UDS field. This approach has the advantage that the stellar masses, redshifts and survey volumes are estimated in an identical manner to those employed for the SMGs.  
We select those field galaxies that have redshifts lying in the 16--84$^{\rm th}$ percentile range of the AS2UDS redshift distribution ($z $\,$=$\,1.8--3.4). To ensure we
have robust stellar mass estimates, we limit the field sample to galaxies with the best photometry and SED fits with a reduced $\chi^2 <$\,4.   We determine the influence of these cuts on the
resulting sample size and increase the normalisation of the field sample by a factor of 1.35 to correct for this. The UDS field catalogue is selected in the $K$-band, with a 3$\sigma$ limit of $K=$\,25.7\,mag, which roughly corresponds to $M_\ast \sim$\,5$\times$10$^9$\,M$_\odot$ at $z\sim$\,3 for typical star-formation histories. Therefore, for the field, we construct the stellar mass function above this stellar mass threshold to avoid incompleteness. We sum the number of galaxies in each stellar mass bin and divide by the volume defined by the span of their redshifts. 

We calculate the SMG stellar mass function in an equivalent manner and then calculate the duty cycle of the SMGs by comparing the visibility time from \S \,\ref{formation} to the age spanning a given redshift slice ($\Delta T_{z}$): duty-cycle correction as $\Delta T_{z}/T_{\rm vis}$.  For the redshift range of $z=$\,1.8--3.4, we find a median visibility time of 490\,$\pm$\,20\,Myrs  and corresponding to duty correction factor of 5.1$\times$ which we apply to the SMG mass function. The uncertainties for both field and SMG stellar mass functions were obtained by re-sampling the stellar mass and redshift probability distributions and taking the 16--84$^{\rm th}$ percentile range as the 1-$\sigma$ error. 

Overlaying the corrected SMG
mass function on the field  in Fig.\,\ref{fig:mass_func}a, we see  that  at lower stellar masses, galaxies which have passed through an SMG-phase would account for  only a modest fraction of the total space density, e.g.\ $\sim$\,30 per cent of galaxies above $M_\ast $\,$=$\,3\,$\times$\,10$^{10}$\,M$_{\odot}$. This fraction increases to $\sim$\,100 per cent at $M_\ast $\,$\gtrsim$\,3\,$\times$\,10$^{11}$\,M$_{\odot}$, indicating that all of the galaxies above this mass are likely to have experienced an SMG-phase in the course of their evolution. The results from the EAGLE simulation presented in \cite{2019mcalpine} indicates that effectively all galaxies at $z\sim$\,0 in the simulation with stellar
masses above $M_\ast =$\,2\,$\times$\,10$^{11}$\,M$_{\odot}$ experienced a ULIRG-like phase where their star-formation rate exceeded $\sim$\,100\,M$_\odot$\,yr$^{-1}$. This result is consistent with our finding as there is little evolution of the stellar mass function of these galaxies in this mass range since $z\lesssim$\,1.5 \citep{2020kawinwanichakij}.

\subsection{Co-moving star-formation rate density}

To investigate the contribution of SMGs to the total star-formation rate density (SFRD) as a function of redshift we make use of the predicted star-formation rates of AS2UDS sources. We calculate the star-formation rate density for two sub-sets of our sample SMGs: those SMGs with $S_{870}\geq$\,3.6\,mJy (which is the limit of the parent survey) and those with $S_{870}\geq$\,1\,mJy SMGs. For the $S_{870}\geq$\,3.6\,mJy sub-sample we correct for the incompleteness using the number counts from \citealt{2017geach}. We also correct the estimated star-formation rate of an SMG within its redshift PDF to account for the variation as a function of redshift in the inferred star-formation rate.
We correct the number of $S_{870}\geq$\,1\,mJy sources from our survey to match the expected number counts to this flux limit. We derive these using the ALMA 1.13-mm counts in the GOODS-S field from \citet{2018hatsukade} and a factor of 1.8 to convert the 1.13-mm flux densities to 870$\mu$m.

The resulting star-formation rate density of the AS2UDS SMGs is shown in the Fig.~\ref{fig:mass_func}b. For comparison we overlay the combined optical and infrared star-formation rate density from \cite{2014madau}, this represents the total star-formation rate density in the Universe at $z\lesssim$\,3, above which it is constrained only by surveys in the UV (unobscured sources). This comparison demonstrates that the activity of SMGs peaks at $z \sim$\,3, higher than the peak of the \cite{2014madau} SFRD at $z\sim$\,2. This suggests that more massive and obscured galaxies are more active at earlier times. Fig.~\ref{fig:mass_func}b also shows that contribution to the total star-formation rate density increases steeply from $z\sim$\,1 with the peak contribution being $\sim$\,15 per cent at $z\sim$\,3 for the $S_{870} \geq$\,3.6\,mJy sub-sample or $\sim$\,60 per cent for sources brighter than $S_{870}$=\,1\,mJy. This indicates that roughly half of the star-formation rate density at $z\sim$\,3 arises in ULIRG-luminosity sources and this population appears to decline only slowly across the 1\,Gyr from $z\sim$\,3 to $z\sim$\,6.

\begin{figure*}
  \includegraphics[width=\textwidth]{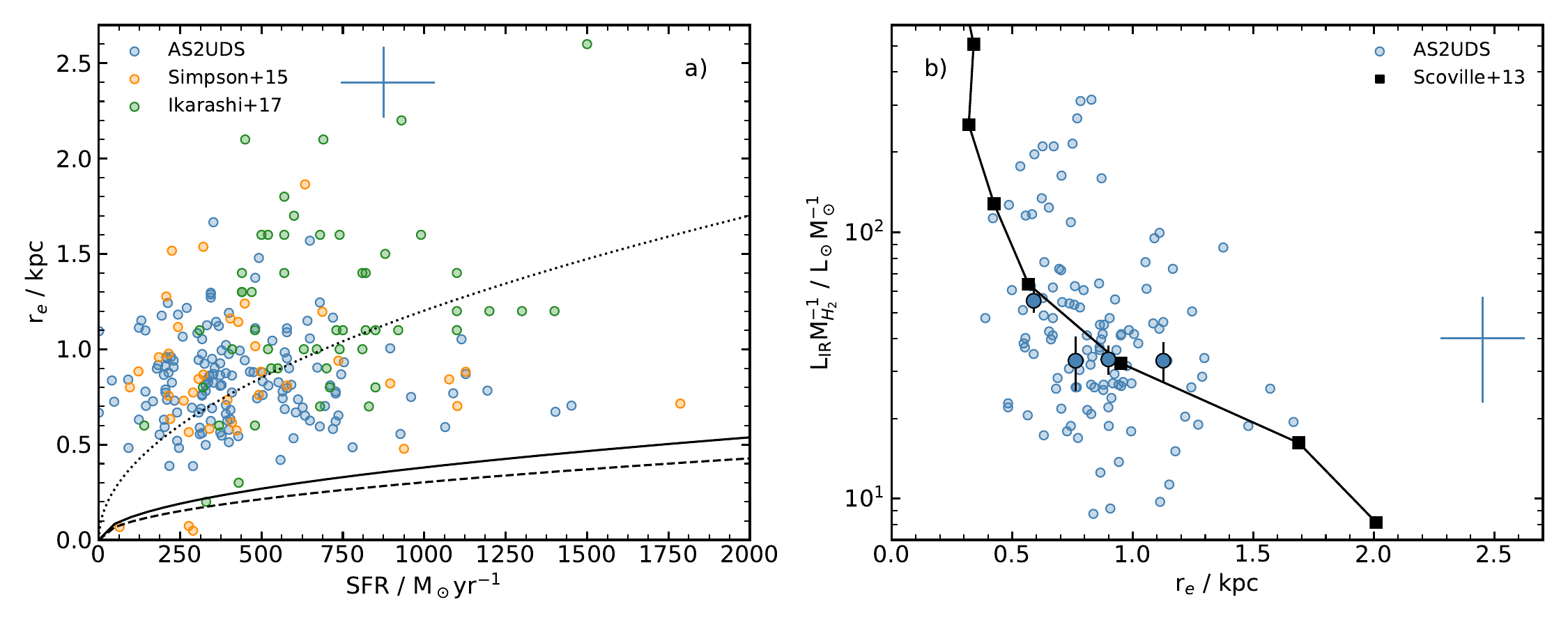}
  \caption{\textbf{(a)} The variation of 870-$\mu$m dust continuum size from \protect\citep{2019gullberg} with star-formation rate for the AS2UDS SMGs. We also plot the sizes derived from lower-resolution ALMA continuum observations of samples of
SMGs in the UDS field at 870$\mu$m  from \protect\cite{2015simpsonb} and 1.1-mm by \protect\cite{2017ikarashi} (the SFR for these are derived from our {\sc magphys} analysis of these galaxies).
We see a  weak trend of increasing size at higher star-formation rates, with significant scatter.   We compare this trend to the boundary expected from the estimated Eddington limit following \protect\cite{2011andrews&thompson} for $f_{\rm gas}$=\,1, which is shown as the solid line. The dotted line indicates the 0.1$\times$ this Eddington limit, which roughly goes through the median of our sample. The dashed line indicates the Eddington limit for $f_{\rm gas}=$\,0.4.
We see that very few of the SMGs have sizes which are compact enough for them to approach the Eddington limit at their star-formation rate. On average the AS2UDS have sizes around ten times larger, and thus areas approaching two orders of magnitude larger than an Eddington-limited system with their star-formation rate, suggesting that this fundamental feedback process will not quench their activity.
\textbf{(b)} 870-$\mu$m dust continuum size as a function of far-infrared luminosity-to-gas mass ratio for SMGs with at least one SPIRE detection. Large circles are the median values binned by radius of $\sim$30 sources and the median error for individual sources is indicated at the top of the panel. We overlay the model of an optically-thick dust cloud from the \protect\cite{2013scoville}. The AS2UDS SMGs have a similar trend to this model, however with scatter below the model at low luminosity-to-mass ratios and above the prediction for high ratios. The overall agreement between model and AS2UDS sizes at a given radius indicates that SMGs are possibly homogenous, and thus homologous, systems in the far-infrared, behaving as expected for a single dust cloud.}
\label{fig:sizes}
\end{figure*}

\subsection{The scale of far-infrared emission in SMGs}

Finally, we wish to investigate what we can learn about the conditions and structure of the star-forming regions of SMGs from our sample. For this, we will employ sizes for a sub-set of our SMGs which have been derived from the analysis of high-resolution dust continuum observations of 154 SMGs in A2SUDS by \citealt{2019gullberg}. This work exploits the fact that the Cycle 3 observations for AS2UDS were obtained with ALMA in an extended configuration which yielded
a synthesised beam with a FWHM of 0.18$''$ ($\sim$\,1\,kpc). \citealt{2019gullberg} undertook extensive testing and analysis of the constraints on the 
sizes, profiles and shapes of SMGs provided by these high-resolution 870-$\mu$m maps.  On the basis of these tests they restricted their analysis to only
the highest angular resolution data available and in addition, applied a further cut that the sources had to be detected in a 0.5$''$-tapered map with a
 signal-to-noise of SNR\,$>$\,8. This ensured that the resulting profile and shape measurements were unbiased and of sufficient quality to be useful.  The resulting sizes have median errors of just 20 per cent for a sample of 154 SMGs broadly representative of the full population of AS2UDS SMGs. \citealt{2019gullberg} measure a number of profile, shape and size parameters for the SMGs from fits to the $uv$ amplitudes in Fourier space and also image-plane fits to the reconstructed maps. They show that dust emission from typical SMGs are well-fit by exponential profiles described by a Sersic model with $n$\,$\sim$\,1 \citep[see also][]{2016hodge,2017simpson,2019hodge}. In the analysis here we make use of the circularised effective radii derived from fits to the $uv$ amplitudes for the sources adopting $n$\,$=$\,1 Sersic profiles.
We then convert these angular sizes into physical units using the photometric redshifts determined for the individual sources. The median physical size, expressed as $R_e$, of the sample is 0.83$\pm$0.01\,kpc. 

\subsubsection{Star-formation conditions in SMGs} \label{eddington}

The high median star formation of the AS2UDS sources (see Fig. \ref{fig:distribution}) may be a result of SMGs behaving as Eddington-limited starbursts \citep{2011andrews&thompson}, where the radiation pressure from massive stars is sufficient to quench further activity. To test this, we plot the 870-$\mu$m physical sizes versus the star-formation rate for the SMGs in Fig.~\ref{fig:sizes}a. For comparison, we overlay two other studies which employed similar signal-to-noise ALMA observations, but at lower resolution (0.3$''$--0.7$''$ FWHM), of samples of SMGs in the UDS field:  an   
870-$\mu$m sizes from the AS2UDS-pilot study of bright SMGs by  \cite{2015simpsona}, which have a median effective radius of $R_e $\,$=$\,0.79$\pm$0.05\,kpc, and a sub-set of AS2UDS SMGs detected using the AzTEC camera on the ASTE telescope and followed-up with ALMA at 1.1-mm by \cite{2017ikarashi}, which yield a median effective radius of $R_e $\,$=$\,1.1$\pm$0.1\,kpc. Even though these two samples use lower resolution observations, they recover similar distributions in terms of the physical sizes of the SMGs.

 We observe no strong trend in size with star-formation rate and so we now  test whether the star-formation activity in SMGs is affected by their approaching the  Eddington luminosity limit for their observed sizes and star-formation rates.  We follow \cite{2011andrews&thompson} who demonstrated that the balance of radiation pressure from star formation, with self gravity results in a maximum star-formation rate  surface density (in units of M$_\odot$\,yr$^{-1}$\,kpc$^{-2}$) of: 

\begin{equation}
\mu^{\rm max}_{\rm SFR} = 11 f_{\rm gas}^{-0.5}\delta_{\rm GDR},
\label{eq:sfr_max}
\end{equation}

where $f_{\rm gas}$ is the gas fraction in the star-forming region and $\delta_{\rm GDR}$ is the gas-to-dust ratio which (as mentioned in \S~\ref{dust_gas} we adopt 100). The equation assumes optically-thick dust emission and no heating from an AGN. 
Our estimated galaxy-integrated gas fractions from Fig.~\ref{fig:pars_z}d are $f_{\rm gas}\sim$\,0.4 and we see no significant variation in this with redshift. However,
the available near-infrared imaging suggests that the stellar mass component, which is used to estimate $f_{\rm gas}$, is likely to be more extended than the size of the dust continuum emission, potentially by a factor of $\sim$\,4$\times$
 (\citealt{2017simpson,2015ikarashi,2019lang,2019gullberg}. Due to uncertainty in the calculated gas fraction, we take a conservative approach and adopt a gas fraction of unity, which provides the lower limit of the star-formation rate surface density (for comparison, the Eddington limit assuming $f_{\rm gas}$=\,0.4 is shown in Fig.~\ref{fig:sizes}a).
 
 The resulting maximum star-formation rate surface density predicted by the model is 1,100\,M$_\odot$yr$^{-1}$\,kpc$^{-2}$ and we show this line in Fig.~\ref{fig:sizes}a.
Comparing the AS2UDS SMGs to this line, we see that very few have sizes which are compact enough for them to approach the Eddington limit for their observed star-formation rates. On average the AS2UDS have sizes around ten times larger than an Eddington-limited system with their same star-formation rate, indicating that averaged on kpc-scales the radiation pressure from the on-going star formation in these
systems is not sufficient to quench their activity. However, if the star-forming region for a given galaxy is ``clumpy'' on small scales \citep{2011swinbank,2011danielson,2013memendez}, then individual regions on sub-kpc scales may be Eddington limited. 

\subsubsection{Structure of the far-infrared regions of SMGs} \label{sizes}

We now turn to examine the possible structure of the far-infrared luminous component of the SMGs. As suggested in \S \ref{eddington}. the conditions of star formation are dependent on whether the far-infrared sources have a single homogenous dust cloud structure or are ``clumpy'' systems. We investigate this by comparing our results to an optically-thick model from \cite{2013scoville}.

As outlined in \cite{2013scoville}, internally heated far-infrared sources are described by two properties: the luminosity of the central heating source and total dust mass in the surrounding envelope. Thus this structure can be characterised by a single parameter: their luminosity-to-mass ratio. The far-infrared SEDs can be calculated for this using a temperature profile, which is estimated from a combination of optically-thin dust emission at the inner region (where $T_{\rm d} \propto r^{-0.42}$), optically-thick (where $T_{\rm d} \propto r^{-0.5}$ as the photons heating the grains and those that are emitted by the grains have similar wavelength distributions) at intermediate radii and optically-thin again at larger radii \citep{2013scoville}. The inner radius is taken at 1\,pc (where $T_{\rm d}$=1000\,K, close to the dust sublimation limit) and the outer radius is taken at 2\,kpc, which is roughly appropriate for SMGs \citep{2019gullberg}. In the plot the effective radius of the model is defined as radius of the shell producing the largest fraction of the overall luminosity for each of the $L/M_{ISM}$ values, where $M_{\rm ISM}=M_{\rm HI}+M_{\rm H_2}$. A full description can be found in \cite{2013scoville}. In Fig.~\ref{fig:sizes}b we show results for an optically-thick, radiative transfer modelling of the dust emission for a $r^{-1}$ dust density distribution (they found similar results were found with other reasonable power laws) from \cite{2013scoville}. They set the luminosity as 10$^{12}$\,L$_\odot$ and vary the total dust mass in the range of 10$^7$--10$^9$\,M$_\odot$ (both appropriate for our sample) and calculate the effective radius as the radius of the shell producing the majority of the overall luminosity. In our analysis we assume a dust-to-gas ratio of 100 to estimate the gas mass of SMGs, which is comparable to the M$_{ISM}$ definition of the model used in \cite{2013scoville} who adopt a ratio of $\sim$100.

As mentioned above, the model in Fig.~\ref{fig:sizes}b assumes SMGs are described by a single homogeneous optically-thick dust cloud. In order to compare how the effective radius of SMGs varies with the luminosity-to-mass ratio, we overlay the results derived for the 122 AS2UDS sources which have 870$\mu$m size information and at least one SPIRE detection. We split the sources by radius into four bins to assess the broad trends. We see an overall trend of decreasing dust continuum size with increasing luminosity-to-mass ratio. The observations broadly agree with the ratio of size to $L/M_{ISM}$ in the \cite{2013scoville} model, suggesting that the dust emission from our SMG sample is, on average, consistent with a homogeneous and homologous population of centrally-illuminated dust clouds. The scatter we observe could partly be due to the variation of profiles in the clouds.

We note the structure of the dust clouds in SMGs did not necessarily have to follow this trend: some studies have claimed ``clumpy" structure of the star-forming regions in SMGs \citep{2011swinbank,2011danielson,2013memendez}. If the structure of dust clouds in the SMGs was indeed ``clumpy", the radial extent of the emission would be higher for a given luminosity-to-mass ratio. 
Thus, from their far-infrared emission, SMGs appear to, on average, behave as a sample of sources with very similar structures where the emission is consistent with a central source (starburst) illuminating a surrounding dust/gas reservoir with relatively similar sizes, densities and profiles.

\section{Conclusions} \label{conc}

In this study, we investigated the physical properties of 707
ALMA-identified sub-millimetre galaxies from the AS2UDS survey \citep{2019stach}, with flux densities in the
range $S_{870}$\,$=$\,0.6--13.6\,mJy (with a median of 3.7\,mJy). We fit spectral energy distribution models to the available photometry in 22 bands (from the UV to radio wavelengths) of each SMG  using {\sc magphys}, deriving physical properties
such as their photometric redshifts, stellar and dust masses and far-infrared luminosities. 
Our homogeneously selected survey with uniform photometric coverage allows us to construct sub-samples (including an unbiased luminosity-selected sub-sample) to investigate the evolutionary behaviour of this population. Our main findings are:

\noindent$\bullet$~~ For a sample with a median 870-$\mu$m flux density
of $S_{870}$\,$=$\,3.6\,mJy, $\sim$80 per cent of the galaxies are detected
in the extremely deep $K$-band data available for the UKIDSS UDS field (3-$\sigma$ limit of $K$\,$=$\,25.7).
This demonstrates that $\sim$20 per cent of SMGs are undetectable in very deep optical/near-infrared observations and hence, that there exists a significant population of strongly star-forming, but strongly dust-obscured
galaxies missed by surveys in those wavebands. 

\noindent$\bullet$~~ 
The redshift distribution of our full sample of SMGs has a median of
$z$\,$=$\,2.61\,$\pm$\,0.08 with a 68$^{\rm th}$ percentile range of $z$\,$=$\,1.8--3.4, which is consistent with 
results for smaller samples of
SMGs in other fields using photometric or spectroscopic redshifts.  Those SMGs
which are undetected in the $K$-band appear to preferentially lie at higher redshifts, with
$z$\,$=$\,3.0\,$\pm$\,0.1, while SMGs
which are detected at 1.4\,GHz lie at redshift comparable to the median of the whole SMG population,
$z$\,$=$\,2.5$\pm$\,0.1.
The volume density of SMGs has a
distribution which is log-normal, peaking $\sim$\,2.4\,Gyr after the Big
Bang with the 16--84$^{\rm th}$ percentile range of 1.8--4.5\,Gyr. The inferred formation age distribution peaks at $\sim$1.8\,Gyr ($z\sim$\,3.5).     

\noindent$\bullet$~~ 
The SMG redshift distribution can be reproduced by a simple model describing the growth of halos through a characteristic halo mass, of $\sim$\,6\,$\times$\,10$^{12}$\,M$_\odot$, combined with an increasing molecular gas fraction at higher redshifts. This model suggests that SMGs may represent efficient collapse occurring in the most massive halos that can host such activity. For a  dark matter halo mass of 6\,$\times$\,10$^{12}$\,M$_\odot$ at $z \sim$\,2.6, the median descendent halo mass at $z \sim$\,0 is $\gtrsim$\,10$^{13}$\,M$_\odot$, which is consistent with these galaxies evolving into 2--4\,$L^\ast$ ellipticals at the present day.

\noindent$\bullet$~~ 
Our 870-$\mu$m selected sample most closely resembles a 
sample selected on dust mass, with a ratio of dust mass in M$_{\odot}$
to 870-$\mu$m flux of 
$\log_{10}(M_{\rm d}[M_\odot])\,=\,(1.20\,\pm\,0.03)\,\times \log_{10}(S_{\rm
870}[{\rm mJy}])\,+\,8.16\,\pm\,0.02$.  There
is a weaker correlation between 870-$\mu$m and  
far-infrared luminosity (or star-formation rate, with
SFR\,$\propto$\,$S_{870}^{0.42\pm0.06}$).  The median dust mass of
our sample is $M_{
  d}$\,$=$\,(6.8\,$\pm$\,0.3)\,$\times$\,10$^8$\,M$_\odot$.  Adopting a gas-to-dust ratio of $\delta_{\rm
  GDR}$\,$=$\,100, this implies a median molecular mass of $M_{\rm
  H2}$\,$\sim$\,7\,$\times$\,10$^{10}$\,M$_\odot$.  The median far-infrared
luminosity of the SMGs in our sample is $L_{\rm
  IR}$\,$=$\,(2.88\,$\pm$\,0.09)\,$\times$\,10$^{12}$\,L$_\odot$ and, with a median star-formation rate of
SFR\,=\,236\,$\pm$\,8\,M$_\odot$\,yr$^{-1}$ (68$^{\rm th}$ percentile range of SFR\,=\,113--481\,M$_\odot$\,yr$^{-1}$), suggests a gas
depletion times of approximately 150\,Myr (or an SMG-phase lifetime of
$\sim$\,300\,Myr assuming that, on average, we are witnessing the
SMG halfway through its peak star-formation rate  phase). The characteristic gas
depletion timescale declines by a factor of $\sim$\,2--3\,$\times$ across
$z$\,$=$\,1--4 the trend being driven by an increase in far-infrared luminosity with redshift in our sample as a result of selection effects.

\noindent$\bullet$~~ 
The average mass produced since the start of the SMG-phase (the last $\sim$\,150\,Myr) assuming a constant star-formation rate compared to the total stellar mass has a median of $M_{150\rm Myr}/M_\ast\,\sim$\,0.3. Therefore, for an average SMG, $\sim$\,30 per cent of the current stellar mass was formed in the last 150\,Myr, and by the end of the SMG-phase these systems are likely to roughly double their pre-existing stellar masses.

\noindent$\bullet$~~  
For SMGs with well-constrained
far-infrared SEDs, we show that the median characteristic dust temperature 
for our sample is $T^{\rm MBB}_{\rm d}$\,$=$\,30.4\,$\pm$\,0.3\,K with a 68$^{\rm th}$ percentile range of $T^{\rm MBB}_{\rm d}$\,$=$\,25.7--37.3\,K, with a trend of increasing
temperature with luminosity.  
With a $L_{\rm IR}$-complete sample
across $z$\,$=$\,1.5--4 we are able to exclude the covariance with redshift. We see no evidence for a variation of dust temperature with redshift at fixed luminosity in this sub-sample, suggesting that previous claims of such behaviour are a result of luminosity evolution in the samples employed.
However, we note
that there is an apparent offset in dust temperature between our high-redshift
sample and ULIRGs at $z<$\,1, with the 
high-redshift SMGs being 3\,$\pm$\,1,K cooler at fixed $L_{\rm IR}$, but
this comparison is complicated by the selection function of the local samples.
We suggest the origin of this offset, if real, is likely to be due to the more compact
dust distributions in the ULIRG population at $z<$\,1.

\noindent$\bullet$~~  
We find that gas mass fraction of the SMGs evolves weakly from $\sim$\,30 per cent at $z\sim$\,1.5 to $\sim$\,55 per cent at
$z\sim$\,5. These gas fractions are similar to those suggested for
other high-redshift star-forming populations from mass and gas-selected samples.
We note that the gas mass fraction of SMGs is similar to that estimated 
in an identical manner for {\it Herschel}-detected ULIRGs with comparable star-formation rate at $z<$\,1 from the
GAMA survey:   $\sim$\,35 per cent.  Thus the primary differences we infer
for ULIRGs at $z<$\,1 is a much lower space density and more compact ISM
distribution than those at  $z\gg$\,1.

\noindent$\bullet$~~ We find that the median stellar mass of the SMGs is $M_\ast $\,$=$\,(12.6\,$\pm$\,0.5)\,$\times$\,$10^{10}$\,M$_\odot$ with a 16--68$^{\rm th}$ percentile range of $M_\ast $\,$=$\,(3.5--26.9)\,$\times$\,$10^{10}$\,M$_\odot$. The typical mass does not
evolve strongly with redshift, varying by $<$\,10 per cent over the redshift
range $z$\,$=$\,1--4, although the star-formation rates for our sample
increase by a factor $\sim$\,3 over this same range (driven by the luminosity-redshift trend from the selection).  
In terms of the specific star-formation rate (SFR/$M_\ast$), we see that, at $z\sim$\,1,
typical SMGs lie a factor of $\sim$\,6 above the ``main sequence'' (defined by the field population modelled using {\sc magphys} for consistency). By
$z\sim$\,4 SMGs lie a factor of two above the ``main sequence'', due to the strong evolution of sSFR of the ``main sequence''. 

\noindent$\bullet$~~ By comparing to the stellar mass function of massive field galaxies, and accounting for the duty cycle of SMGs due to gas-depletion, we show that above a stellar mass of $M_\ast>$\,3\,$\times$\,10$^{10}$\,M$_\odot$, $\sim$\,30 per cent of all galaxies at
$z\sim$\,1.8--3.4 (the quartile range of our sample) have gone through
a sub-millimetre-luminous phase, rising to $\sim$\,100 per cent at
$M_\ast\gtrsim$\,3\,$\times$\,10$^{11}$\,M$_\odot$. This is in good agreement with the predictions of simulations.

\noindent$\bullet$~~ We also show
that the volume density of massive,  gas-rich galaxies from our survey is $\sim$\,3$\times$10$^{-4}$\,Mpc$^{-3}$ for galaxies with H$_2$ masses of $\sim$\,10$^{11}$\,M$_\odot$ at $z \sim$\,2.6 and that extrapolating to lower masses this broadly agrees with results from recent blind surveys for CO-emitters with ALMA and JVLA. Thus panoramic sub-millimetre surveys provide an efficient route to identify and study the most massive gas-rich galaxies at high redshifts.

\noindent$\bullet$~~ The contribution of 870-$\mu$m selected SMGs to the total star-formation rate density in the Universe increases steeply with redshift from $z\sim$\,1, with the peak contribution being $\sim$\,15 per cent at $z\sim$\,3 for the $S_{870} \geq$\,3.6\,mJy sub-sample and $\sim$\,60 per cent for SMGs brighter than $S_{870}=$\,1\,mJy. Thus, roughly half of the star-formation rate density at $z\sim$\,3 arises in ULIRG-luminosity sources and the star formation contribution from this population appears to decline only slowly across the 1\,Gyr from $z\sim$\,3 to $z\sim$\,6.

\noindent$\bullet$~~ Finally, we investigate the scale of the rest-frame far-infrared emission in SMGs.  We determine that the star-formation rate in the SMGs is
significantly sub-Eddington, with a typical Eddington ratio of
$\sim$\,0.1. We find that the far-infrared spectral
energy distributions of SMGs are consistent with a modified blackbody model which has an optical depth ($\tau$) of unity at $\lambda_0\geq$\,100\,$\mu$m, and the 870-$\mu$m sizes of SMGs are broadly consistent with them acting as a homologous population of centrally illuminated dust clouds.

Our analysis underlines the fundamental connection between the population of
gas-rich, strongly star-forming galaxies at high redshifts and the formation phase of the most massive galaxy populations over cosmic time.   We suggest that the characteristics of these short-lived, but very active systems represent events where massive halos (with characteristic total masses of $\sim$\,6\,$\times$\,10$^{12}$\,M$_\odot$) with high gas fractions transform their large gas reservoirs into stars on a few dynamical times.  Analysis of the dust continuum morphologies of AS2UDS and ALMA observations of other SMG samples suggests that the continuum emission arises from bar-like structures with diameters of
$\sim$\,2--3\,kpc in more extended gas disks, which suggests that their strong evolution is likely driven by dynamical perturbations of marginally stable gas disks
(\citealt[][]{2019hodge,2019gullberg}).  

\section{Acknowledgements}

The authors thank the anonymous referee for their helpful and insightful comments which have improved the paper. UD acknowledges the support of STFC studentship (ST/R504725/1). The Durham co-authors acknowledge support from STFC (ST/P000541/1). The authors thank John Helly and Lydia Heck for help with HPC.
The ALMA data used in this paper were obtained under programs ADS/JAO.ALMA\#2012.1.00090.S, \#2015.1.01528.S and \#2016.1.00434.S. ALMA is a partnership of ESO (representing its member states), NSF (USA) and NINS (Japan), together with NRC (Canada) and NSC and ASIAA (Taiwan), in cooperation with the Republic of Chile. The Joint ALMA Observatory is operated by ESO, AUI/NRAO, and NAOJ.
This work used the DiRAC@Durham facility managed by the Institute for
Computational Cosmology on behalf of the STFC DiRAC HPC Facility
(www.dirac.ac.uk). 
The equipment was funded by BEIS capital funding
via STFC capital grants ST/K00042X/1, ST/P002293/1, ST/R002371/1 and
ST/S002502/1, Durham University and STFC operations grant
ST/R000832/1. DiRAC is part of the National e-Infrastructure.
We extend our gratitude to the staff at UKIRT for their tireless efforts in ensuring the success of the UKIDSS UDS project.
EdC gratefully acknowledges the Australian Research Council as the recipient of a Future Fellowship (project FT150100079). JLW acknowledges support from an STFC Ernest Rutherford Fellowship (ST/P004784/1 and ST/P004784/2).
All the data required for this project is available through the relevant archives.

%%%%%%%%%%%%%%%%%%%% REFERENCES %%%%%%%%%%%%%%%%%%

\bibliographystyle{mnras}
\bibliography{} 

\begin{thebibliography}{}
\makeatletter
\relax
\def\mn@urlcharsother{\let\do\@makeother \do\$\do\&\do\#\do\^\do\_\do\%\do\~}
\def\mn@doi{\begingroup\mn@urlcharsother \@ifnextchar [ {\mn@doi@}
  {\mn@doi@[]}}
\def\mn@doi@[#1]#2{\def\@tempa{#1}\ifx\@tempa\@empty \href
  {http://dx.doi.org/#2} {doi:#2}\else \href {http://dx.doi.org/#2} {#1}\fi
  \endgroup}
\def\mn@eprint#1#2{\mn@eprint@#1:#2::\@nil}
\def\mn@eprint@arXiv#1{\href {http://arxiv.org/abs/#1} {{\tt arXiv:#1}}}
\def\mn@eprint@dblp#1{\href {http://dblp.uni-trier.de/rec/bibtex/#1.xml}
  {dblp:#1}}
\def\mn@eprint@#1:#2:#3:#4\@nil{\def\@tempa {#1}\def\@tempb {#2}\def\@tempc
  {#3}\ifx \@tempc \@empty \let \@tempc \@tempb \let \@tempb \@tempa \fi \ifx
  \@tempb \@empty \def\@tempb {arXiv}\fi \@ifundefined
  {mn@eprint@\@tempb}{\@tempb:\@tempc}{\expandafter \expandafter \csname
  mn@eprint@\@tempb\endcsname \expandafter{\@tempc}}}

\bibitem[\protect\citeauthoryear{{An} et~al.,}{{An} et~al.}{2018}]{2018an}
{An} F.~X.,  et~al., 2018, \mn@doi [\apj] {10.3847/1538-4357/aacdaa}, \href
  {https://ui.adsabs.harvard.edu/abs/2018ApJ...862..101A} {862, 101}

\bibitem[\protect\citeauthoryear{{Andrews} \& {Thompson}}{{Andrews} \&
  {Thompson}}{2011}]{2011andrews&thompson}
{Andrews} B.~H.,  {Thompson} T.~A.,  2011, \mn@doi [\apj]
  {10.1088/0004-637X/727/2/97}, \href
  {https://ui.adsabs.harvard.edu/abs/2011ApJ...727...97A} {727, 97}

\bibitem[\protect\citeauthoryear{{Aravena} et~al.,}{{Aravena}
  et~al.}{2019}]{2019aravena}
{Aravena} M.,  et~al., 2019, arXiv e-prints, \href
  {https://ui.adsabs.harvard.edu/abs/2019arXiv190309162A} {p. arXiv:1903.09162}

\bibitem[\protect\citeauthoryear{{Baes}, {Verstappen}, {De Looze}, {Fritz},
  {Saftly}, {Vidal P{\'e}rez}, {Stalevski}  \& {Valcke}}{{Baes}
  et~al.}{2011}]{2011baes}
{Baes} M.,  {Verstappen} J.,  {De Looze} I.,  {Fritz} J.,  {Saftly} W.,  {Vidal
  P{\'e}rez} E.,  {Stalevski} M.,   {Valcke} S.,  2011, \mn@doi [\apjs]
  {10.1088/0067-0049/196/2/22}, \href
  {https://ui.adsabs.harvard.edu/abs/2011ApJS..196...22B} {196, 22}

\bibitem[\protect\citeauthoryear{{Barger}, {Cowie}, {Sanders}, {Fulton},
  {Taniguchi}, {Sato}, {Kawara}  \& {Okuda}}{{Barger}
  et~al.}{1998}]{1998barger}
{Barger} A.~J.,  {Cowie} L.~L.,  {Sanders} D.~B.,  {Fulton} E.,  {Taniguchi}
  Y.,  {Sato} Y.,  {Kawara} K.,   {Okuda} H.,  1998, \mn@doi [\nat]
  {10.1038/28338}, \href
  {https://ui.adsabs.harvard.edu/abs/1998Natur.394..248B} {394, 248}

\bibitem[\protect\citeauthoryear{{Battisti} et~al.,}{{Battisti}
  et~al.}{2019}]{2019battisti}
{Battisti} A.~J.,  et~al., 2019, arXiv e-prints, \href
  {https://ui.adsabs.harvard.edu/abs/2019arXiv190800771B} {p. arXiv:1908.00771}

\bibitem[\protect\citeauthoryear{{Baugh}, {Lacey}, {Frenk}, {Granato}, {Silva},
  {Bressan}, {Benson}  \& {Cole}}{{Baugh} et~al.}{2005}]{2005baugh}
{Baugh} C.~M.,  {Lacey} C.~G.,  {Frenk} C.~S.,  {Granato} G.~L.,  {Silva} L.,
  {Bressan} A.,  {Benson} A.~J.,   {Cole} S.,  2005, \mn@doi [\mnras]
  {10.1111/j.1365-2966.2004.08553.x}, \href
  {https://ui.adsabs.harvard.edu/abs/2005MNRAS.356.1191B} {356, 1191}

\bibitem[\protect\citeauthoryear{{Behroozi}, {Wechsler}  \&
  {Conroy}}{{Behroozi} et~al.}{2013}]{2013behroozi}
{Behroozi} P.~S.,  {Wechsler} R.~H.,   {Conroy} C.,  2013, \mn@doi [\apj]
  {10.1088/0004-637X/770/1/57}, \href
  {https://ui.adsabs.harvard.edu/abs/2013ApJ...770...57B} {770, 57}

\bibitem[\protect\citeauthoryear{{B{\'e}thermin}, {Dole}, {Lagache}, {Le
  Borgne}  \& {Penin}}{{B{\'e}thermin} et~al.}{2011}]{2011bethermin}
{B{\'e}thermin} M.,  {Dole} H.,  {Lagache} G.,  {Le Borgne} D.,   {Penin} A.,
  2011, \mn@doi [\aap] {10.1051/0004-6361/201015841}, \href
  {https://ui.adsabs.harvard.edu/abs/2011A&A...529A...4B} {529, A4}

\bibitem[\protect\citeauthoryear{{Biggs} et~al.,}{{Biggs}
  et~al.}{2011}]{2011biggs}
{Biggs} A.~D.,  et~al., 2011, \mn@doi [\mnras]
  {10.1111/j.1365-2966.2010.18132.x}, \href
  {https://ui.adsabs.harvard.edu/abs/2011MNRAS.413.2314B} {413, 2314}

\bibitem[\protect\citeauthoryear{{Blain} \& {Longair}}{{Blain} \&
  {Longair}}{1993}]{1993blain&longair}
{Blain} A.~W.,  {Longair} M.~S.,  1993, \mn@doi [\mnras]
  {10.1093/mnras/264.2.509}, \href
  {https://ui.adsabs.harvard.edu/abs/1993MNRAS.264..509B} {264, 509}

\bibitem[\protect\citeauthoryear{{Blain}, {Smail}, {Ivison}, {Kneib}  \&
  {Frayer}}{{Blain} et~al.}{2002}]{2002blain}
{Blain} A.~W.,  {Smail} I.,  {Ivison} R.~J.,  {Kneib} J.~P.,   {Frayer} D.~T.,
  2002, \mn@doi [\physrep] {10.1016/S0370-1573(02)00134-5}, \href
  {https://ui.adsabs.harvard.edu/abs/2002PhR...369..111B} {369, 111}

\bibitem[\protect\citeauthoryear{{Boselli} et~al.,}{{Boselli}
  et~al.}{2010}]{2010boselli}
{Boselli} A.,  et~al., 2010, \mn@doi [\pasp] {10.1086/651535}, \href
  {https://ui.adsabs.harvard.edu/abs/2010PASP..122..261B} {122, 261}

\bibitem[\protect\citeauthoryear{{Bothwell} et~al.,}{{Bothwell}
  et~al.}{2013}]{2013bothwell}
{Bothwell} M.~S.,  et~al., 2013, \mn@doi [\mnras] {10.1093/mnras/sts562}, \href
  {https://ui.adsabs.harvard.edu/abs/2013MNRAS.429.3047B} {429, 3047}

\bibitem[\protect\citeauthoryear{{Bourne} et~al.,}{{Bourne}
  et~al.}{2016}]{2016bourne}
{Bourne} N.,  et~al., 2016, \mn@doi [\mnras] {10.1093/mnras/stw1654}, \href
  {https://ui.adsabs.harvard.edu/abs/2016MNRAS.462.1714B} {462, 1714}

\bibitem[\protect\citeauthoryear{{Brisbin} et~al.,}{{Brisbin}
  et~al.}{2017}]{2017brisbin}
{Brisbin} D.,  et~al., 2017, \mn@doi [\aap] {10.1051/0004-6361/201730558},
  \href {https://ui.adsabs.harvard.edu/abs/2017A&A...608A..15B} {608, A15}

\bibitem[\protect\citeauthoryear{{Bruzual} \& {Charlot}}{{Bruzual} \&
  {Charlot}}{2003}]{2003bruzual&charlot}
{Bruzual} G.,  {Charlot} S.,  2003, \mn@doi [\mnras]
  {10.1046/j.1365-8711.2003.06897.x}, \href
  {https://ui.adsabs.harvard.edu/abs/2003MNRAS.344.1000B} {344, 1000}

\bibitem[\protect\citeauthoryear{{Burgarella}, {Nanni}, {Hirashita}, {Theule},
  {Inoue}  \& {Takeuchi}}{{Burgarella} et~al.}{2020}]{2020burgarella}
{Burgarella} D.,  {Nanni} A.,  {Hirashita} H.,  {Theule} P.,  {Inoue} A.~K.,
  {Takeuchi} T.~T.,  2020, arXiv e-prints, \href
  {https://ui.adsabs.harvard.edu/abs/2020arXiv200201858B} {p. arXiv:2002.01858}

\bibitem[\protect\citeauthoryear{{Calura} et~al.,}{{Calura}
  et~al.}{2017}]{2017calura}
{Calura} F.,  et~al., 2017, \mn@doi [\mnras] {10.1093/mnras/stw2749}, \href
  {https://ui.adsabs.harvard.edu/abs/2017MNRAS.465...54C} {465, 54}

\bibitem[\protect\citeauthoryear{{Calzetti}, {Armus}, {Bohlin}, {Kinney},
  {Koornneef}  \& {Storchi-Bergmann}}{{Calzetti} et~al.}{2000}]{2000calzetti}
{Calzetti} D.,  {Armus} L.,  {Bohlin} R.~C.,  {Kinney} A.~L.,  {Koornneef} J.,
   {Storchi-Bergmann} T.,  2000, \mn@doi [\apj] {10.1086/308692}, \href
  {https://ui.adsabs.harvard.edu/abs/2000ApJ...533..682C} {533, 682}

\bibitem[\protect\citeauthoryear{{Camps} \& {Baes}}{{Camps} \&
  {Baes}}{2015}]{2015camps&baes}
{Camps} P.,  {Baes} M.,  2015, \mn@doi [Astronomy and Computing]
  {10.1016/j.ascom.2014.10.004}, \href
  {https://ui.adsabs.harvard.edu/abs/2015A&C.....9...20C} {9, 20}

\bibitem[\protect\citeauthoryear{{Camps} et~al.,}{{Camps}
  et~al.}{2018}]{2018camps}
{Camps} P.,  et~al., 2018, \mn@doi [\apjs] {10.3847/1538-4365/aaa24c}, \href
  {https://ui.adsabs.harvard.edu/abs/2018ApJS..234...20C} {234, 20}

\bibitem[\protect\citeauthoryear{{Casey} et~al.,}{{Casey}
  et~al.}{2012}]{2012casey}
{Casey} C.~M.,  et~al., 2012, \mn@doi [\apj] {10.1088/0004-637X/761/2/140},
  \href {https://ui.adsabs.harvard.edu/abs/2012ApJ...761..140C} {761, 140}

\bibitem[\protect\citeauthoryear{{Casey}, {Narayanan}  \& {Cooray}}{{Casey}
  et~al.}{2014}]{2014casey}
{Casey} C.~M.,  {Narayanan} D.,   {Cooray} A.,  2014, \mn@doi [\physrep]
  {10.1016/j.physrep.2014.02.009}, \href
  {https://ui.adsabs.harvard.edu/abs/2014PhR...541...45C} {541, 45}

\bibitem[\protect\citeauthoryear{{Chabrier}}{{Chabrier}}{2003}]{2003chabrier}
{Chabrier} G.,  2003, \mn@doi [\pasp] {10.1086/376392}, \href
  {https://ui.adsabs.harvard.edu/abs/2003PASP..115..763C} {115, 763}

\bibitem[\protect\citeauthoryear{{Chapman}, {Blain}, {Smail}  \&
  {Ivison}}{{Chapman} et~al.}{2005}]{2005chapman}
{Chapman} S.~C.,  {Blain} A.~W.,  {Smail} I.,   {Ivison} R.~J.,  2005, \mn@doi
  [\apj] {10.1086/428082}, \href
  {https://ui.adsabs.harvard.edu/abs/2005ApJ...622..772C} {622, 772}

\bibitem[\protect\citeauthoryear{{Charlot} \& {Fall}}{{Charlot} \&
  {Fall}}{2000}]{2000charlot&fall}
{Charlot} S.,  {Fall} S.~M.,  2000, \mn@doi [\apj] {10.1086/309250}, \href
  {https://ui.adsabs.harvard.edu/abs/2000ApJ...539..718C} {539, 718}

\bibitem[\protect\citeauthoryear{{Chen} et~al.,}{{Chen}
  et~al.}{2016}]{2016chen}
{Chen} C.-C.,  et~al., 2016, \mn@doi [ApJ] {10.3847/0004-637X/831/1/91}, \href
  {https://ui.adsabs.harvard.edu/abs/2016ApJ...831...91C} {831, 91}

\bibitem[\protect\citeauthoryear{{Clemens} et~al.,}{{Clemens}
  et~al.}{2013}]{2013clemens}
{Clemens} M.~S.,  et~al., 2013, \mn@doi [\mnras] {10.1093/mnras/stt760}, \href
  {https://ui.adsabs.harvard.edu/abs/2013MNRAS.433..695C} {433, 695}

\bibitem[\protect\citeauthoryear{{Clements} et~al.,}{{Clements}
  et~al.}{2008}]{2008clements}
{Clements} D.~L.,  et~al., 2008, \mn@doi [\mnras]
  {10.1111/j.1365-2966.2008.13172.x}, \href
  {https://ui.adsabs.harvard.edu/abs/2008MNRAS.387..247C} {387, 247}

\bibitem[\protect\citeauthoryear{{Clements} et~al.,}{{Clements}
  et~al.}{2018}]{2018clements}
{Clements} D.~L.,  et~al., 2018, \mn@doi [\mnras] {10.1093/mnras/stx3227},
  \href {https://ui.adsabs.harvard.edu/abs/2018MNRAS.475.2097C} {475, 2097}

\bibitem[\protect\citeauthoryear{{Condon}}{{Condon}}{1992}]{1992condon}
{Condon} J.~J.,  1992, \mn@doi [\araa] {10.1146/annurev.aa.30.090192.003043},
  \href {https://ui.adsabs.harvard.edu/abs/1992ARA&A..30..575C} {30, 575}

\bibitem[\protect\citeauthoryear{{Coppin} et~al.,}{{Coppin}
  et~al.}{2009}]{2009coppin}
{Coppin} K.~E.~K.,  et~al., 2009, \mn@doi [\mnras]
  {10.1111/j.1365-2966.2009.14700.x}, \href
  {https://ui.adsabs.harvard.edu/abs/2009MNRAS.395.1905C} {395, 1905}

\bibitem[\protect\citeauthoryear{{Cowie}, {Gonz{\'a}lez-L{\'o}pez}, {Barger},
  {Bauer}, {Hsu}  \& {Wang}}{{Cowie} et~al.}{2018}]{2018cowie}
{Cowie} L.~L.,  {Gonz{\'a}lez-L{\'o}pez} J.,  {Barger} A.~J.,  {Bauer} F.~E.,
  {Hsu} L.~Y.,   {Wang} W.~H.,  2018, \mn@doi [\apj]
  {10.3847/1538-4357/aadc63}, \href
  {https://ui.adsabs.harvard.edu/abs/2018ApJ...865..106C} {865, 106}

\bibitem[\protect\citeauthoryear{{Crain} et~al.,}{{Crain}
  et~al.}{2015}]{2015crain}
{Crain} R.~A.,  et~al., 2015, \mn@doi [\mnras] {10.1093/mnras/stv725}, \href
  {https://ui.adsabs.harvard.edu/abs/2015MNRAS.450.1937C} {450, 1937}

\bibitem[\protect\citeauthoryear{{Daddi}, {Cimatti}, {Renzini}, {Fontana},
  {Mignoli}, {Pozzetti}, {Tozzi}  \& {Zamorani}}{{Daddi}
  et~al.}{2004}]{2004daddi}
{Daddi} E.,  {Cimatti} A.,  {Renzini} A.,  {Fontana} A.,  {Mignoli} M.,
  {Pozzetti} L.,  {Tozzi} P.,   {Zamorani} G.,  2004, \mn@doi [\apj]
  {10.1086/425569}, \href
  {https://ui.adsabs.harvard.edu/abs/2004ApJ...617..746D} {617, 746}

\bibitem[\protect\citeauthoryear{{Daddi} et~al.,}{{Daddi}
  et~al.}{2007}]{2007daddi}
{Daddi} E.,  et~al., 2007, \mn@doi [\apj] {10.1086/521818}, \href
  {https://ui.adsabs.harvard.edu/abs/2007ApJ...670..156D} {670, 156}

\bibitem[\protect\citeauthoryear{{Danielson} et~al.,}{{Danielson}
  et~al.}{2011}]{2011danielson}
{Danielson} A.~L.~R.,  et~al., 2011, \mn@doi [\mnras]
  {10.1111/j.1365-2966.2010.17549.x}, \href
  {https://ui.adsabs.harvard.edu/abs/2011MNRAS.410.1687D} {410, 1687}

\bibitem[\protect\citeauthoryear{{Danielson} et~al.,}{{Danielson}
  et~al.}{2017}]{2017danielson}
{Danielson} A.~L.~R.,  et~al., 2017, \mn@doi [\apj] {10.3847/1538-4357/aa6caf},
  \href {https://ui.adsabs.harvard.edu/abs/2017ApJ...840...78D} {840, 78}

\bibitem[\protect\citeauthoryear{{Dav{\'e}}, {Finlator}, {Oppenheimer},
  {Fardal}, {Katz}, {Kere{\v{s}}}  \& {Weinberg}}{{Dav{\'e}}
  et~al.}{2010}]{2010dave}
{Dav{\'e}} R.,  {Finlator} K.,  {Oppenheimer} B.~D.,  {Fardal} M.,  {Katz} N.,
  {Kere{\v{s}}} D.,   {Weinberg} D.~H.,  2010, \mn@doi [\mnras]
  {10.1111/j.1365-2966.2010.16395.x}, \href
  {https://ui.adsabs.harvard.edu/abs/2010MNRAS.404.1355D} {404, 1355}

\bibitem[\protect\citeauthoryear{{Decarli} et~al.,}{{Decarli}
  et~al.}{2016}]{2016decarli_gas}
{Decarli} R.,  et~al., 2016, \mn@doi [\apj] {10.3847/1538-4357/833/1/70}, \href
  {https://ui.adsabs.harvard.edu/abs/2016ApJ...833...70D} {833, 70}

\bibitem[\protect\citeauthoryear{{Decarli} et~al.,}{{Decarli}
  et~al.}{2019}]{2019decarli}
{Decarli} R.,  et~al., 2019, arXiv e-prints, \href
  {https://ui.adsabs.harvard.edu/abs/2019arXiv190309164D} {p. arXiv:1903.09164}

\bibitem[\protect\citeauthoryear{{Donley} et~al.,}{{Donley}
  et~al.}{2012}]{2012donley}
{Donley} J.~L.,  et~al., 2012, \mn@doi [\apj] {10.1088/0004-637X/748/2/142},
  \href {https://ui.adsabs.harvard.edu/abs/2012ApJ...748..142D} {748, 142}

\bibitem[\protect\citeauthoryear{{Draine}}{{Draine}}{2009}]{2009draine}
{Draine} B.~T.,  2009, in {Henning} T.,  {Gr{\"u}n} E.,   {Steinacker} J.,
  eds,  ASP Conf. Ser. Vol. 414, Cosmic Dust - Near and Far. p.~453 (\mn@eprint
  {arXiv} {0903.1658})

\bibitem[\protect\citeauthoryear{{Driver} et~al.,}{{Driver}
  et~al.}{2018}]{2018driver}
{Driver} S.~P.,  et~al., 2018, \mn@doi [\mnras] {10.1093/mnras/stx2728}, \href
  {https://ui.adsabs.harvard.edu/abs/2018MNRAS.475.2891D} {475, 2891}

\bibitem[\protect\citeauthoryear{{Dunlop}}{{Dunlop}}{2011}]{2011dunlop}
{Dunlop} J.~S.,  2011, in {Wang} W.,  {Lu} J.,  {Luo} Z.,  {Yang} Z.,  {Hua}
  H.,   {Chen} Z.,  eds,  ASP Conf. Ser. Vol. 446, Galaxy Evolution: Infrared
  to Millimeter Wavelength Perspective. p.~209

\bibitem[\protect\citeauthoryear{{Dunlop} et~al.,}{{Dunlop}
  et~al.}{2017}]{2017dunlop}
{Dunlop} J.~S.,  et~al., 2017, \mn@doi [\mnras] {10.1093/mnras/stw3088}, \href
  {https://ui.adsabs.harvard.edu/abs/2017MNRAS.466..861D} {466, 861}

\bibitem[\protect\citeauthoryear{{Dunne} et~al.,}{{Dunne}
  et~al.}{2011}]{2011dunne}
{Dunne} L.,  et~al., 2011, \mn@doi [\mnras] {10.1111/j.1365-2966.2011.19363.x},
  \href {https://ui.adsabs.harvard.edu/abs/2011MNRAS.417.1510D} {417, 1510}

\bibitem[\protect\citeauthoryear{{Eales}, {Lilly}, {Gear}, {Dunne}, {Bond},
  {Hammer}, {Le F{\`e}vre}  \& {Crampton}}{{Eales} et~al.}{1999}]{1999eales}
{Eales} S.,  {Lilly} S.,  {Gear} W.,  {Dunne} L.,  {Bond} J.~R.,  {Hammer} F.,
  {Le F{\`e}vre} O.,   {Crampton} D.,  1999, \mn@doi [\apj] {10.1086/307069},
  \href {https://ui.adsabs.harvard.edu/abs/1999ApJ...515..518E} {515, 518}

\bibitem[\protect\citeauthoryear{{Farrah} et~al.,}{{Farrah}
  et~al.}{2006}]{2006farrah}
{Farrah} D.,  et~al., 2006, \mn@doi [\apjl] {10.1086/503769}, \href
  {https://ui.adsabs.harvard.edu/abs/2006ApJ...641L..17F} {641, L17}

\bibitem[\protect\citeauthoryear{{Farrah} et~al.,}{{Farrah}
  et~al.}{2008}]{2008farrah}
{Farrah} D.,  et~al., 2008, \mn@doi [\apj] {10.1086/529485}, \href
  {https://ui.adsabs.harvard.edu/abs/2008ApJ...677..957F} {677, 957}

\bibitem[\protect\citeauthoryear{{Franceschini}, {Toffolatti}, {Mazzei},
  {Danese}  \& {de Zotti}}{{Franceschini} et~al.}{1991}]{1991franceschini}
{Franceschini} A.,  {Toffolatti} L.,  {Mazzei} P.,  {Danese} L.,   {de Zotti}
  G.,  1991, \aaps, \href
  {https://ui.adsabs.harvard.edu/abs/1991A&AS...89..285F} {89, 285}

\bibitem[\protect\citeauthoryear{{Franco} et~al.,}{{Franco}
  et~al.}{2018}]{2018franco}
{Franco} M.,  et~al., 2018, \mn@doi [\aap] {10.1051/0004-6361/201832928}, \href
  {https://ui.adsabs.harvard.edu/abs/2018A&A...620A.152F} {620, A152}

\bibitem[\protect\citeauthoryear{{Frayer}, {Ivison}, {Scoville}, {Yun},
  {Evans}, {Smail}, {Blain}  \& {Kneib}}{{Frayer} et~al.}{1998}]{1998frayer}
{Frayer} D.~T.,  {Ivison} R.~J.,  {Scoville} N.~Z.,  {Yun} M.,  {Evans} A.~S.,
  {Smail} I.,  {Blain} A.~W.,   {Kneib} J.~P.,  1998, \mn@doi [\apjl]
  {10.1086/311639}, \href
  {https://ui.adsabs.harvard.edu/abs/1998ApJ...506L...7F} {506, L7}

\bibitem[\protect\citeauthoryear{{Geach}, {Smail}, {Moran}, {MacArthur},
  {Lagos}  \& {Edge}}{{Geach} et~al.}{2011}]{2011geach}
{Geach} J.~E.,  {Smail} I.,  {Moran} S.~M.,  {MacArthur} L.~A.,  {Lagos} C.
  d.~P.,   {Edge} A.~C.,  2011, \mn@doi [ApJ] {10.1088/2041-8205/730/2/L19},
  \href {https://ui.adsabs.harvard.edu/abs/2011ApJ...730L..19G} {730, L19}

\bibitem[\protect\citeauthoryear{{Geach} et~al.,}{{Geach}
  et~al.}{2017}]{2017geach}
{Geach} J.~E.,  et~al., 2017, \mn@doi [\mnras] {10.1093/mnras/stw2721}, \href
  {https://ui.adsabs.harvard.edu/abs/2017MNRAS.465.1789G} {465, 1789}

\bibitem[\protect\citeauthoryear{{Gerhard}, {Kronawitter}, {Saglia}  \&
  {Bender}}{{Gerhard} et~al.}{2001}]{2001gerhard}
{Gerhard} O.,  {Kronawitter} A.,  {Saglia} R.~P.,   {Bender} R.,  2001, \mn@doi
  [\aj] {10.1086/319940}, \href
  {https://ui.adsabs.harvard.edu/abs/2001AJ....121.1936G} {121, 1936}

\bibitem[\protect\citeauthoryear{{Greve} et~al.,}{{Greve}
  et~al.}{2005}]{2005greve}
{Greve} T.~R.,  et~al., 2005, \mn@doi [\mnras]
  {10.1111/j.1365-2966.2005.08979.x}, \href
  {https://ui.adsabs.harvard.edu/abs/2005MNRAS.359.1165G} {359, 1165}

\bibitem[\protect\citeauthoryear{{Gruppioni} \& {Pozzi}}{{Gruppioni} \&
  {Pozzi}}{2019}]{2019gruppioni}
{Gruppioni} C.,  {Pozzi} F.,  2019, \mn@doi [\mnras] {10.1093/mnras/sty3278},
  \href {https://ui.adsabs.harvard.edu/abs/2019MNRAS.483.1993G} {483, 1993}

\bibitem[\protect\citeauthoryear{{Gruppioni} et~al.,}{{Gruppioni}
  et~al.}{2013}]{2013gruppioni}
{Gruppioni} C.,  et~al., 2013, \mn@doi [\mnras] {10.1093/mnras/stt308}, \href
  {https://ui.adsabs.harvard.edu/abs/2013MNRAS.432...23G} {432, 23}

\bibitem[\protect\citeauthoryear{{Gullberg} et~al.,}{{Gullberg}
  et~al.}{2019}]{2019gullberg}
{Gullberg} B.,  et~al., 2019, \mn@doi [\mnras] {10.1093/mnras/stz2835}, \href
  {https://ui.adsabs.harvard.edu/abs/2019MNRAS.490.4956G} {490, 4956}

\bibitem[\protect\citeauthoryear{{Hainline}, {Blain}, {Smail}, {Alexand er},
  {Armus}, {Chapman}  \& {Ivison}}{{Hainline} et~al.}{2011}]{2011hainline}
{Hainline} L.~J.,  {Blain} A.~W.,  {Smail} I.,  {Alexand er} D.~M.,  {Armus}
  L.,  {Chapman} S.~C.,   {Ivison} R.~J.,  2011, \mn@doi [\apj]
  {10.1088/0004-637X/740/2/96}, \href
  {https://ui.adsabs.harvard.edu/abs/2011ApJ...740...96H} {740, 96}

\bibitem[\protect\citeauthoryear{{Harwit} \& {Pacini}}{{Harwit} \&
  {Pacini}}{1975}]{1975harwit&pacini}
{Harwit} M.,  {Pacini} F.,  1975, \mn@doi [ApJ] {10.1086/181913}, \href
  {https://ui.adsabs.harvard.edu/abs/1975ApJ...200L.127H} {200, L127}

\bibitem[\protect\citeauthoryear{{Hatsukade} et~al.,}{{Hatsukade}
  et~al.}{2016}]{2016hatsukade}
{Hatsukade} B.,  et~al., 2016, \mn@doi [\pasj] {10.1093/pasj/psw026}, \href
  {https://ui.adsabs.harvard.edu/abs/2016PASJ...68...36H} {68, 36}

\bibitem[\protect\citeauthoryear{{Hatsukade} et~al.,}{{Hatsukade}
  et~al.}{2018}]{2018hatsukade}
{Hatsukade} B.,  et~al., 2018, \mn@doi [\pasj] {10.1093/pasj/psy104}, \href
  {https://ui.adsabs.harvard.edu/abs/2018PASJ...70..105H} {70, 105}

\bibitem[\protect\citeauthoryear{{Hayward}, {Kere{\v{s}}}, {Jonsson},
  {Narayanan}, {Cox}  \& {Hernquist}}{{Hayward} et~al.}{2011}]{2011hayward}
{Hayward} C.~C.,  {Kere{\v{s}}} D.,  {Jonsson} P.,  {Narayanan} D.,  {Cox}
  T.~J.,   {Hernquist} L.,  2011, \mn@doi [\apj] {10.1088/0004-637X/743/2/159},
  \href {https://ui.adsabs.harvard.edu/abs/2011ApJ...743..159H} {743, 159}

\bibitem[\protect\citeauthoryear{{Helou}, {Soifer}  \&
  {Rowan-Robinson}}{{Helou} et~al.}{1985}]{1985helou}
{Helou} G.,  {Soifer} B.~T.,   {Rowan-Robinson} M.,  1985, \mn@doi [\apjl]
  {10.1086/184556}, \href
  {https://ui.adsabs.harvard.edu/abs/1985ApJ...298L...7H} {298, L7}

\bibitem[\protect\citeauthoryear{{Hickox} et~al.,}{{Hickox}
  et~al.}{2012}]{2012hickox}
{Hickox} R.~C.,  et~al., 2012, \mn@doi [\mnras]
  {10.1111/j.1365-2966.2011.20303.x}, \href
  {https://ui.adsabs.harvard.edu/abs/2012MNRAS.421..284H} {421, 284}

\bibitem[\protect\citeauthoryear{{Hodge} et~al.,}{{Hodge}
  et~al.}{2013}]{2013hodge}
{Hodge} J.~A.,  et~al., 2013, \mn@doi [\apj] {10.1088/0004-637X/768/1/91},
  \href {https://ui.adsabs.harvard.edu/abs/2013ApJ...768...91H} {768, 91}

\bibitem[\protect\citeauthoryear{{Hodge} et~al.,}{{Hodge}
  et~al.}{2016}]{2016hodge}
{Hodge} J.~A.,  et~al., 2016, \mn@doi [\apj] {10.3847/1538-4357/833/1/103},
  \href {https://ui.adsabs.harvard.edu/abs/2016ApJ...833..103H} {833, 103}

\bibitem[\protect\citeauthoryear{{Hodge} et~al.,}{{Hodge}
  et~al.}{2019}]{2019hodge}
{Hodge} J.~A.,  et~al., 2019, \mn@doi [\apj] {10.3847/1538-4357/ab1846}, \href
  {https://ui.adsabs.harvard.edu/abs/2019ApJ...876..130H} {876, 130}

\bibitem[\protect\citeauthoryear{{Hughes} et~al.,}{{Hughes}
  et~al.}{1998}]{1998hughes}
{Hughes} D.~H.,  et~al., 1998, \mn@doi [\nat] {10.1038/28328}, \href
  {https://ui.adsabs.harvard.edu/abs/1998Natur.394..241H} {394, 241}

\bibitem[\protect\citeauthoryear{{Ikarashi} et~al.,}{{Ikarashi}
  et~al.}{2011}]{2011ikarashi}
{Ikarashi} S.,  et~al., 2011, \mn@doi [\mnras]
  {10.1111/j.1365-2966.2011.18918.x}, \href
  {https://ui.adsabs.harvard.edu/abs/2011MNRAS.415.3081I} {415, 3081}

\bibitem[\protect\citeauthoryear{{Ikarashi} et~al.,}{{Ikarashi}
  et~al.}{2015}]{2015ikarashi}
{Ikarashi} S.,  et~al., 2015, \mn@doi [\apj] {10.1088/0004-637X/810/2/133},
  \href {https://ui.adsabs.harvard.edu/abs/2015ApJ...810..133I} {810, 133}

\bibitem[\protect\citeauthoryear{{Ikarashi} et~al.,}{{Ikarashi}
  et~al.}{2017}]{2017ikarashi}
{Ikarashi} S.,  et~al., 2017, \mn@doi [\apjl] {10.3847/2041-8213/aa9572}, \href
  {https://ui.adsabs.harvard.edu/abs/2017ApJ...849L..36I} {849, L36}

\bibitem[\protect\citeauthoryear{{Iono} et~al.,}{{Iono}
  et~al.}{2009}]{2009iono}
{Iono} D.,  et~al., 2009, \mn@doi [\apj] {10.1088/0004-637X/695/2/1537}, \href
  {https://ui.adsabs.harvard.edu/abs/2009ApJ...695.1537I} {695, 1537}

\bibitem[\protect\citeauthoryear{{Ivison}, {Smail}, {Le Borgne}, {Blain},
  {Kneib}, {Bezecourt}, {Kerr}  \& {Davies}}{{Ivison}
  et~al.}{1998}]{1998ivison}
{Ivison} R.~J.,  {Smail} I.,  {Le Borgne} J.~F.,  {Blain} A.~W.,  {Kneib}
  J.~P.,  {Bezecourt} J.,  {Kerr} T.~H.,   {Davies} J.~K.,  1998, \mn@doi
  [\mnras] {10.1046/j.1365-8711.1998.01677.x}, \href
  {https://ui.adsabs.harvard.edu/abs/1998MNRAS.298..583I} {298, 583}

\bibitem[\protect\citeauthoryear{{Ivison} et~al.,}{{Ivison}
  et~al.}{2002}]{2002ivison}
{Ivison} R.~J.,  et~al., 2002, \mn@doi [\mnras]
  {10.1046/j.1365-8711.2002.05900.x}, \href
  {https://ui.adsabs.harvard.edu/abs/2002MNRAS.337....1I} {337, 1}

\bibitem[\protect\citeauthoryear{{Ivison} et~al.,}{{Ivison}
  et~al.}{2007}]{2007ivison}
{Ivison} R.~J.,  et~al., 2007, \mn@doi [\mnras]
  {10.1111/j.1365-2966.2007.12044.x}, \href
  {https://ui.adsabs.harvard.edu/abs/2007MNRAS.380..199I} {380, 199}

\bibitem[\protect\citeauthoryear{{Ivison} et~al.,}{{Ivison}
  et~al.}{2010}]{2010ivison}
{Ivison} R.~J.,  et~al., 2010, \mn@doi [\mnras]
  {10.1111/j.1365-2966.2009.15918.x}, \href
  {https://ui.adsabs.harvard.edu/abs/2010MNRAS.402..245I} {402, 245}

\bibitem[\protect\citeauthoryear{{Karim} et~al.,}{{Karim}
  et~al.}{2013}]{2013karim}
{Karim} A.,  et~al., 2013, \mn@doi [MNRAS] {10.1093/mnras/stt196}, \href
  {https://ui.adsabs.harvard.edu/abs/2013MNRAS.432....2K} {432, 2}

\bibitem[\protect\citeauthoryear{{Kawinwanichakij} et~al.,}{{Kawinwanichakij}
  et~al.}{2020}]{2020kawinwanichakij}
{Kawinwanichakij} L.,  et~al., 2020, arXiv e-prints, \href
  {https://ui.adsabs.harvard.edu/abs/2020arXiv200207189K} {p. arXiv:2002.07189}

\bibitem[\protect\citeauthoryear{{Kennicutt}}{{Kennicutt}}{1998}]{1998kennicutt}
{Kennicutt} Robert~C. J.,  1998, \mn@doi [\araa]
  {10.1146/annurev.astro.36.1.189}, \href
  {https://ui.adsabs.harvard.edu/abs/1998ARA&A..36..189K} {36, 189}

\bibitem[\protect\citeauthoryear{{Lagos}, {Baugh}, {Lacey}, {Benson}, {Kim}  \&
  {Power}}{{Lagos} et~al.}{2011}]{2011lagos}
{Lagos} C. D.~P.,  {Baugh} C.~M.,  {Lacey} C.~G.,  {Benson} A.~J.,  {Kim}
  H.-S.,   {Power} C.,  2011, \mn@doi [MNRAS]
  {10.1111/j.1365-2966.2011.19583.x}, \href
  {https://ui.adsabs.harvard.edu/abs/2011MNRAS.418.1649L} {418, 1649}

\bibitem[\protect\citeauthoryear{{Laigle} et~al.,}{{Laigle}
  et~al.}{2016}]{2016laigle}
{Laigle} C.,  et~al., 2016, \mn@doi [\apjs] {10.3847/0067-0049/224/2/24}, \href
  {https://ui.adsabs.harvard.edu/abs/2016ApJS..224...24L} {224, 24}

\bibitem[\protect\citeauthoryear{{Lang} et~al.,}{{Lang}
  et~al.}{2019}]{2019lang}
{Lang} P.,  et~al., 2019, \mn@doi [\apj] {10.3847/1538-4357/ab1f77}, \href
  {https://ui.adsabs.harvard.edu/abs/2019ApJ...879...54L} {879, 54}

\bibitem[\protect\citeauthoryear{{Lawrence} et~al.,}{{Lawrence}
  et~al.}{2007}]{2007lawrence}
{Lawrence} A.,  et~al., 2007, \mn@doi [\mnras]
  {10.1111/j.1365-2966.2007.12040.x}, \href
  {https://ui.adsabs.harvard.edu/abs/2007MNRAS.379.1599L} {379, 1599}

\bibitem[\protect\citeauthoryear{{Lee}, {Ferguson}, {Somerville}, {Wiklind}  \&
  {Giavalisco}}{{Lee} et~al.}{2010}]{2010lee}
{Lee} S.-K.,  {Ferguson} H.~C.,  {Somerville} R.~S.,  {Wiklind} T.,
  {Giavalisco} M.,  2010, \mn@doi [\apj] {10.1088/0004-637X/725/2/1644}, \href
  {https://ui.adsabs.harvard.edu/abs/2010ApJ...725.1644L} {725, 1644}

\bibitem[\protect\citeauthoryear{{Liang}, {Feldmann}, {Faucher-Gigu{\`e}re},
  {Kere{\v{s}}}, {Hopkins}, {Hayward}, {Quataert}  \& {Scoville}}{{Liang}
  et~al.}{2018}]{2018liang}
{Liang} L.,  {Feldmann} R.,  {Faucher-Gigu{\`e}re} C.-A.,  {Kere{\v{s}}} D.,
  {Hopkins} P.~F.,  {Hayward} C.~C.,  {Quataert} E.,   {Scoville} N.~Z.,  2018,
  \mn@doi [\mnras] {10.1093/mnrasl/sly071}, \href
  {https://ui.adsabs.harvard.edu/abs/2018MNRAS.478L..83L} {478, L83}

\bibitem[\protect\citeauthoryear{{Lilly}, {Eales}, {Gear}, {Hammer}, {Le
  F{\`e}vre}, {Crampton}, {Bond}  \& {Dunne}}{{Lilly} et~al.}{1999}]{1999lilly}
{Lilly} S.~J.,  {Eales} S.~A.,  {Gear} W. K.~P.,  {Hammer} F.,  {Le F{\`e}vre}
  O.,  {Crampton} D.,  {Bond} J.~R.,   {Dunne} L.,  1999, \mn@doi [\apj]
  {10.1086/307310}, \href
  {https://ui.adsabs.harvard.edu/abs/1999ApJ...518..641L} {518, 641}

\bibitem[\protect\citeauthoryear{{Lindner} et~al.,}{{Lindner}
  et~al.}{2011}]{2011lindner}
{Lindner} R.~R.,  et~al., 2011, \mn@doi [\apj] {10.1088/0004-637X/737/2/83},
  \href {https://ui.adsabs.harvard.edu/abs/2011ApJ...737...83L} {737, 83}

\bibitem[\protect\citeauthoryear{{Madau}}{{Madau}}{1995}]{1995madau}
{Madau} P.,  1995, \mn@doi [\apj] {10.1086/175332}, \href
  {https://ui.adsabs.harvard.edu/abs/1995ApJ...441...18M} {441, 18}

\bibitem[\protect\citeauthoryear{Madau \& Dickinson}{Madau \&
  Dickinson}{2014}]{2014madau}
Madau P.,  Dickinson M.,  2014, \mn@doi [ARA&A]
  {10.1146/annurev-astro-081811-125615}, 52, 415

\bibitem[\protect\citeauthoryear{{Magnelli} et~al.,}{{Magnelli}
  et~al.}{2010}]{2010magnelli}
{Magnelli} B.,  et~al., 2010, \mn@doi [\aap] {10.1051/0004-6361/201014616},
  \href {https://ui.adsabs.harvard.edu/abs/2010A&A...518L..28M} {518, L28}

\bibitem[\protect\citeauthoryear{{Magnelli} et~al.,}{{Magnelli}
  et~al.}{2012}]{2012magnelli}
{Magnelli} B.,  et~al., 2012, \mn@doi [\aap] {10.1051/0004-6361/201118312},
  \href {https://ui.adsabs.harvard.edu/abs/2012A&A...539A.155M} {539, A155}

\bibitem[\protect\citeauthoryear{{Magnelli} et~al.,}{{Magnelli}
  et~al.}{2013}]{2013magnelli}
{Magnelli} B.,  et~al., 2013, \mn@doi [\aap] {10.1051/0004-6361/201321371},
  \href {https://ui.adsabs.harvard.edu/abs/2013A&A...553A.132M} {553, A132}

\bibitem[\protect\citeauthoryear{{McAlpine} et~al.,}{{McAlpine}
  et~al.}{2019}]{2019mcalpine}
{McAlpine} S.,  et~al., 2019, \mn@doi [\mnras] {10.1093/mnras/stz1692}, \href
  {https://ui.adsabs.harvard.edu/abs/2019MNRAS.tmp.1653M} {p.~1653}

\bibitem[\protect\citeauthoryear{{Men{\'e}ndez-Delmestre}
  et~al.,}{{Men{\'e}ndez-Delmestre} et~al.}{2009}]{2009menendez}
{Men{\'e}ndez-Delmestre} K.,  et~al., 2009, \mn@doi [\apj]
  {10.1088/0004-637X/699/1/667}, \href
  {https://ui.adsabs.harvard.edu/abs/2009ApJ...699..667M} {699, 667}

\bibitem[\protect\citeauthoryear{{Men{\'e}ndez-Delmestre}, {Blain}, {Swinbank},
  {Smail}, {Ivison}, {Chapman}  \& {Gon{\c{c}}alves}}{{Men{\'e}ndez-Delmestre}
  et~al.}{2013}]{2013memendez}
{Men{\'e}ndez-Delmestre} K.,  {Blain} A.~W.,  {Swinbank} M.,  {Smail} I.,
  {Ivison} R.~J.,  {Chapman} S.~C.,   {Gon{\c{c}}alves} T.~S.,  2013, \mn@doi
  [\apj] {10.1088/0004-637X/767/2/151}, \href
  {https://ui.adsabs.harvard.edu/abs/2013ApJ...767..151M} {767, 151}

\bibitem[\protect\citeauthoryear{{Miettinen} et~al.,}{{Miettinen}
  et~al.}{2017}]{2017miettinen}
{Miettinen} O.,  et~al., 2017, \mn@doi [\aap] {10.1051/0004-6361/201730762},
  \href {https://ui.adsabs.harvard.edu/abs/2017A&A...606A..17M} {606, A17}

\bibitem[\protect\citeauthoryear{{Mu{\~n}oz Arancibia} et~al.,}{{Mu{\~n}oz
  Arancibia} et~al.}{2018}]{2018munoz}
{Mu{\~n}oz Arancibia} A.~M.,  et~al., 2018, \mn@doi [\aap]
  {10.1051/0004-6361/201732442}, \href
  {https://ui.adsabs.harvard.edu/abs/2018A&A...620A.125M} {620, A125}

\bibitem[\protect\citeauthoryear{{Narayanan} et~al.,}{{Narayanan}
  et~al.}{2015}]{2015narayanan}
{Narayanan} D.,  et~al., 2015, \mn@doi [\nat] {10.1038/nature15383}, \href
  {https://ui.adsabs.harvard.edu/abs/2015Natur.525..496N} {525, 496}

\bibitem[\protect\citeauthoryear{{Oliver} et~al.,}{{Oliver}
  et~al.}{2012}]{2012oliver}
{Oliver} S.~J.,  et~al., 2012, \mn@doi [\mnras]
  {10.1111/j.1365-2966.2012.20912.x}, \href
  {https://ui.adsabs.harvard.edu/abs/2012MNRAS.424.1614O} {424, 1614}

\bibitem[\protect\citeauthoryear{{Overzier}, {R{\"o}ttgering}, {Rengelink}  \&
  {Wilman}}{{Overzier} et~al.}{2003}]{2003overzier}
{Overzier} R.~A.,  {R{\"o}ttgering} H.~J.~A.,  {Rengelink} R.~B.,   {Wilman}
  R.~J.,  2003, \mn@doi [\aap] {10.1051/0004-6361:20030527}, \href
  {https://ui.adsabs.harvard.edu/abs/2003A&A...405...53O} {405, 53}

\bibitem[\protect\citeauthoryear{{Pettini}, {Rix}, {Steidel}, {Adelberger},
  {Hunt}  \& {Shapley}}{{Pettini} et~al.}{2002}]{2002Pettini}
{Pettini} M.,  {Rix} S.~A.,  {Steidel} C.~C.,  {Adelberger} K.~L.,  {Hunt}
  M.~P.,   {Shapley} A.~E.,  2002, \mn@doi [\apj] {10.1086/339355}, \href
  {https://ui.adsabs.harvard.edu/abs/2002ApJ...569..742P} {569, 742}

\bibitem[\protect\citeauthoryear{{Planck Collaboration} et~al.,}{{Planck
  Collaboration} et~al.}{2011}]{2011planck}
{Planck Collaboration} et~al., 2011, \mn@doi [\aap]
  {10.1051/0004-6361/201116479}, \href
  {https://ui.adsabs.harvard.edu/abs/2011A&A...536A..19P} {536, A19}

\bibitem[\protect\citeauthoryear{{Pope} et~al.,}{{Pope}
  et~al.}{2006}]{2006pope}
{Pope} A.,  et~al., 2006, \mn@doi [\mnras] {10.1111/j.1365-2966.2006.10575.x},
  \href {https://ui.adsabs.harvard.edu/abs/2006MNRAS.370.1185P} {370, 1185}

\bibitem[\protect\citeauthoryear{{Pope} et~al.,}{{Pope}
  et~al.}{2008}]{2008pope}
{Pope} A.,  et~al., 2008, \mn@doi [\apj] {10.1086/592739}, \href
  {https://ui.adsabs.harvard.edu/abs/2008ApJ...689..127P} {689, 127}

\bibitem[\protect\citeauthoryear{{Puget}, {Abergel}, {Bernard}, {Boulanger},
  {Burton}, {Desert}  \& {Hartmann}}{{Puget} et~al.}{1996}]{1996puget}
{Puget} J.~L.,  {Abergel} A.,  {Bernard} J.~P.,  {Boulanger} F.,  {Burton}
  W.~B.,  {Desert} F.~X.,   {Hartmann} D.,  1996, \aap, \href
  {https://ui.adsabs.harvard.edu/abs/1996A&A...308L...5P} {308, L5}

\bibitem[\protect\citeauthoryear{{Riechers} et~al.,}{{Riechers}
  et~al.}{2013}]{2013riechers}
{Riechers} D.~A.,  et~al., 2013, \mn@doi [\nat] {10.1038/nature12050}, \href
  {https://ui.adsabs.harvard.edu/abs/2013Natur.496..329R} {496, 329}

\bibitem[\protect\citeauthoryear{{Riechers} et~al.,}{{Riechers}
  et~al.}{2019}]{2019riechers}
{Riechers} D.~A.,  et~al., 2019, \mn@doi [\apj] {10.3847/1538-4357/aafc27},
  \href {https://ui.adsabs.harvard.edu/abs/2019ApJ...872....7R} {872, 7}

\bibitem[\protect\citeauthoryear{{Roseboom} et~al.,}{{Roseboom}
  et~al.}{2012}]{2012roseboom}
{Roseboom} I.~G.,  et~al., 2012, \mn@doi [\mnras]
  {10.1111/j.1365-2966.2011.19827.x}, \href
  {https://ui.adsabs.harvard.edu/abs/2012MNRAS.419.2758R} {419, 2758}

\bibitem[\protect\citeauthoryear{{Ross} et~al.,}{{Ross}
  et~al.}{2009}]{2009ross}
{Ross} N.~P.,  et~al., 2009, \mn@doi [\apj] {10.1088/0004-637X/697/2/1634},
  \href {https://ui.adsabs.harvard.edu/abs/2009ApJ...697.1634R} {697, 1634}

\bibitem[\protect\citeauthoryear{{Sanders} \& {Mirabel}}{{Sanders} \&
  {Mirabel}}{1996}]{1996sanders&mirabel}
{Sanders} D.~B.,  {Mirabel} I.~F.,  1996, \mn@doi [\araa]
  {10.1146/annurev.astro.34.1.749}, \href
  {https://ui.adsabs.harvard.edu/abs/1996ARA&A..34..749S} {34, 749}

\bibitem[\protect\citeauthoryear{{Sanders}, {Soifer}, {Elias}, {Neugebauer}  \&
  {Matthews}}{{Sanders} et~al.}{1988}]{1988sanders}
{Sanders} D.~B.,  {Soifer} B.~T.,  {Elias} J.~H.,  {Neugebauer} G.,
  {Matthews} K.,  1988, \mn@doi [\apjl] {10.1086/185155}, \href
  {https://ui.adsabs.harvard.edu/abs/1988ApJ...328L..35S} {328, L35}

\bibitem[\protect\citeauthoryear{{Schaye} et~al.,}{{Schaye}
  et~al.}{2015}]{2015schaye}
{Schaye} J.,  et~al., 2015, \mn@doi [\mnras] {10.1093/mnras/stu2058}, \href
  {https://ui.adsabs.harvard.edu/abs/2015MNRAS.446..521S} {446, 521}

\bibitem[\protect\citeauthoryear{{Schechter}}{{Schechter}}{1976}]{1976schechter}
{Schechter} P.,  1976, \mn@doi [\apj] {10.1086/154079}, \href
  {https://ui.adsabs.harvard.edu/abs/1976ApJ...203..297S} {203, 297}

\bibitem[\protect\citeauthoryear{{Scoville}}{{Scoville}}{2013}]{2013scoville}
{Scoville} N.~Z.,  2013, {Evolution of star formation and gas, Eds J.
  Falcon-Barroso \& J.H. Knapen, Cambridge University Press}.
p.~491

\bibitem[\protect\citeauthoryear{{Scoville} et~al.,}{{Scoville}
  et~al.}{2014}]{2014scoville}
{Scoville} N.,  et~al., 2014, \mn@doi [\apj] {10.1088/0004-637X/783/2/84},
  \href {https://ui.adsabs.harvard.edu/abs/2014ApJ...783...84S} {783, 84}

\bibitem[\protect\citeauthoryear{{Simpson} et~al.,}{{Simpson}
  et~al.}{2012}]{2012simpson}
{Simpson} J.~M.,  et~al., 2012, \mn@doi [\mnras]
  {10.1111/j.1365-2966.2012.21941.x}, \href
  {https://ui.adsabs.harvard.edu/abs/2012MNRAS.426.3201S} {426, 3201}

\bibitem[\protect\citeauthoryear{{Simpson}, {Westoby}, {Arumugam}, {Ivison},
  {Hartley}  \& {Almaini}}{{Simpson} et~al.}{2013}]{2013simpson}
{Simpson} C.,  {Westoby} P.,  {Arumugam} V.,  {Ivison} R.,  {Hartley} W.,
  {Almaini} O.,  2013, \mn@doi [\mnras] {10.1093/mnras/stt940}, \href
  {https://ui.adsabs.harvard.edu/abs/2013MNRAS.433.2647S} {433, 2647}

\bibitem[\protect\citeauthoryear{{Simpson} et~al.,}{{Simpson}
  et~al.}{2014}]{2014simpson}
{Simpson} J.~M.,  et~al., 2014, \mn@doi [\apj] {10.1088/0004-637X/788/2/125},
  \href {https://ui.adsabs.harvard.edu/abs/2014ApJ...788..125S} {788, 125}

\bibitem[\protect\citeauthoryear{{Simpson} et~al.,}{{Simpson}
  et~al.}{2015a}]{2015simpsona}
{Simpson} J.~M.,  et~al., 2015a, \mn@doi [\apj] {10.1088/0004-637X/799/1/81},
  \href {https://ui.adsabs.harvard.edu/abs/2015ApJ...799...81S} {799, 81}

\bibitem[\protect\citeauthoryear{{Simpson} et~al.,}{{Simpson}
  et~al.}{2015b}]{2015simpsonb}
{Simpson} J.~M.,  et~al., 2015b, \mn@doi [\apj] {10.1088/0004-637X/807/2/128},
  \href {https://ui.adsabs.harvard.edu/abs/2015ApJ...807..128S} {807, 128}

\bibitem[\protect\citeauthoryear{{Simpson} et~al.,}{{Simpson}
  et~al.}{2017}]{2017simpson}
{Simpson} J.~M.,  et~al., 2017, \mn@doi [\apj] {10.3847/1538-4357/aa65d0},
  \href {https://ui.adsabs.harvard.edu/abs/2017ApJ...839...58S} {839, 58}

\bibitem[\protect\citeauthoryear{{Smail}, {Ivison}  \& {Blain}}{{Smail}
  et~al.}{1997}]{1997smail}
{Smail} I.,  {Ivison} R.~J.,   {Blain} A.~W.,  1997, \mn@doi [\apjl]
  {10.1086/311017}, \href
  {https://ui.adsabs.harvard.edu/abs/1997ApJ...490L...5S} {490, L5}

\bibitem[\protect\citeauthoryear{{Smail}, {Ivison}, {Kneib}, {Cowie}, {Blain},
  {Barger}, {Owen}  \& {Morrison}}{{Smail} et~al.}{1999}]{1999smail}
{Smail} I.,  {Ivison} R.~J.,  {Kneib} J.~P.,  {Cowie} L.~L.,  {Blain} A.~W.,
  {Barger} A.~J.,  {Owen} F.~N.,   {Morrison} G.,  1999, \mn@doi [\mnras]
  {10.1046/j.1365-8711.1999.02819.x}, \href
  {https://ui.adsabs.harvard.edu/abs/1999MNRAS.308.1061S} {308, 1061}

\bibitem[\protect\citeauthoryear{{Smail}, {Chapman}, {Blain}  \&
  {Ivison}}{{Smail} et~al.}{2004}]{2004smail}
{Smail} I.,  {Chapman} S.~C.,  {Blain} A.~W.,   {Ivison} R.~J.,  2004, \mn@doi
  [\apj] {10.1086/424896}, \href
  {https://ui.adsabs.harvard.edu/abs/2004ApJ...616...71S} {616, 71}

\bibitem[\protect\citeauthoryear{{Smail}, {Sharp}, {Swinbank}, {Akiyama},
  {Ueda}, {Foucaud}, {Almaini}  \& {Croom}}{{Smail} et~al.}{2008}]{2008smail}
{Smail} I.,  {Sharp} R.,  {Swinbank} A.~M.,  {Akiyama} M.,  {Ueda} Y.,
  {Foucaud} S.,  {Almaini} O.,   {Croom} S.,  2008, \mn@doi [\mnras]
  {10.1111/j.1365-2966.2008.13579.x}, \href
  {https://ui.adsabs.harvard.edu/abs/2008MNRAS.389..407S} {389, 407}

\bibitem[\protect\citeauthoryear{{Smith} et~al.,}{{Smith}
  et~al.}{2013}]{2013smith}
{Smith} D.~J.~B.,  et~al., 2013, \mn@doi [\mnras] {10.1093/mnras/stt1737},
  \href {https://ui.adsabs.harvard.edu/abs/2013MNRAS.436.2435S} {436, 2435}

\bibitem[\protect\citeauthoryear{{Springel} et~al.,}{{Springel}
  et~al.}{2005}]{2005springel}
{Springel} V.,  et~al., 2005, \mn@doi [\nat] {10.1038/nature03597}, \href
  {https://ui.adsabs.harvard.edu/abs/2005Natur.435..629S} {435, 629}

\bibitem[\protect\citeauthoryear{{Stach} et~al.,}{{Stach}
  et~al.}{2018}]{2018stach}
{Stach} S.~M.,  et~al., 2018, \mn@doi [\apj] {10.3847/1538-4357/aac5e5}, \href
  {https://ui.adsabs.harvard.edu/abs/2018ApJ...860..161S} {860, 161}

\bibitem[\protect\citeauthoryear{{Stach} et~al.,}{{Stach}
  et~al.}{2019}]{2019stach}
{Stach} S.~M.,  et~al., 2019, \mn@doi [\mnras] {10.1093/mnras/stz1536}, \href
  {https://ui.adsabs.harvard.edu/abs/2019MNRAS.487.4648S} {487, 4648}

\bibitem[\protect\citeauthoryear{{Strandet} et~al.,}{{Strandet}
  et~al.}{2016}]{2016strandet}
{Strandet} M.~L.,  et~al., 2016, \mn@doi [\apj] {10.3847/0004-637X/822/2/80},
  \href {https://ui.adsabs.harvard.edu/abs/2016ApJ...822...80S} {822, 80}

\bibitem[\protect\citeauthoryear{{Swinbank}, {Chapman}, {Smail}, {Lindner},
  {Borys}, {Blain}, {Ivison}  \& {Lewis}}{{Swinbank}
  et~al.}{2006}]{2006swinbank}
{Swinbank} A.~M.,  {Chapman} S.~C.,  {Smail} I.,  {Lindner} C.,  {Borys} C.,
  {Blain} A.~W.,  {Ivison} R.~J.,   {Lewis} G.~F.,  2006, \mn@doi [\mnras]
  {10.1111/j.1365-2966.2006.10673.x}, \href
  {https://ui.adsabs.harvard.edu/abs/2006MNRAS.371..465S} {371, 465}

\bibitem[\protect\citeauthoryear{{Swinbank} et~al.,}{{Swinbank}
  et~al.}{2011}]{2011swinbank}
{Swinbank} A.~M.,  et~al., 2011, \mn@doi [\apj] {10.1088/0004-637X/742/1/11},
  \href {https://ui.adsabs.harvard.edu/abs/2011ApJ...742...11S} {742, 11}

\bibitem[\protect\citeauthoryear{{Swinbank} et~al.,}{{Swinbank}
  et~al.}{2012}]{2012swinbank}
{Swinbank} A.~M.,  et~al., 2012, \mn@doi [\mnras]
  {10.1111/j.1365-2966.2012.22048.x}, \href
  {https://ui.adsabs.harvard.edu/abs/2012MNRAS.427.1066S} {427, 1066}

\bibitem[\protect\citeauthoryear{{Swinbank} et~al.,}{{Swinbank}
  et~al.}{2014}]{2014swinbank}
{Swinbank} A.~M.,  et~al., 2014, \mn@doi [\mnras] {10.1093/mnras/stt2273},
  \href {https://ui.adsabs.harvard.edu/abs/2014MNRAS.438.1267S} {438, 1267}

\bibitem[\protect\citeauthoryear{{Symeonidis}, {Page}  \&
  {Seymour}}{{Symeonidis} et~al.}{2011}]{2011symeonidis}
{Symeonidis} M.,  {Page} M.~J.,   {Seymour} N.,  2011, \mn@doi [\mnras]
  {10.1111/j.1365-2966.2010.17735.x}, \href
  {https://ui.adsabs.harvard.edu/abs/2011MNRAS.411..983S} {411, 983}

\bibitem[\protect\citeauthoryear{{Symeonidis} et~al.,}{{Symeonidis}
  et~al.}{2013}]{2013symeonidis}
{Symeonidis} M.,  et~al., 2013, \mn@doi [\mnras] {10.1093/mnras/stt330}, \href
  {https://ui.adsabs.harvard.edu/abs/2013MNRAS.431.2317S} {431, 2317}

\bibitem[\protect\citeauthoryear{{Tacconi} et~al.,}{{Tacconi}
  et~al.}{2018}]{2018tacconi}
{Tacconi} L.~J.,  et~al., 2018, \mn@doi [\apj] {10.3847/1538-4357/aaa4b4},
  \href {https://ui.adsabs.harvard.edu/abs/2018ApJ...853..179T} {853, 179}

\bibitem[\protect\citeauthoryear{{Tasca} et~al.,}{{Tasca}
  et~al.}{2015}]{2015tasca}
{Tasca} L.~A.~M.,  et~al., 2015, \mn@doi [\aap] {10.1051/0004-6361/201425379},
  \href {https://ui.adsabs.harvard.edu/abs/2015A&A...581A..54T} {581, A54}

\bibitem[\protect\citeauthoryear{{Toft} et~al.,}{{Toft}
  et~al.}{2014}]{2014toft}
{Toft} S.,  et~al., 2014, \mn@doi [\apj] {10.1088/0004-637X/782/2/68}, \href
  {https://ui.adsabs.harvard.edu/abs/2014ApJ...782...68T} {782, 68}

\bibitem[\protect\citeauthoryear{{Umehata} et~al.,}{{Umehata}
  et~al.}{2018}]{2018umehata}
{Umehata} H.,  et~al., 2018, \mn@doi [\pasj] {10.1093/pasj/psy065}, \href
  {https://ui.adsabs.harvard.edu/abs/2018PASJ...70...65U} {70, 65}

\bibitem[\protect\citeauthoryear{{Vlahakis}, {Dunne}  \& {Eales}}{{Vlahakis}
  et~al.}{2005}]{2005vlahakis}
{Vlahakis} C.,  {Dunne} L.,   {Eales} S.,  2005, \mn@doi [\mnras]
  {10.1111/j.1365-2966.2005.09666.x}, \href
  {https://ui.adsabs.harvard.edu/abs/2005MNRAS.364.1253V} {364, 1253}

\bibitem[\protect\citeauthoryear{{Vlahakis}, {Eales}  \& {Dunne}}{{Vlahakis}
  et~al.}{2007}]{2007vlahakis}
{Vlahakis} C.,  {Eales} S.,   {Dunne} L.,  2007, \mn@doi [\mnras]
  {10.1111/j.1365-2966.2007.12007.x}, \href
  {https://ui.adsabs.harvard.edu/abs/2007MNRAS.379.1042V} {379, 1042}

\bibitem[\protect\citeauthoryear{{Wall}, {Pope}  \& {Scott}}{{Wall}
  et~al.}{2008}]{2008wall}
{Wall} J.~V.,  {Pope} A.,   {Scott} D.,  2008, \mn@doi [\mnras]
  {10.1111/j.1365-2966.2007.12547.x}, \href
  {https://ui.adsabs.harvard.edu/abs/2008MNRAS.383..435W} {383, 435}

\bibitem[\protect\citeauthoryear{{Walter} et~al.,}{{Walter}
  et~al.}{2016}]{2016walter}
{Walter} F.,  et~al., 2016, \mn@doi [\apj] {10.3847/1538-4357/833/1/67}, \href
  {https://ui.adsabs.harvard.edu/abs/2016ApJ...833...67W} {833, 67}

\bibitem[\protect\citeauthoryear{{Wang} et~al.,}{{Wang}
  et~al.}{2013}]{2013wang}
{Wang} S.~X.,  et~al., 2013, \mn@doi [\apj] {10.1088/0004-637X/778/2/179},
  \href {https://ui.adsabs.harvard.edu/abs/2013ApJ...778..179W} {778, 179}

\bibitem[\protect\citeauthoryear{{Wardlow} et~al.,}{{Wardlow}
  et~al.}{2011}]{2011wardlow}
{Wardlow} J.~L.,  et~al., 2011, \mn@doi [\mnras]
  {10.1111/j.1365-2966.2011.18795.x}, \href
  {https://ui.adsabs.harvard.edu/abs/2011MNRAS.415.1479W} {415, 1479}

\bibitem[\protect\citeauthoryear{{Wei{\ss}} et~al.,}{{Wei{\ss}}
  et~al.}{2009}]{2009weiss}
{Wei{\ss}} A.,  et~al., 2009, \mn@doi [ApJ] {10.1088/0004-637X/707/2/1201},
  \href {https://ui.adsabs.harvard.edu/abs/2009ApJ...707.1201W} {707, 1201}

\bibitem[\protect\citeauthoryear{{Wei{\ss}} et~al.,}{{Wei{\ss}}
  et~al.}{2013}]{2013weiss}
{Wei{\ss}} A.,  et~al., 2013, \mn@doi [\apj] {10.1088/0004-637X/767/1/88},
  \href {https://ui.adsabs.harvard.edu/abs/2013ApJ...767...88W} {767, 88}

\bibitem[\protect\citeauthoryear{{Whitaker}, {Kriek}, {van Dokkum}, {Bezanson},
  {Brammer}, {Franx}  \& {Labb{\'e}}}{{Whitaker} et~al.}{2012}]{2012whitaker}
{Whitaker} K.~E.,  {Kriek} M.,  {van Dokkum} P.~G.,  {Bezanson} R.,  {Brammer}
  G.,  {Franx} M.,   {Labb{\'e}} I.,  2012, \mn@doi [\apj]
  {10.1088/0004-637X/745/2/179}, \href
  {https://ui.adsabs.harvard.edu/abs/2012ApJ...745..179W} {745, 179}

\bibitem[\protect\citeauthoryear{{White} \& {Rees}}{{White} \&
  {Rees}}{1978}]{1978white&rees}
{White} S.~D.~M.,  {Rees} M.~J.,  1978, \mn@doi [\mnras]
  {10.1093/mnras/183.3.341}, \href
  {https://ui.adsabs.harvard.edu/abs/1978MNRAS.183..341W} {183, 341}

\bibitem[\protect\citeauthoryear{{Wilkinson} et~al.,}{{Wilkinson}
  et~al.}{2017}]{2017wilkinson}
{Wilkinson} A.,  et~al., 2017, \mn@doi [MNRAS] {10.1093/mnras/stw2405}, \href
  {https://ui.adsabs.harvard.edu/abs/2017MNRAS.464.1380W} {464, 1380}

\bibitem[\protect\citeauthoryear{{Wilson}, {Rangwala}, {Glenn}, {Maloney},
  {Spinoglio}  \& {Pereira-Santaella}}{{Wilson} et~al.}{2014}]{2014wilson}
{Wilson} C.~D.,  {Rangwala} N.,  {Glenn} J.,  {Maloney} P.~R.,  {Spinoglio} L.,
    {Pereira-Santaella} M.,  2014, \mn@doi [\apjl]
  {10.1088/2041-8205/789/2/L36}, \href
  {https://ui.adsabs.harvard.edu/abs/2014ApJ...789L..36W} {789, L36}

\bibitem[\protect\citeauthoryear{{Yun}, {Reddy}  \& {Condon}}{{Yun}
  et~al.}{2001}]{2001yun}
{Yun} M.~S.,  {Reddy} N.~A.,   {Condon} J.~J.,  2001, \mn@doi [\apj]
  {10.1086/323145}, \href
  {https://ui.adsabs.harvard.edu/abs/2001ApJ...554..803Y} {554, 803}

\bibitem[\protect\citeauthoryear{{Zehavi} et~al.,}{{Zehavi}
  et~al.}{2011}]{2011zehavi}
{Zehavi} I.,  et~al., 2011, \mn@doi [\apj] {10.1088/0004-637X/736/1/59}, \href
  {https://ui.adsabs.harvard.edu/abs/2011ApJ...736...59Z} {736, 59}

\bibitem[\protect\citeauthoryear{{Zhang}, {Romano}, {Ivison}, {Papadopoulos}
  \& {Matteucci}}{{Zhang} et~al.}{2018}]{2018zhang}
{Zhang} Z.-Y.,  {Romano} D.,  {Ivison} R.~J.,  {Papadopoulos} P.~P.,
  {Matteucci} F.,  2018, \mn@doi [\nat] {10.1038/s41586-018-0196-x}, \href
  {https://ui.adsabs.harvard.edu/abs/2018Natur.558..260Z} {558, 260}

\bibitem[\protect\citeauthoryear{{Zhukovska}, {Henning}  \&
  {Dobbs}}{{Zhukovska} et~al.}{2018}]{2018zhukovska}
{Zhukovska} S.,  {Henning} T.,   {Dobbs} C.,  2018, \mn@doi [\apj]
  {10.3847/1538-4357/aab438}, \href
  {https://ui.adsabs.harvard.edu/abs/2018ApJ...857...94Z} {857, 94}

\bibitem[\protect\citeauthoryear{{da Cunha}, {Charlot}  \& {Elbaz}}{{da Cunha}
  et~al.}{2008}]{2008dacunha}
{da Cunha} E.,  {Charlot} S.,   {Elbaz} D.,  2008, \mn@doi [\mnras]
  {10.1111/j.1365-2966.2008.13535.x}, \href
  {https://ui.adsabs.harvard.edu/abs/2008MNRAS.388.1595D} {388, 1595}

\bibitem[\protect\citeauthoryear{{da Cunha} et~al.,}{{da Cunha}
  et~al.}{2015}]{2015dacunha}
{da Cunha} E.,  et~al., 2015, \mn@doi [\apj] {10.1088/0004-637X/806/1/110},
  \href {https://ui.adsabs.harvard.edu/abs/2015ApJ...806..110D} {806, 110}

\bibitem[\protect\citeauthoryear{{van der Kruit}}{{van der
  Kruit}}{1971}]{1971vanderkruit}
{van der Kruit} P.~C.,  1971, \aap, \href
  {https://ui.adsabs.harvard.edu/abs/1971A&A....15..110V} {15, 110}

\bibitem[\protect\citeauthoryear{{van der Kruit}}{{van der
  Kruit}}{1973}]{1973vanderkruit}
{van der Kruit} P.~C.,  1973, \aap, \href
  {https://ui.adsabs.harvard.edu/abs/1973A&A....29..263V} {29, 263}

\makeatother
\end{thebibliography}

\appendix

\section{Supporting material} \label{appendixA}

The following supporing material is available at MNRAS online. \\
{\bf Table A1.} The AS2UDS catalog containing the results for all 707 SMGs from this study. The catalog includes all of the photometry used to fit the SEDs and the {\sc magphys} outputs for each of the sources. This includes photometric redshift, stellar mass, SFR, A$_{\rm V}$, mass-weighted age, far-infrared luminosity, dust mass, modified blackbody temperatures and their associated uncertainties (16--84$^{\rm th}$ percentile range values). \\
{\bf Fig. A1.} A figure showing the observed photometry and best-fit {\sc
magphys} model SEDs for all 707 ALMA SMGs.\\
{\bf Fig. A2.} A figure showing the results of a {\sc magphys} analysis on EAGLE simulated galaxies, specifically stellar mass, SFR, dust mass, dust temperature and mass-weighted age.

\bsp	% typesetting comment
\label{lastpage}
\end{document}